\documentclass[11pt]{article}
\usepackage[utf8]{inputenc}
\usepackage{amsmath, amsfonts, amssymb, amsthm,  graphicx, mathtools, enumerate, float}
\usepackage[round]{natbib}
\usepackage[CJKbookmarks=true,
            bookmarksnumbered=true,  
			bookmarksopen=true,
			colorlinks=true,
			citecolor=blue,
			linkcolor=blue,
			anchorcolor=red,
			urlcolor=blue]{hyperref}
\usepackage{hyperref}
\usepackage{array}
\usepackage{thmtools}
\usepackage{thm-restate}
\usepackage{mathrsfs}
\usepackage{todonotes}
\usepackage{soul}

\newtheorem{remark}{Remark}[section]
\newtheorem{lemma}{Lemma}[section]
\definecolor{darkorange}{RGB}{211, 84, 0}

\usepackage{caption}
\usepackage{subcaption}

\usepackage[blocks, affil-it]{authblk}
\usepackage{thmtools,}
\usepackage{thm-restate}
\usepackage{mathrsfs}
\usepackage{todonotes}
\usepackage[letterpaper, left=1.2truein, right=1.2truein, top = 1.2truein, bottom = 1.2truein]{geometry}
\usepackage[ruled, vlined, lined, commentsnumbered]{algorithm2e}
\usepackage{booktabs}
\usepackage{ulem}
\renewcommand{\H}{\mathcal{H}}
\newcommand{\argmin}{\textnormal{argmin}}
\newcommand{\argmax}{\textnormal{argmax}}

\newcommand{\lfdr}{\textnormal{lfdr}}
\renewcommand{\P}{\mathbb{P}}
\newcommand{\E}{\mathbb{E}}
\newcommand{\bFDR}{\textnormal{bFDR}}
\newcommand{\FDR}{\textnormal{FDR}}

\usepackage{listings}

\newcommand{\ind}[1]{{\mathbf{1}\{#1\}}}
\newcommand{\range}[1]{[#1]}%{\llbracket #1\rrbracket}
\newcommand{\R}{\mathbb{R}}
\renewcommand{\E}{\ensuremath{\mathbb E}}

\renewcommand{\P}{\ensuremath{\mathbb P}}
\newcommand{\cH}{{\mathcal{H}}}

\newcommand{\mbf}{\mathbf}

\renewcommand{\FDR}{\mathrm{FDR}}

\renewcommand{\bFDR}{\mathrm{bFDR}}
\newcommand{\BH}{\mathrm{BH}}
\newcommand{\SL}{\mathrm{SL}}
\newcommand{\SLG}{\mathrm{SLG}}
\newcommand{\SLgap}{\mathrm{SLG}}
\newcommand{\SLC}{\mathrm{SLC}}
\newcommand{\SLCp}{\mathrm{SLC+}}
\newcommand{\SLCpp}{\mathrm{SLC\hspace{-1mm}+\hspace{-1mm}+}}
\newcommand{\ASLC}{\mathrm{ASLC}}
\newcommand{\ASLCp}{\mathrm{ASLC+}}
\newcommand{\ASLCpp}{\mathrm{ASLC\hspace{-1mm}+\hspace{-1mm}+}}
\newcommand{\ASL}{\mathrm{ASL}}

\def\EE{\mathbb{E}}

\def\PP{\mathbb{P}}

\def\calS{\mathcal{S}}

\def\1{\mathbbm{1}}

\renewcommand{\argmax}{\mathop{\mathrm{argmax}}}
\renewcommand{\argmin}{\mathop{\mathrm{argmin}}}

\newcommand{\msub}{s}%{\ell_0}%{\underline{m}}

\newtheorem{assumption}{Assumption}
\newtheorem{proposition}{Proposition}
\newtheorem{theorem}{Theorem}
\newtheorem{definition}{Definition}
\newtheorem{corollary}{Corollary}

\title{Reliable conformal novelty detection at the decision boundary}

\author[1]{Zijun Gao}
\author[2]{Etienne Roquain}
\author[3]{Daniel Xiang}
\affil[1]{University of Southern California}
\affil[2]{Sorbonne University}
\affil[3]{University of Chicago}
\date{\today}

\begin{document}

\maketitle

\begin{abstract}
Novelty detection via conformal $p$-values and BH procedure \citep{BH1995} provides distribution-free global false discovery rate (FDR) control \citep{bates2023testing,marandon2024adaptive}. We present here fundamental limits of this approach by showing that it does not produce reliable detection at the decision boundary. 
We study boundary false discovery rate (bFDR), the probability that the least extreme reported novelty is in fact a null observation. We first show that the support line (SL) procedure \citep{soloff2024edge}, controlling the bFDR in the continuous independent framework, fails to control the bFDR in the conformal case. 
We then present several modifications of the SL procedure that restore reliability at the decision boundary, by controlling the bFDR, with specific improvements in situations where many novelties are expected (adaptive procedures) and when the calibration sample is too small with respect to the test sample (subsampled procedures).
Numerical experiments with both synthetic and real data  support our  findings and show the relevance of the new proposed approach. 
\end{abstract}

\smallskip


\section{Motivation and contributions}\label{sec:intro}

Novelty detection is a classical problem in machine learning, aiming to
identify test examples that differ from a 'null' reference population
\citep{bishop1994novelty,pimentel2014review}. 
Standard methods \citep{tax1999support,scholkopf2001estimating} typically provide decision rules for identifying
novelties, but do not in general provide finite-sample guarantees for the false discoveries, which may be dangerous when incorrectly flagging an example as a novelty is costly for the practitioner. 

Conformal inference provides a flexible framework to provide distribution-free statistical error guarantees to any machine learning algorithms  \citep{vovk2005algorithmic,angelopoulos2021gentle}. In particular, combining the Benjamini--Hochberg (BH) procedure \citep{BH1995} with conformal $p$-values has been shown to guarantee rigorous control of the false discovery rate (FDR), that is, of the expected proportion of errors among the declared novelties \citep{bates2023testing,marandon2024adaptive}.
FDR is a criterion coming from large-scale multiple testing \citep{benjamini2010discovering}, which became very popular as it is highly interpretable and permissive in exploratory research where the user wants to identify most 'possible' signal by allowing some prescribed portion of errors.

\begin{figure}[tbp]
  \centering
    \includegraphics[clip, trim = 0cm 0cm 0cm 0cm, width = 1\textwidth]{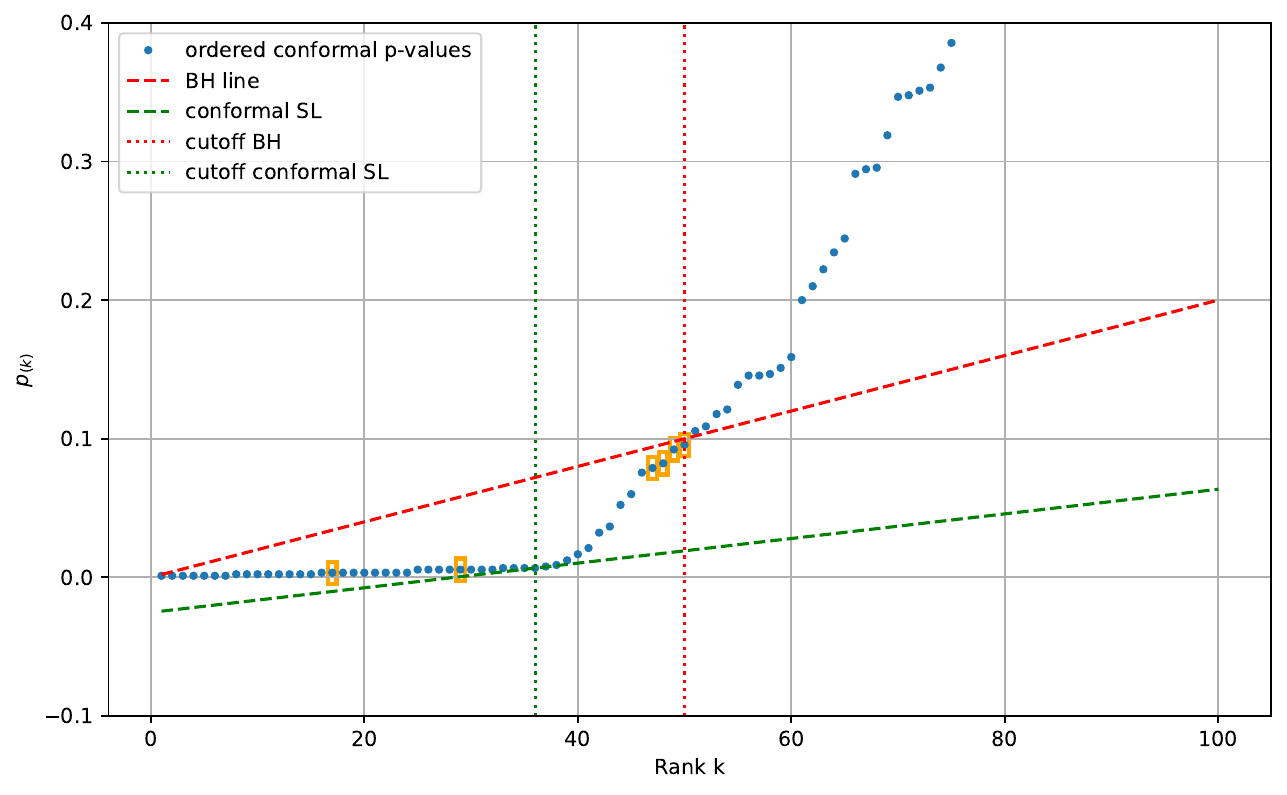}\\
    \includegraphics[clip, trim = 0cm 0cm 0cm 0cm, width = 0.8\textwidth]{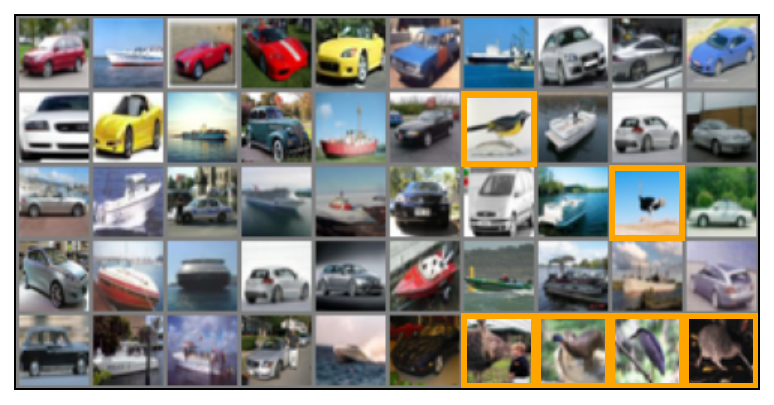}
  \caption{Limitation of the BH/FDR paradigm and the new conformal support line procedure: the null class (here `bird') is fairly separated from alternative classes (here `car' and `boat') and the BH procedure `adds' artificial nulls in the rejection set, likely near the boundary, to maintain an FDR close to the nominal level. The new suggested procedure (SLC) solves this issue by controlling the boundary FDR in a conformal setting. Top: illustration of BH and SLC procedures for one data generation of the \texttt{CIFAR-10} dataset with $n=899$, $m=100$, $m_0=50$ null labels `bird', $m_1=50$ labels `car' or `boat'. Level $\alpha=0.2$. Bottom: the images corresponding to the $50$ rejections of the BH procedure. Each false discovery is depicted by an orange rectangle in both displays. 
}
  \label{figillustration}
\end{figure}

However, controlling the FDR is not exempt from drawbacks and potential misinterpretations. Figure~\ref{figillustration} presents  the BH--novelty detection procedure in action with the \texttt{CIFAR-10} data set (\url{https://www.cs.toronto.edu/~kriz/cifar.html}), for which the null class is 'bird' while novelties are 'car' or 'boat'. On this example, while the proportion of false discovery $6/50=0.12$ remains below the target $\alpha=0.2$, it is clear that the procedure does not provide reliable novelties at the boundary. This phenomenon is not due to the data-generation process used here: Figure~\ref{fig:simulation.cifar10.threshold} in Section~\ref{sec:xp} shows in a similar setting that the whole distribution of the false discovery proportion is shifted around the BH--decision boundary. 
More generally, when inliers and novelties are well separated, instead of identifying only novelties, BH may also reject some true nulls while still satisfying the prescribed FDR level. This is not a failure of FDR control: 
rather, the numerous ``certain'' discoveries make additional false discoveries relatively inexpensive in terms of the FDR criterion. From a practical perspective, however, such ``free-riders'' provide no benefit and should be avoided: they enter the rejection set because the FDR can absorb them, rather than because their rejection provides meaningful evidence of novelty. 
Along the same lines, and perhaps more formally,  when $\alpha$ is equal to the proportion of null examples in the test sample ($\pi_0$), the FDR constraint becomes vacuous: the procedure  rejecting all hypotheses controls the FDR at level $\alpha=\pi_0$, regardless of the observed measurements.

To guard against free-riders near the decision boundary, one can instead aim to control the probability that the last rejection is a false discovery. This idea was developed by \cite{soloff2024edge}, who proposed the support-line (SL) procedure and proved that it controls this quantity at level $\pi_0\alpha$ under independent, continuously distributed $p$-values; we refer to it here as the boundary FDR (bFDR). The machine-learning/conformal framework raises a fundamental question: does the control hold for conformal $p$-values? We show that it does not. Conformal $p$-values are discrete and share the same calibration sample: this dependence and discreteness have a specific finite-sample effect at the rejection boundary. In particular, the standard SL procedure can violate its nominal bFDR guarantee for conformal $p$-values. We show that this failure can be quantified by a remainder term of order $m/(n+1)$, where $n$ is the calibration sample size and $m$ is the test sample size. 

In this work, we develop a family of boundary-aware procedures that explicitly accounts for the conformal framework. Our main contributions are as follows.
\begin{itemize}
    \item[(i)] {\it Characterization of bFDR failure.} We show that the bound $\operatorname{bFDR}(\mathrm{SL})\leq \pi_0\alpha$ does not hold in general for conformal p-values. Instead, we establish the upper bound  $\bFDR(\SL)\leq \pi_0 \alpha + \pi_0 m/(n+1)$, together with a nearly matching lower bound (Section~\ref{sec:bound});
   \item[(ii)] {\it Boundary-valid conformal procedures.} We introduce the Support Line Conformal ($\SLC$) that provides valid $\bFDR$ control at level $ \pi_0 \alpha$ (Section~\ref{secSLC}). We further develop an adaptive version, ($\ASLC$), that incorporates an estimator of the proportion of null observations $\pi_0$, and can reduce the conservativeness of $\SLC$ (Section~\ref{sec:adaptive});
   \item[(iii)] {\it Large test sets relative to the calibration sample.} When $m/(n+1)$ is not small, $\SLC$ is overly conservative or even prevent any rejection. We address this regime through random subsampling, leading to $\SLCp$ and $\ASLCp$, which both retain finite-sample bFDR guarantees while improving power in this regime. We also study stabilized multiple-subsampling variants $\SLCpp$ and $\ASLCpp$ and introduce a monotonicity assumption required for their guarantees     (Section~\ref{sec:supsampling});
    \item[(iv)] {\it Strengthen interpretation of bFDR and SLC}: under this monotonicity assumption, the bFDR \eqref{equbFDR} can be seen as a worst case, meaning that bFDR control not only provides a guarantee at the boundary but also for all individual items below the boundary in this case. Consequently, boundary-FDR control provides a form of individual reliability. We further show that SLC admits an empirical-Bayes interpretation through an isotonic estimate of the local false discovery rate  (Section~\ref{sec:monotonicity}).   
    \item[(v)] {\it Empirical evaluation.} We evaluate the proposed procedures on synthetic conformal scores and on CIFAR-10 novelty detection. The experiments illustrate the failure of SL, the role of the calibration-to-test ratio, and the power gains obtained by adaptive and subsampled procedures 
(Section~\ref{sec:xp}). 
\end{itemize}
{A summary of all procedures introduced in the paper is provided in Table~\ref{tab:method.summary}.  Our practical recommendation is to use SLC by default, with ASLC as an improvement if $\pi_{0}$ is expected to be not close to $1$. If the ratio $m/(n+1)$ is of larger order than $\alpha$, we recommend the subsampling variants (A)SLC+. Further stabilization can be obtained with (A)SLC++/(A)SLC++/2, but they should be used with care because their guarantee is more narrowed: the bFDR guarantee for (A)SLC++ is only empirical for small values of $\alpha$, while the one of (A)SLC++/2 requires a monotonicity assumption (Assumption~\ref{ass-scoredensity}).  

\begin{table}[h!]
\centering
\caption{
Summary of the proposed methods (recommended are in bold). The notation (A)X denotes the non-adaptive variant X when $C=1$ and its adaptive counterpart AX when $C=\hat{\pi}_0$, for the null proportion
estimator $\hat{\pi}_0$ defined in \eqref{equpi0hat}. 
The parameter $\alpha$ is the target bFDR level, the adjusted level is
the value used in place of $\alpha$ in the support line procedure, see \eqref{equkchap-conf} and 
$s\in[m]$ is the subsample size. 
A checkmark $\checkmark$ indicates finite-sample bFDR control under the assumptions
stated in the paper. The symbol $\approx$ indicates empirical control,
in particular for small values of $\alpha$. 
}
\begin{tabular}{lcccc}
\toprule
Method & Adjusted level & Subsampling & Control & Recommended use \\
\midrule
(A)SL       & $\alpha/C$ &
---      & $\times$      & Baseline \\
\midrule
{\bf (A)SLC}      & $\left(\alpha/C - m/(n+1)\right)_+$ &
---      & $\checkmark$  & Default \\
{\bf (A)SLC+}     & $\left(\alpha/C - s/(n+1)\right)_+$ &
Single   & $\checkmark$  & $m/(n+1)\gtrsim \alpha$ \\
\midrule
(A)SLC++    & $\left(\alpha/C - s/(n+1)\right)_+$ &
Multiple & $\approx$     & Stabilization \\
(A)SLC++/2  & $\left(\alpha/(2C) - s/(n+1)\right)_+$ &
Multiple & $\checkmark$ if \eqref{ass-scoredensity}  & Stab. with guarantee \\
\bottomrule
\end{tabular}
\label{tab:method.summary}
\end{table}

\section{Setting}\label{sec:setting}
We consider the 
``split'' (or ``inductive'') conformal setting \citep{papadopoulos2002inductive}, for which the user has at hand:
\begin{itemize}
     \item a ``training'' (or ``calibration'') sample $(X_{i})_{1\leq i\leq n}$ composed of $n$ null examples  which are random variables valued in $\mathcal{X}$ sharing a common distribution $P_{0}$, which is arbitrary and unknown;
    \item a ``test'' sample $(X_{n+i})_{1\leq i\leq m}$ composed of a mixture between null examples and novelties. Formally, the random variables $X_{n+i}$ are valued in $\mathcal{X}$ with $X_{n+i}\sim P_0$ for $i\in \cH_{0}$, where $\cH_{0}\subseteq \range{m}$ is a fixed index set corresponding to the ``nulls'';
    \item a score function $x \in \mathcal{X}\mapsto S(x)\in \R $ that measures how the example $x$ deviates from a null. %
    \end{itemize}
Above, the distribution of $X_{n+i}$ is left arbitrary for the novelties, that is, for the index  $i\in \cH_1:=\range{m}\backslash \cH_0$. We denote $m_0=|\cH_0|$ and $\pi_0=m_0/m$ (resp. $m_1=m-m_0$ and $\pi_1=m_1/m$) the number and proportion of nulls (resp. novelties).
The score function can be built using for instance, a one-class classifier \citep{bates2023testing} (only based on nulls), or a two-class classifier \citep{liang2022integrative,marandon2024adaptive} (based both on nulls and novelties). 
We assume that the score function has been built with independent data and thus is considered as fixed here. Here, larger score values are expected to provide stronger evidence for rejection. 
We consider the observed scores $S_j=S(X_j)$, $j\in \range{n+m}$, and will rely on a classical exchangeability assumption. 

\begin{assumption}\label{assglobal}
    \mbox{The scores $(S_{1},\dots,S_{n}, S_{n+i}, i\in \cH_{0})$ are exchangeable conditionally on}\\\mbox{ $(S_{n+i}, i\in \cH_{1})$. In addition, $(S_{j})_{j\in \range{n+m}}$ is a vector with no ties almost surely.}% \dx{(among finite scores)}.}
\end{assumption}
{Typically, Assumption~\ref{assglobal} is satisfied if all measurements $(X_{i})_{ i\in [n+m]}$ are independent and the $S_j$'s are continuous random variables.}
  Key quantities are the conformal $p$-values 
\begin{equation}
    \label{equ-conformalpvalues}
  p_{i} = (n+1)^{-1} \bigg(1+\sum_{j=1}^{n} \ind{S_{j}\geq S_{n+i}}\bigg),\:\:\: i\in \range{m}.
\end{equation}
Under Assumption~\ref{assglobal}, a simple exchangeability argument shows that the null conformal $p$-values are marginally super-uniform, that is,
\begin{equation}
    \label{equ-superunif}
    \forall i \in \cH_0,\:\: \forall t \in [0,1], \:\P(p_i\leq t)\leq t.
\end{equation}
The conformal $p$-values are nevertheless dependent, because they use the same calibration sample. This dependency structure can however be well described (see Lemma~\ref{lemmaPropConformalFull} in appendix) and turns out to be PRDS on $\cH_0$ (positively regressively dependent on each one of a subset, \citealp{BY2001}) \citep{bates2023testing,marandon2024adaptive}.
Hence, applying the \cite{BH1995} procedure to the conformal $p$-values \eqref{equ-conformalpvalues} allows to control the false discovery rate (FDR), which is the average proportion of false discoveries in the declared novelties. More formally, let us define
\begin{equation*}
    \FDR(R) = \E \Bigg(\frac{\sum_{i\in R} (1-H_i)  }{1\vee |R|}\Bigg),
\end{equation*}
where $H_i:=\ind{i\in \cH_1}$, $i\in [m]$, corresponds to the indices of  novelties and $R\subseteq [m]$ is the novelty detection procedure, identified  as the subset of indices $i$ such that $X_{n+i}$ is declared as being a novelty by the procedure $R$.
The BH procedure works by sorting the conformal $p$-values in increasing order. Since they all belong to the grid $\ell/(n+1)$, $\ell\in [n+1]$, there can be ties in the conformal $p$-values set, and this should be handled in a specific way. For this, let us consider the permutation $\sigma$ (unique a.s. from Assumption~\ref{assglobal})  of $[m]$ such that 
$
S_{\sigma(1)}> S_{\sigma(2)} > \dots > S_{\sigma(m)}
$
(almost surely) are the ordered values of $(S_{n+i}, i\in [m])$, 
so that 
$
p_{\sigma(1)}\leq p_{\sigma(2)} \leq \dots \leq p_{\sigma(m)}.
$
The BH procedure applied on conformal $p$-values works by considering $\hat{k}:=\max\{k\in \{0,1,\dots,m\}\::\:p_{\sigma(k)}\leq \alpha k/m\}$ and by rejecting $\BH=\{i\in [m]\::\: p_i\leq \alpha\hat{k}/m \}$. By \cite{bates2023testing}, the following result holds
\begin{equation}
    \label{equFDR-control}
    \FDR(\BH) \leq (m_0/m) \alpha,
\end{equation}
for any distribution of nulls and novelty and any sample sizes $n,m\geq 1$, provided that Assumption~\ref{assglobal} is satisfied. {In addition, \eqref{equFDR-control} is an equality if $(n+1)\alpha/m$ is a positive integer \citep{marandon2024adaptive}.}

In the case of independent, continuous (non-conformal) $p$-values, \cite{soloff2024edge,xiang2025frequentist}  introduced the {\it boundary FDR} (bFDR), defined for any top-$\hat{k}$ procedure $R=\{i\in [m]\::\: S_{n+i}\geq S_{\sigma(\hat{k})}\}$ as 
\begin{equation}
    \label{equbFDR}
    \bFDR(R) = \P ( H_{\sigma(\hat{k})}=0  ).
\end{equation}
Above, we set $S_0=\infty$, $\sigma(0) := 0$ and $H_{0}:=1$ to handle the case where $\hat{k}=0$ (no rejection).
We emphasize that the vector $(H_i)_{i\in [m]}$ is deterministic in the bFDR criterion \eqref{equbFDR}: the randomness is only due to the index $\sigma(\hat{k})$ of the last rejection. Hence, as is the case with FDR, the bFDR is a purely frequentist criterion. 
To control this criterion, \cite{soloff2024edge} introduced the support line (SL) procedure, which at level $\alpha\in (0,1)$ corresponds to the choice
\begin{equation}\label{equkchap}
\hat{k} =  \argmin_{{k} \in [0,m]} \{ p_{\sigma(k)} - \alpha k/m\},
\end{equation}
with the convention $p_{\sigma(0)} := 0$ and which is well defined for $p$-values that have no ties. When the $p$-values are independent and uniform under the null, they show that $\bFDR(\SL)= \pi_0\alpha$.

\section{Main results}

In this section, we show that the SL procedure fails in the conformal setting and introduce our novel procedures SLC and ASLC. 

\subsection{Bounds for SL procedure}\label{sec:bound}

First note that, in the conformal setting, the $\argmin$ in \eqref{equkchap} might be undefined, because there can be $k\neq k'$ such that $p_{\sigma(k)} - \alpha k/m=p_{\sigma(k')} - \alpha k'/m$ (when $\alpha(n+1)/m$ is an integer). Hence, we extend the SL threshold definition \eqref{equkchap} by considering the maximum of the elements of the $\argmin$:
\begin{equation}\label{equkchap-conf}
\hat{k} = \max \bigg\{\argmin_{{k} \in [0,m]} \{ p_{\sigma(k)} - \alpha k/m\}\bigg\},
\end{equation}
with the convention $p_{\sigma(0)} := 0$. 

Note that  $\bFDR(\SL) \leq \alpha m_{0}/m=\alpha$ is true when $m_0=m$ by the Simes inequality, which is correct in this case by using the FDR control proved in \cite{bates2023testing} (under the full null).
However, the following negative result shows that the bFDR control does not hold in general when $m_0<m$.

\begin{proposition}\label{prop:counterex}
Assume $m_0<m$ and consider a level $\alpha\in (0,1)$ with $\alpha \ge 1 / \{(n+1)(1-m_0/m)\}$. Then there exists a data generation process satisfying Assumption~\ref{assglobal} so that $\bFDR(\SL) \geq  m_{0}/(m_0 + n)$. 
\end{proposition}
The proof is given in Section~\ref{sec:proofprop:counterex}.  This lower bound exceeds $\alpha m_0 / m$ whenever $\alpha < m/(m_0 + n)$. Hence, for $1 / \{(n+1)(1-m_0/m)\}\leq \alpha < m/(m_0 + n)$, the inequality $\bFDR(\SL) \leq \alpha m_{0}/m$ does not hold in general. 
Figure~\ref{fig:counter.example} demonstrates a specific counter example.

\begin{remark}
To achieve the lower bound, we consider in the proof the situation where the scores under the alternatives are  larger than those under the null (almost surely). It is an analogue to the so-called ``Dirac-Uniform'' configuration \citep{finner2009false} in the conformal setting.
\end{remark}

Our second result fills the gap put forward in Proposition~\ref{prop:counterex} by proposing a valid bound.

\begin{theorem}\label{th:bFDR}
For any $\alpha\in (0,1)$, under Assumption~\ref{assglobal},  the SL procedure at level $\alpha\in (0,1)$ corresponding to \eqref{equkchap-conf} is such that
\begin{align}
\label{rel:bFDR2}
 \mathrm{bFDR}(\mathrm{SL}) &
 \leq  \frac{\alpha m_{0}}{m} + \frac{m_{0}}{n+1}.
\end{align}
\end{theorem}

The proof is provided in Section~\ref{sec:proofth:bFDR}.

\begin{remark}\label{rem:improvinglb}
Proposition~\ref{prop:lower-bound} in Section~\ref{sec:morelbforub} shows that the upper bound \eqref{rel:bFDR2} is tight when $m_{0}=1$ and $(n+1)\alpha/m$ is an integer. Proposition~\ref{prop:lower-bound}  also presents the improved bFDR lower bound  $1-\prod_{j=0}^{\lfloor(n+1)\alpha/m\rfloor} \frac{n-j}{m_0+n-j}$, which is shown to closely match the upper bound \eqref{rel:bFDR2} when $\alpha\pi_{0}$ is small (see Remark~\ref{rem:match} therein).
\end{remark}

\begin{remark}\label{rem:randomized}
In the independent case, considering discrete $p$-values can already cause bFDR control violation, see Section~\ref{sec:violationindep}. Hence, one might legitimately ask whether the bFDR violation put forward in the conformal case is only due to the discrete nature of the conformal $p$-values, or if their dependence structure plays a role.  It turns out that dependence matters: a version of the  SL procedure used with randomized $p$-values, which are continuous by nature, still violates the bFDR control, see Section~\ref{sec:randpvalues}.
\end{remark}

\subsection{SLC procedure}\label{secSLC}

To control the bFDR at level $\alpha$, Theorem~\ref{th:bFDR} suggests the following modification of~SL.
\begin{definition}\label{def:SLC}
   The support line conformal $(\SLC)$ procedure  at level $\alpha\in (0,1)$ is defined as making the same discoveries as the $\SL$ procedure taken at level 
    $\alpha-m/(n+1)>0$ if $1/(n+1)< \alpha/m$, and no discovery otherwise.
 \end{definition}

 In other words, the $\SLC$ procedure corresponds to the top-$\hat{k}$ procedure with
 \begin{equation*}
\hat{k} = \max \bigg\{\argmin_{{k} \in [0,m]} \left\{ p_{\sigma(k)} - k(\alpha/m - 1/(n+1))_+ \right\}\bigg\},
\end{equation*}
with the convention $p_{\sigma(0)} := 0$. 

\begin{corollary}\label{cor:bFDR}
   For any $\alpha\in (0,1)$, under Assumption~\ref{assglobal},  the $\SLC$ procedure at level $\alpha\in (0,1)$ (Definition~\ref{def:SLC}) is such that
$
 \bFDR(\SLC) \leq \alpha m_{0}/m.
$
\end{corollary}

This follows directly from Theorem~\ref{th:bFDR} because if $\alpha>m/(n+1)$, then $ \bFDR(\SLC)\leq (m_0/m) (\alpha - m/(n+1)) + (m_0/m) m/(n+1) = (m_0/m) \alpha$. 

{An illustration of the $\SLC$ procedure is provided in Figure~\ref{fig:ecdf-maj}. Interestingly, we also show that the SLC procedure corresponds to an empirical Bayes procedure that uses an isotonic estimator of the local FDR, see Proposition~\ref{prop:iso-gren-equivalence} in Section~\ref{sec:bayes}.}

Let us also mention that we introduce another procedure controlling the bFDR in Section~\ref{secSL3} (SLG). Since it is generally less efficient than $\SLC$, we have postponed it in the appendix.

\subsection{Adaptive SLC procedure}\label{sec:adaptive}

The SLC procedure  controls the bFDR at level $\pi_0 \alpha$, not $\alpha$, and the additional factor can introduce conservativeness. To solve this issue, we present here an adaptive variant that maintains the desired bFDR control. We develop here a Storey-type procedure that estimates $\pi_0$ by
\begin{equation}
    \label{equpi0hat}
    \hat{\pi}_0 = \frac{1+\sum_{i=1}^m \ind{p_i\geq (s_0+1)/(n+1)}}{m(1-(s_0+1)/(n+1))}
\end{equation}
for some parameter $s_0\in [0,n-1]$. 
The adaptive SL procedure at level $\alpha$ (and with a $s_0$-Storey estimator \cite{Storey2003}) is defined by
\begin{equation}\label{equkchap-conf-adapt}
\hat{k} = \max \bigg\{\argmin_{{k} \in [0,m]: p_{\sigma(k)}\leq s_0/(n+1)} \left\{ p_{\sigma(k)} - \frac{\alpha k}{m \hat{\pi}_0}\right\}\bigg\}.
\end{equation}
Compared to \eqref{equkchap-conf} note that $\hat{k}$ in \eqref{equkchap-conf-adapt} has the advantage of using $ \hat{\pi}_0$ as an estimate of $\pi_0$, which is potentially much smaller than $1$ and is likely to make $\hat{k}$ larger. 
The counterpart, however, is that we need to cap $k$ by imposing $p_{\sigma(k)}\leq s_0/(n+1)$.

\begin{theorem}\label{th:bFDR-adapt}
For any $\alpha\in (0,1)$, under Assumption~\ref{assglobal},  the adaptive SL procedure at level $\alpha\in (0,1)$ and using any $s_0\in [0,n-1]$, denoted by ASL and corresponding to \eqref{equkchap-conf-adapt} is such that
\begin{align}
 \mathrm{bFDR}(\ASL) &
 \leq  \alpha + \frac{m_{0}}{n+1}\label{rel:bFDR2-adapt}.
\end{align}
\end{theorem}

The proof is provided in Section~\ref{sec:proofth:bFDR-adapt}. It motivates the introduction of the following procedure.

\begin{definition}\label{def:ASLC}
   The adaptive support line conformal $(\ASLC)$ procedure  at level $\alpha \in~\hspace{-0.5em}(0,1)$ and using any $s_0\in [0,n-1]$, is defined as making the same discoveries as $\SL$ procedure  at level 
    $\alpha/\hat{\pi}_0-m/(n+1)>0$ if $\hat{\pi}_0/(n+1)< \alpha/m$ and no discovery otherwise.
 \end{definition}

 In other words, the $\ASLC$ procedure corresponds to the top-$\hat{k}$ procedure with
{ \begin{equation*}
\hat{k} = \max \bigg\{\argmin_{{k} \in [0,m]: p_{\sigma(k)}\leq s_0/(n+1)} \{ p_{\sigma(k)} - k(\alpha/(m\hat{\pi}_0) - 1/(n+1))_+ \}\bigg\},
\end{equation*}
}
with the convention $p_{\sigma(0)} := 0$.
Note that when $\hat{\pi}_0\leq 1$ (which is generally the case), ASLC is slightly more powerful than the procedure $\ASL$ taken at level $\alpha-m/(n+1)$. 

\begin{theorem}\label{cor:bFDR-adapt}
   For any $\alpha\in (0,1)$, under Assumption~\ref{assglobal},  the ASLC procedure at level $\alpha\in (0,1)$ using any $s_0\in [0,n-1]$ (Definition~\ref{def:ASLC}) is such that
$
 \bFDR(\ASLC) \leq \alpha .
$    
\end{theorem}

Theorem~\ref{cor:bFDR-adapt} is proved in Section~\ref{sec:proofcor:bFDR-adapt}.

\section{Enhancing power via subsampling under limited calibration data}\label{sec:supsampling}

If $\alpha/m < 1/(n+1)$ (as for $m=n+1$ for instance), the $\SLC$ procedure makes no rejection and thus has no power. 
Since $\alpha$ is typically pre-fixed, meeting the condition $\alpha/m \geq 1/(n+1)$ requires either to  enlarge $n$ by obtaining additional calibration data, or to reduce $m$ the number of test data. This section deals with the case where the first solution is impractical and examines the reduction of $m$ via subsampling.

\subsection{Single subsampling}

\begin{definition}
    \label{def:subsampling}
    The single-subsampled version $\SLCp$ of the $\SLC$ procedure 
    at level $\alpha\in (0,1)$ and with subsample size $\msub\in [m]$  
    is obtained as follows:
    \begin{itemize}
        \item Draw a subsample $\calS$ of \(|\calS|=\msub\) units uniformly randomly without replacement from the test scores $(S_{n+i},i\in [m])$;
        \item Apply the $\SLC$ procedure at level $\alpha$ in restriction to \(\calS\), that is, consider
    \begin{equation}\label{equkchap-conf-subsample}
         \hat{k}_{\calS} = \max \bigg\{\argmin_{0\leq k\leq s} \big\{ p_{\sigma_{\calS}(k)} - k\big(\alpha/s -1/(n+1)\big)_+ \big\}\bigg\},
        \end{equation}
        where $\sigma_{\calS}$ is the unique permutation of $\calS$ ordering the test scores $(S_{n+i},i\in \calS)$ in decreasing order ($\sigma_{\calS}(0)=0$); 
        \item Compute the threshold $S_{\sigma_{\calS}(\hat{k}_{\calS})}$ ($S_0=+\infty$);
        \item Finally reject $\SLCp=\{i\in [m]\::\: S_{n+i}\geq S_{\sigma_{\calS}(\hat{k}_{\calS})}\}$  (make no rejection if $\hat{k}_{\calS}=0$).
    \end{itemize}
\end{definition}
While subsampling randomizes the procedure, the $\SLCp$ procedure can only make more rejections than the original procedures when $1/(n+1)> \alpha/m$, because $\SLC$ does not make any rejections 
in that case. The next result shows that $\SLCp$ maintains bFDR control.

\begin{theorem}\label{th-subsampling}
    Under Assumption~\ref{assglobal}, the single-subsampled version $\SLCp$ of the $\SLC$ procedure at level $\alpha\in (0,1)$ (Definition~\ref{def:subsampling}) is such that 
    $\bFDR(\SLCp) \leq \alpha m_{0}/m$,
    where the probability appearing in the bFDR expression is also taken marginally with respect to the subsampling process.
\end{theorem}

The proof is provided in Section~\ref{sec:proof:sub}.

\subsection{Multiple subsampling}

While the performance of SLC is enhanced by single subsampling, SLC+ is randomized so may be unstable in practice: the result might change from a subsample generation to another. We solve this issue here by proposing multiple subsampling, which stabilizes the methods.

\begin{definition}
    \label{def:multisubsampling}
    The multiple-subsampled version $\SLCpp$ of $\SLC$  
    at level $\alpha\in (0,1)$ and with subsample size $\msub\in [m]$  
    is obtained as follows:
  \begin{itemize}
        \item Draw $B$ subsamples $\calS_1,\dots,\calS_B$ of \(|\calS|=\msub\) uunits independently and uniformly at random without replacement from the test scores $(S_{n+i},i\in [m])$;
        \item For each subsample $\calS_b$, $b\in [B]$, get $R_b$ the rejection set of the single-subsampled  $\SLCp$ method that uses $\calS_b$ as the subsample (Definition~\ref{def:subsampling}) and let $r_b=|R_b|$, $b\in [B]$;
        \item Consider 
        $r_{(\lceil B/2\rceil)}$ the empirical median of $(r_b,b\in [B])$,  for $r_{(1)}\geq \dots \geq r_{(B)}$  the ordered $(r_{b},b\in [B])$;
        \item Finally reject $\SLCpp=\{i\in [m]\::\: S_{n+i}\geq S_{\sigma(r_{(\lceil B/2\rceil)})}\}$ (make no rejection if  $r_{(\lceil B/2\rceil)}=0$).
    \end{itemize}
\end{definition}

To obtain a bFDR controlling result for the multiple-sampled method, we strengthen Assumption~\ref{assglobal}.

\begin{assumption}\label{ass-scoredensity}
    \mbox{$(S_{1},\dots,S_{n}, S_{n+i}, i\in \cH_{0})$ are iid with common density\footnote{These densities are taken with respect to the Lebesgue measure on $\R$.} $f_0$ }\\\mbox{and are independent of $(S_{n+i}, i\in \cH_{1})$, which  contains independent variables }\\
    \mbox{with each $S_{n+i}$ having a density $f_i$.}
    \mbox{In addition, for each $i\in \cH_{1}$, the ratio $f_i/f_0$}\\
     \mbox{is nondecreasing, in the sense that for all $x\leq y$, $f_i(x)f_0(y)\leq f_i(y)f_0(x)$.}
\end{assumption}

Assumption~\ref{ass-scoredensity} is a monotonicity property indicating that alternative scores are larger than the null scores (in a stochastic sense). While it reinforces the interest in the bFDR criterion (as discussed in Section~\ref{sec:monotonicity}), it also justifies the use of the multiple-sampled method, as shown by the following result.

\begin{theorem}\label{th-multisubsampling}
    Under Assumption~\ref{ass-scoredensity}, the multi-subsampled version $\SLCpp$ of $\SLC$ at level $\alpha\in (0,1)$ (Definition~\ref{def:multisubsampling}) is such that 
    $\bFDR(\SLCpp) \leq 2\alpha m_{0}/m$, 
    where the probability appearing in the bFDR expression is also taken marginally with respect to the subsampling process.
\end{theorem}

The proof is provided in Section~\ref{sec:proof:multisub}. It relies on a monotonicity property given in Theorem~\ref{th-increasing}
 below. The factor $2$ in the bound is not an artifact of the proof, as it is needed in specific conformal settings (see Section~\ref{sec:addxp}). This motivates the introduction of the procedure $\SLCpp/2$, which simply is $\SLCpp$ taken at level $\alpha/2$ and which controls the bFDR at level $\alpha$  by Theorem~\ref{th-multisubsampling}. However, we observed that it can be avoided in most of the settings considered in our numerical experiments (particularly when $\alpha$ is not too large).

\begin{remark}
If instead of the empirical median, we consider $r_{(\lceil \gamma B\rceil)}$ the empirical $\gamma$-quantile of $(r_b,b\in [B])$ in Definition~\ref{def:multisubsampling}, then Theorem~\ref{th-multisubsampling} holds with the bound $\alpha m_{0}/(\gamma m)$. 
\end{remark}

\subsection{Subsampling for the adaptive variant}

We introduce the single-subsampled variant of the adaptive method $\ASLC$ as follows:
\begin{definition}
    \label{def:subsampling-adapt}
    The single-subsampled version $\ASLCp$ of $\ASLC$ 
    at level $\alpha\in (0,1)$ with subsample size $\msub\in [m]$ 
    is obtained as follows:
    \begin{itemize}
        \item Compute $\hat{\pi}_0$ on the whole sample as in \eqref{equpi0hat} with $s_0\in [0,n-1]$; 
        \item Draw a subsample $\calS$ of \(|\calS|=\msub\) units uniformly at random without replacement from the test scores $(S_{n+i},i\in [m])$;
        \item Compute the $\ASLC$ procedure in restriction to \(\calS\), that is,
          \begin{equation}\label{equkchap-conf-adapt-subsample}
         \hat{k}_{\calS} = \max \bigg\{\argmin_{0\leq k\leq s: p_{\sigma_{\calS}(k)}\leq s_0/(n+1)} \big\{ p_{\sigma_{\calS}(k)} - k\big(\alpha/(\hat{\pi}_0s) -1/(n+1)\big)_+ \big\}\bigg\},
        \end{equation}
        where $\sigma_{\calS}$ is the unique permutation of $\calS$ ordering the test scores $(S_{n+i},i\in \calS)$ in decreasing order ($\sigma_{\calS}(0)=0$);
     \item Compute the threshold $S_{\sigma_{\calS}(\hat{k}_{\calS})}$ ($S_0=+\infty$);
        \item Finally reject $\ASLCp=\{i\in [m]\::\: S_{n+i}\geq S_{\sigma_{\calS}(\hat{k}_{\calS})}\}$ (make no rejection if $\hat{k}_{\calS}=0$).
    \end{itemize}
\end{definition}

The multiple-subsampled, stabilized, version is as follows:

\begin{definition}
    \label{def:multisubsampling-adapt}
    The multiple-subsampled version $\ASLCpp$ of $\ASLC$ at level $\alpha\in (0,1)$  with subsample size $\msub\in [m]$ 
    is obtained as follows:
  \begin{itemize}
        \item Draw $B$ subsamples $\calS_1,\dots,\calS_B$ of \(|\calS|=\msub\) units independently and  uniformly randomly without replacement from the test scores $(S_{n+i},i\in [m])$;
        \item For each subsample $\calS_b$, $b\in [B]$, get $R_b$ the rejection set of the single-subsampled  $\ASLCp$ method that uses $\calS_b$ as the subsample (Definition~\ref{def:subsampling-adapt}) and let $r_b=|R_b|$, $b\in [B]$;
        \item Consider 
        $r_{(\lceil B/2\rceil)}$ the empirical median of $(r_b,b\in [B])$, for $r_{(1)}\geq \dots \geq r_{(B)}$  the ordered $(r_{b},b\in [B])$;
        \item Finally reject $\ASLCpp=\{i\in [m]\::\: S_{n+i}\geq S_{\sigma(r_{(\lceil B/2\rceil)})}\}$ (make no rejection if $r_{(\lceil B/2\rceil)}=0$).
    \end{itemize}
\end{definition}

\begin{theorem}\label{th-subsampling-adapt}
    Under Assumption~\ref{assglobal}, the single-subsampled version $\ASLCp$ of $\ASLC$  at level $\alpha\in (0,1)$ (Definition~\ref{def:subsampling-adapt}) is such that 
    $\bFDR(\ASLCp) \leq \alpha .$
    Furthermore, under Assumption~\ref{ass-scoredensity}, the multi-subsampled version $\ASLCpp$ of $\ASLC$  at level $\alpha\in (0,1)$ (Definition~\ref{def:multisubsampling-adapt}) is such that 
    $\bFDR(\ASLCpp) \leq 2\alpha .$
\end{theorem}

The proof is provided in Section~\ref{sec:proofth-subsampling-adapt}. {The comments made for the non-adaptive subsampling variant $\SLCpp$ also apply for $\ASLCpp$: we introduce the procedure $\ASLCpp/2$ corresponding to $\ASLCpp$ taken at level $\alpha/2$, which provably controls the bFDR at level $\alpha$ by Theorem~\ref{th-subsampling-adapt}. Nevertheless, the procedure $\ASLCpp$ essentially provides this control in most practical cases (for a small $\alpha$).}

\begin{remark}[Hyperparameter choice]
These subsampling methods depend on the hyperparameters $(B,s)$ which should be widely used for a practical use. In our experiments, we observed that the results do not depend critically on $B$, provided that it is chosen large enough (e.g., $B\geq 1000$). The choice of $s$ is more important and we recommend the subsample size $s=\max\{\underline{s},\min\{m, \lfloor \rho\alpha(n+1)\rfloor\}\}$, for some subsampling ratio $\rho < 1$ (e.g., $\rho = 1/5$) and minimal subsample size $\underline{s}$ (e.g., $\underline{s}=100$). It offers stability and performance. Obviously, these choices are not optimal in all cases, since the perfect recommendation would depend on the alternative distribution. A systematic empirical study of the stability of the subsampled variants with respect to the choice of $s$ is provided in Section~\ref{sec:addxp.subsampling.proportion}.
\end{remark}

\section{Interpretation under monotonicity}\label{sec:monotonicity}

Returning to a basic principle in statistical hypothesis testing, individual rejections should be justifiable based on the strength of evidence observed for each hypothesis.
Multiple testing procedures that summarize the type I error of their discovery set in an average-case sense
are susceptible to free-riders (Section \ref{sec:intro}) so they do not necessarily follow this principle and can violate it severely when the signal strength is large. In the BH procedure for example, strong signals produce very small $p$-values that occupy the leading ranks, pushing nulls to larger ranks $k$ where the comparison with $\frac{\alpha k}{m}$ is more lenient. Still, FDR-based procedures are widely used in practice and come with clean finite-sample guarantees at the aggregate level. 
In this section, we show that under monotonicity (Assumption \ref{ass-scoredensity}), the conformal support line procedures also control FDR. In addition, monotonicity ensures that rejections made by the SLC procedure (and of the other variants) can be individually justified.

\subsection{bFDR monotonicity property and FDR control}

Assumption \ref{ass-scoredensity} encodes the intuition that larger scores represent more evidence against the null. In particular, each $p$-value associated with a false null hypothesis has a non-increasing probability mass function. 
Under this assumption, non-null $p$-values tend to be smaller, so that larger $p$-values within the rejection region correspond to less promising discoveries. As a result, the boundary FDR summarizes the false discovery rate of the least promising discovery. 
Control over the bFDR thus implies control over all rejections. This result is recorded below and proved in Section~\ref{secproof:th-increasing}. 
\begin{theorem}\label{th-increasing}
    Under Assumption~\ref{ass-scoredensity}, we have that the function 
  $k\in [m] \mapsto \P(H_{\sigma(k)}=0 \mid (S_j)_{j\in [n]},S_{\sigma(\cdot)})$ is nondecreasing (almost surely), where $S_{\sigma()}=(S_{\sigma(1)},\dots, S_{\sigma(m)})$ are the values $(S_{n+i},i\in [m])$ ordered decreasingly. Consequently, any top-$\hat{k}$ procedure $R=\{i\in [m]\::\: S_{n+i}\geq S_{\sigma(\hat{k})}\}$ is such that 
    \begin{align}\label{th-bFDRLFC}
    \bFDR(R)=\E \Big[ \max_{0\leq k \leq |R|} \P(H_{\sigma(k)}=0 \mid (S_{j})_{j\in[n]},S_{\sigma(\cdot)}) \Big] .
\end{align}
\end{theorem}
The right hand side of the above expression is analogous to the max-lfdr error criterion introduced by \cite{soloff2024edge}, and quantifies the false discovery probability of a top-$\hat{k}$ procedure's least promising discovery.  As a corollary, all bFDR controlling procedures discussed here also control the right hand side above, thus providing individual control across all rejections (in expectation).

A byproduct of  Theorem \ref{th-increasing}, Corollary~\ref{cor:bFDR} and Theorem~\ref{cor:bFDR-adapt} is as follows (proof provided in Section~\ref{proofcor-bFDRimpliesFDR}).

\begin{corollary}\label{cor-bFDRimpliesFDR}
    Under Assumption~\ref{ass-scoredensity}, any top-$\hat{k}$ procedure $R=\{i\in [m]\::\: S_{n+i}\geq S_{\sigma(\hat{k})}\}$ is such that $\FDR(R) \leq \bFDR(R)$.
In particular, we have $\FDR(\SLC)\leq \alpha m_0/m$ and $\FDR(\ASLC) \leq~\alpha$.
\end{corollary}
Although Corollary \ref{cor-bFDRimpliesFDR} implies FDR is controlled by the tuning parameter $\alpha$ of the bFDR procedure ($\SLC$ or $\ASLC$), the actual FDR is typically strictly less than $\alpha$. The relationship is analogous to that between tail and local fdr \cite{efron2005local} of a two-group model, except that $\FDR$ and $\bFDR$ relate specifically to a multiple testing procedure rather than a Bayesian model; see Section 5.1 of \cite{xiang2025frequentist} for more discussions on this point (in the independent case).

\subsection{An empirical Bayes interpretation of SLC}\label{sec:bayes}

Our aim in this section is to present the $\SLC$ procedure as a thresholding rule, rejecting the $i$-th null whenever $\widehat{\lfdr}(p_i) \leq \alpha$ for some natural estimator $\widehat{\lfdr}$ of the local false discovery rate, defined as the probability that $H_i=0$  
given its conformal $p$-value (see \eqref{eq:local-fdr} below). While we have assumed $H_1,\dots,H_m$ to be fixed throughout, for this section only, suppose that the hypothesis indicators and the conformity scores are drawn independently from a two-groups model:
\begin{align*}
    H_i &\sim \text{Bernoulli}(1-\pi_0) \\
    S_i \mid H_i &\sim \begin{cases}
        f_0 \hspace{1em} &\text{if } H_i=0 \\
        f_1 &\text{if } H_i=1,
    \end{cases}
\end{align*}
so that marginally $S_i \sim \pi_0 f_0 + (1-\pi_0) f_1$ for $i=n+1,\dots,n+m$, while $S_1,\dots,S_n \stackrel{\text{iid}}{\sim} f_0$. 

Assumption~\ref{ass-scoredensity} in this setting amounts to the assumption that the density ratio $f_1/f_0$ is nondecreasing\footnote{i.e.~for all $x \leq y$, we have $f_1(x)f_0(y) \leq f_1(y) f_0(x)$}; that is, larger scores tend to be associated with scores from the $f_1$ component, and smaller ones with $f_0$. It is not needed formally (e.g., for the proposition below), but it motivates the lfdr estimate that we will consider.
For conformal $p$-values taking values on the grid $\mathcal{G}_n:=\{\frac{1}{n+1},\frac{2}{n+1},\dots,1\}$, the local false discovery rate (lfdr, \cite{efron2001empirical}) is defined as
\begin{align}
\label{eq:local-fdr}
    \lfdr(t) := \P(H_i = 0 \mid p_i = t) = \frac{\pi_0 \; \ind{t \in \mathcal{G}_n}/(n+1)}{g_n(t)},
\end{align}
where $g_n(t) = \P(p_{n+1}=t)$ is the marginal probability mass function of the $p$-values in the iid two groups model, and is supported on the grid $\mathcal{G}_n$. When the number $m$ of test scores  is large relative to the number  $n$ of calibration scores, $\lfdr(t)$ is roughly the proportion of $p$-values equal to $t$ that correspond to true nulls. 
\begin{figure}[t]
    \centering
    \includegraphics[width=\linewidth]{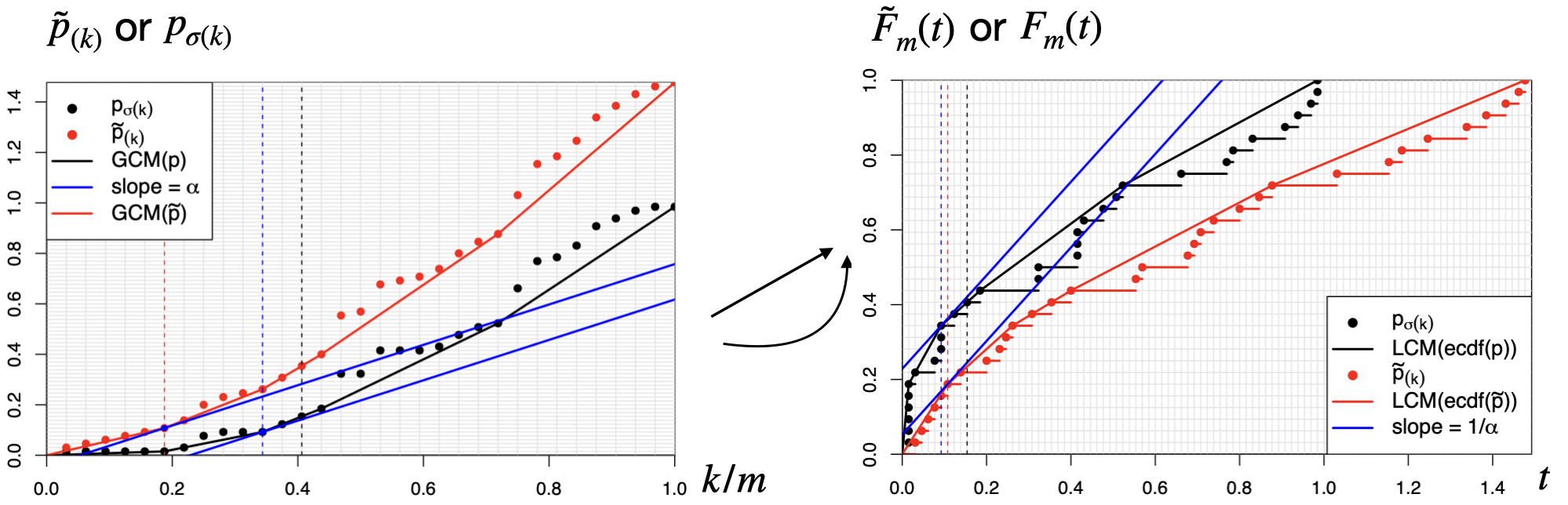}
    \caption{Left: Greatest Convex Minorant (GCM) of $p$-value order statistics (black) and shifted sorted $p$-values $\tilde{p}_{(k)}\coloneqq p_{\sigma(k)}+\frac{k}{n+1}$ (red). Right: Least Concave Majorant (LCM) of $F_m$ (black) and $\tilde{F}_m$ (red). SL makes 11 rejections, while SLC makes 6 rejections, as determined by the points of tangency between the line of slope $\alpha$ (or slope $1/\alpha$) and the GCM (or LCM) of the $p$-values. Here, $m=32,n=64,m_0=16,\alpha=0.8$, $f_1= \text{Uniform}(0.5,1.5)$, and $f_0=\text{Uniform}(0,1)$. Vertical dashed lines indicate the BH$(\alpha/2)$, $\SL(\alpha)$, and $\SLC(\alpha)$ cutoffs, with $13$ (black), $11$ (blue), and $6$ (red) rejections, respectively.
    }
    \label{fig:ecdf-maj}
\end{figure}
Since the null distribution is uniform on $\mathcal{G}_n$, a rough (conservative) proxy for this ratio within a bin $[p_{\sigma(k-1)},p_{\sigma(k)})$ is given by:
\begin{align}
\label{eq:lfdr-raw}
    \widehat{\lfdr}_{\text{raw}}(k) = \frac{(p_{\sigma(k)}-p_{\sigma(k-1)})+1/(n+1)}{(1/m)}, \hspace{1em} k\in [m].
\end{align}
As $n \to \infty$, the above tends to the formula in the continuous case, following from ideas in \cite{soloff2024edge}. To see why the additional term $1/(n+1)$ arises in the conformal setting, consider the shifted $p$-values,
\begin{align*}
    \tilde{p}_{(k)} \coloneqq p_{\sigma(k)} + \frac{k}{n+1}, \hspace{1em} k\in [0,m].
\end{align*}
The shifted $p$-values $\tilde{p}_{(1)}<\dots<\tilde{p}_{(m)}$ ensure that the (strict) ordering induced by the test scores $S_{\sigma(1)} > \dots > S_{\sigma(m)}$ is preserved. This shift resolves the ambiguity due to ties between conformal $p$-values, while maintaining that each $p$-value is a point in the grid of positive integer multiples of $1/(n+1)$. 
Applying the standard formula to the shifted $p$-values gives the equivalent expression $\widehat{\lfdr}_{\text{raw}}(k) = m(\tilde{p}_{(k)}-\tilde{p}_{(k-1)})$ for \eqref{eq:lfdr-raw}.

Nevertheless, inverting $\widehat{\lfdr}_{\text{raw}}(k)\leq \alpha$ in $k$ is not equivalent to the SLC procedure. To draw a formal equivalence, we use an isotonized estimator of the $\lfdr$. More specifically, 
isotonic regression applied to the sequence \eqref{eq:lfdr-raw} gives a smoothed estimate $\widehat{\lfdr}_{\text{iso}}(k)$, equal to the left-hand slope of the greatest convex minorant (GCM) of the order statistic plot $\{(i/m,\tilde{p}_{(i)})\}_{i=0}^m$ at $k/m$, illustrated in the left panel of Figure \ref{fig:ecdf-maj} (red). The number of rejections made by the SLC procedure is equal to the largest $k$ for which $\widehat{\lfdr}_{\text{iso}}(k)\leq \alpha$, as formally stated in Proposition \ref{prop:iso-gren-equivalence} below. 

The SLC procedure can equivalently be understood as thresholding a plug-in version of \eqref{eq:local-fdr} with $\hat{\pi}_0=1$ and $\hat{g}_{m,n}$ the Grenander estimator, i.e.~the monotone MLE \citep{grenander1956theory,jankowski2009estimation}:
\begin{align*}
    \hat{g}_{m,n} = \argmax_{h \in \mathcal H_{n,m} }  
    \bigg\{\prod_{k=1}^{m} h(\tilde{p}_{(k)})\bigg\},
\end{align*}
where $\mathcal H_{n,m}$ is the set of non-increasing pmf on $\{\frac{1}{n+1},\dots,\frac{m+n+1}{n+1}\}$. 
Concretely, $\hat{g}_{m,n}(t)$ is proportional to the 
left-hand 
slope of the least concave majorant (LCM) of the empirical cdf $\tilde{F}_m(t) = \frac{1}{m} \sum_{k=1}^m \ind{\tilde{p}_{(k)} \leq t}$ at $t$,
obtained 
by swapping the axes in the plot of order statistics (right panel of Figure \ref{fig:ecdf-maj}).
The cardinality of the SLC rejection set is the largest $k$ for which $\widehat{\lfdr}_{\text{gren}}(k)\coloneqq \frac{1/(n+1)}{\hat{g}_{m,n}(\tilde{p}_{(k)})} \leq \alpha$. 
We summarize the various equivalent representations of the SLC procedure in Proposition \ref{prop:iso-gren-equivalence} below, which we prove in Section~\ref{sec:proof-prop-iso}. 

\begin{proposition}
\label{prop:iso-gren-equivalence}
    Assume $\alpha/m > 1/(n+1)$ and recall the notation $\tilde{p}_{(k)}\coloneqq p_{\sigma(k)}+\frac{k}{n+1}$ for $k\in [0,m]$. The number of rejections made by the SLC procedure is:
    \begin{align}
    \label{eq:original-SLC}
        \hat{k} &\coloneqq \max\Big\{\argmin_{k\in [0,m]} \Big\{p_{\sigma(k)} - k\; \Big(\frac{\alpha}{m}-\frac{1}{n+1}\Big) \Big\}\Big\} \\
        \label{eq:shifted-SLC}
        &= \max\Big\{\argmin_{k\in [0,m]} \Big\{\tilde{p}_{(k)} - \frac{\alpha k}{m} \Big\} \Big\}\\
        \label{eq:iso-SLC}
        &= \max \Big\{ k\in [0,m] : \widehat{\lfdr}_{\mathrm{iso}}(k) \leq \alpha \Big\} \\
        \label{eq:gren-SLC}
        &= \max \Big\{ k\in [0,m] : \widehat{\lfdr}_{\mathrm{gren}}(k) \leq \alpha\Big\} . 
    \end{align}
\end{proposition}
Note that this result does not use the monotonicity assumption (Assumption~\ref{ass-scoredensity}). Nevertheless, our result shows that the SLC procedure corresponds to an empirical Bayes procedure that uses an isotonic estimator of the lfdr.

\section{Empirical experiments}\label{sec:xp}

In this section, we evaluate SL, SLC, SLC+, ASLC, ASLC+ in Table~\ref{tab:method.summary} over extensive empirical experiments and compare with Holm's procedure\footnote{Holm's procedure controls FWER and thus automatically controls bFDR.}.
In Appendix~\ref{sec:addxp.multiple.subsampling}, we evaluate SLC++, ASLC++; in Appendix~\ref{sec:addxp.SLG}, we implement another SL variant introduced in Appendix~\ref{secSL3}. 
Without further specification, the subsampling proportion is fixed at $0.1$.

\subsection{Simulated data}\label{sec:xp.simulated.data}

By default, we generate null scores and calibration scores following $U(0,1)$, and non-null scores following $U(0.8,1.8)$; all scores are independent. The null proportion is set at $\pi_0 = 0.8$.
We consider three specific settings, and report various summary quantities (bFDR, $|R|/m$, sd$(|R|/m)$) from 1000 repetitions for each setting.
\begin{enumerate}
\item[(a)] $m = 2000$, $n = 4000$: $m$, $n$ are comparable.
\item[(b)] $m = 2000$, $n = 4000$,  non-null scores following $\mathrm{Beta}(30,1)$: null and non-null scores share the same support.
\item[(c)] $m = 200$, $n = 12000$: $n$ is sufficiently large\footnote{$n$ is no smaller than $3m/\alpha$ for all $\alpha$ considered.}. The subsampling proportion is increased to $0.2$.
\end{enumerate}

As shown in Figure~\ref{fig:simulation} panel (c), when $n$ is sufficiently large, SL, SLC, SLC+ behave similarly and approximately control bFDR at level $\pi_0 \alpha$. 
When $n$ is only comparable to $m$, SL may severely violate
the bFDR control at level $\pi_0 \alpha$ (Figure~\ref{fig:simulation}, panel (a)),  SLC yields zero rejections for small $\alpha$
(Figure~\ref{fig:simulation}, panel (a) and (b)), and SLC+ exhibits the most favorable performance, maintaining bFDR control while producing a reasonable number of rejections.
For signals of moderate strength  (Figure~\ref{fig:simulation} panel (b)), Holm’s procedure is substantially less powerful.

As shown in Figure~\ref{fig:simulation.adaptive},
adaptive variants ASLC, ASLC+ can be more powerful than their non-adaptive counterparts SLC, SLC+, but may exhibit increased variability for large $\alpha$ (different subsamples can give different number of rejections each time).

\begin{figure}[h!]
  \centering
  \begin{minipage}{1\textwidth}
    \centering
    \includegraphics[clip, trim = 0cm 0cm 0cm 0cm, width = 0.9\textwidth]{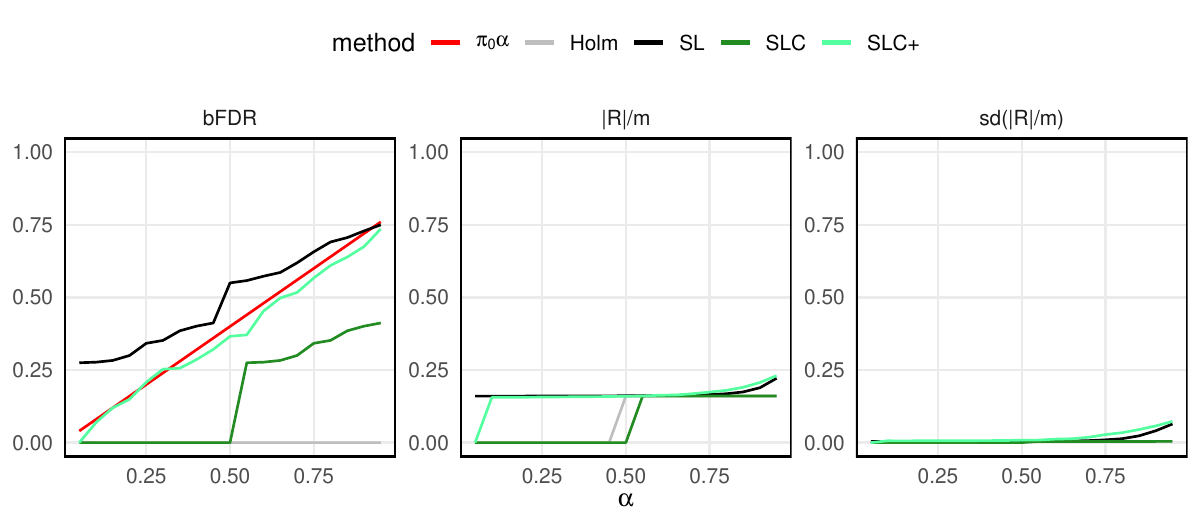}
    \small\text{(a) $m=2000$, $n = 4000$, Non-null scores $\sim U(0.8,1.8)$}
  \end{minipage}
  \begin{minipage}{1\textwidth}
    \centering
    \includegraphics[clip, trim = 0cm 0cm 0cm 1.5cm, width = 0.9\textwidth]{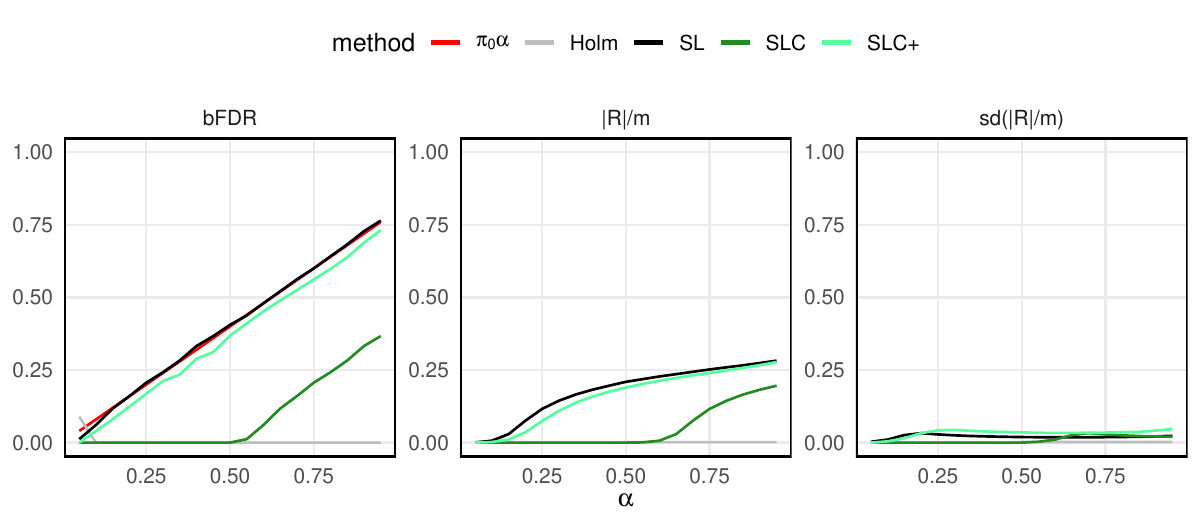}
    \small\text{(b) $m=2000$, $n=4000$, Non-null scores $\sim\mathrm{Beta}(30,1)$}
  \end{minipage}
     \begin{minipage}{1\textwidth}
    \centering
    \includegraphics[clip, trim = 0cm 0cm 0cm 1.5cm, width = 0.9\textwidth]{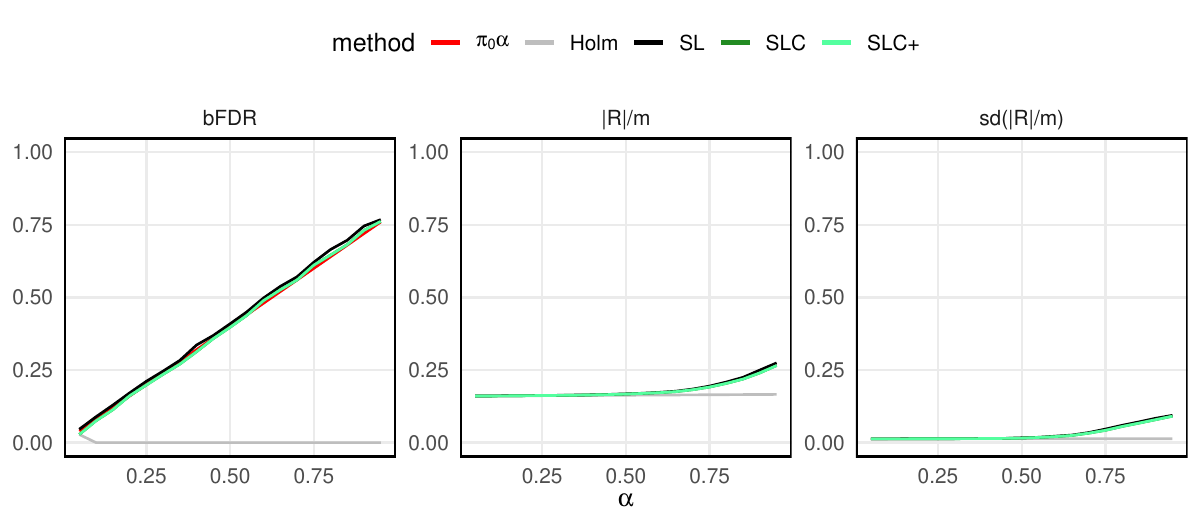}
    \small\text{(c) $m=200$, $n = 12000$, Non-null scores $\sim U(0.8,1.8)$}
  \end{minipage}

  \caption{Comparison of SL, SLC, SLC+ applied to simulated conformal $p$-values. }
  \label{fig:simulation}
\end{figure}

\begin{figure}[tbp]
  \centering
  \begin{minipage}{1\textwidth}
    \centering
    \includegraphics[clip, trim = 0cm 0cm 0cm 0cm, width = 0.9\textwidth]{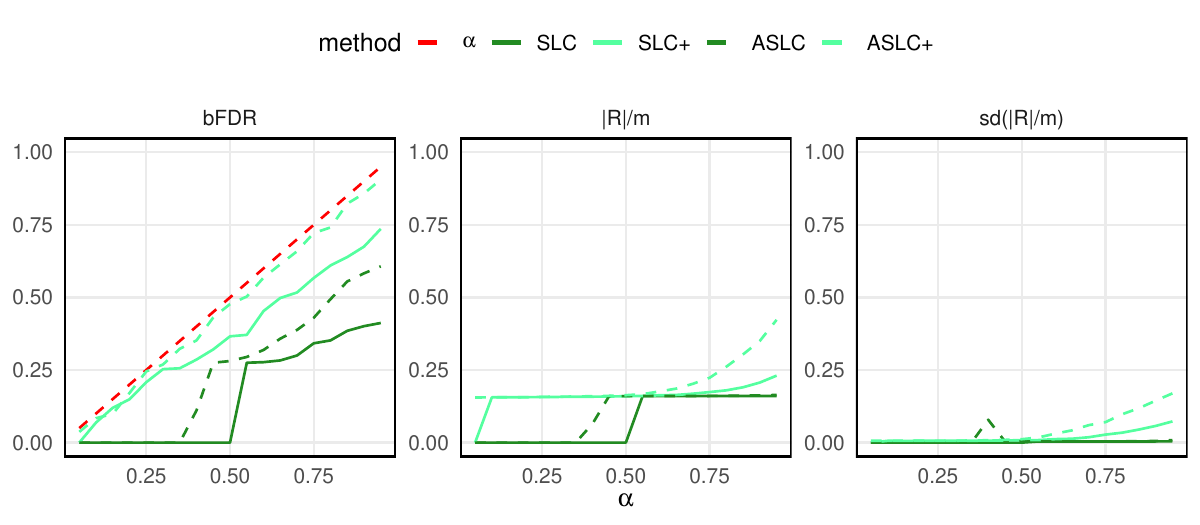}
    \small\text{(a) $m=2000$, $n = 4000$, Non-null scores $\sim U(0.8,1.8)$}
  \end{minipage}
\begin{minipage}{1\textwidth}
    \centering
    \includegraphics[clip, trim = 0cm 0cm 0cm 1.5cm, width = 0.9\textwidth]{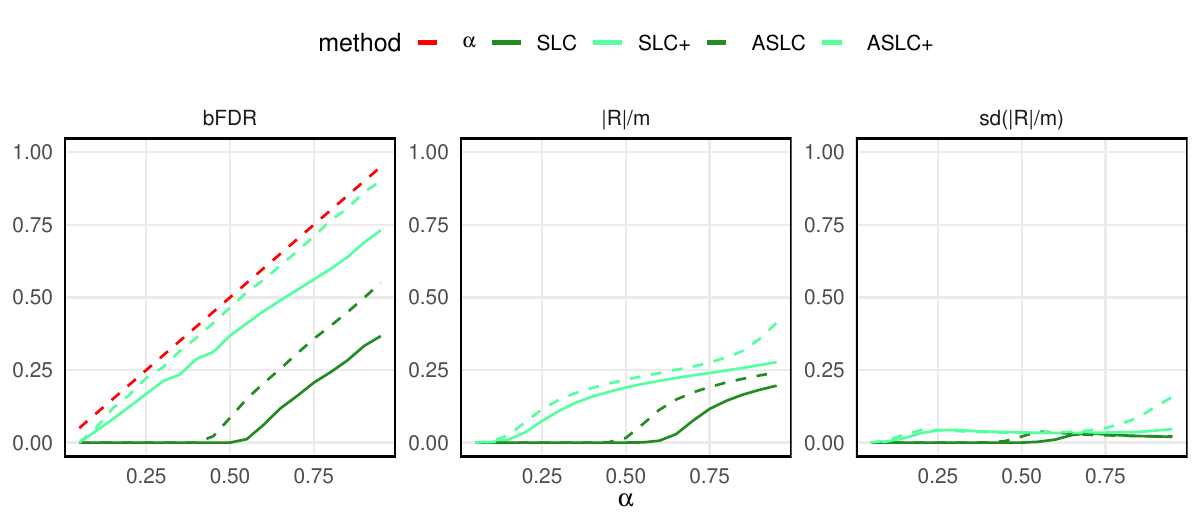}
    \small\text{(b) $m=2000$, $n=4000$, Non-null scores $\sim\mathrm{Beta}(30,1)$}
  \end{minipage}
  \caption{Comparison of SLC, SLC+, and their adaptive variants ASLC, ASLC+ applied to simulated conformal $p$-values.
  }
  \label{fig:simulation.adaptive}
\end{figure}

\begin{figure}[tbp]
  \centering
    \begin{minipage}{0.3\textwidth}
    \centering
    \includegraphics[clip, trim = 0cm 0cm 0cm 0cm, width = 0.95\textwidth]{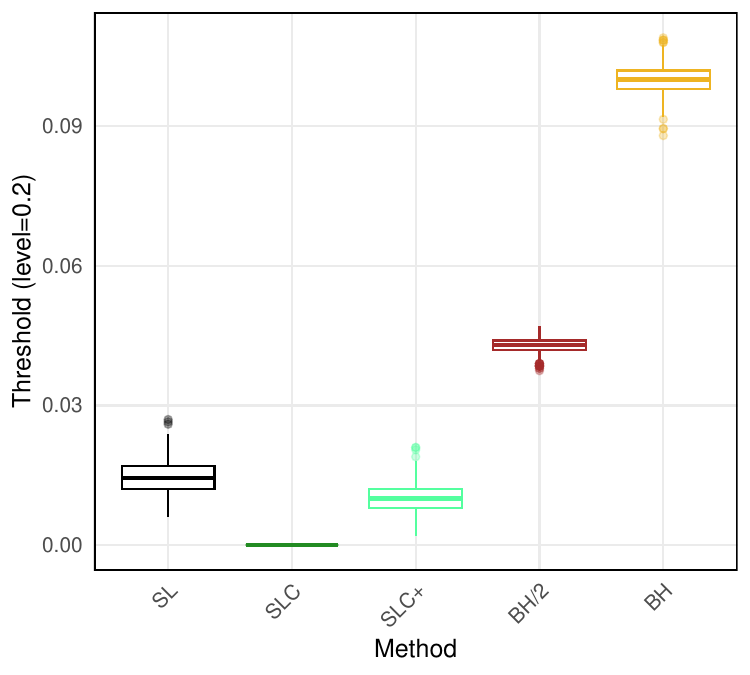}
    \vspace{-0.2cm}
    \begin{center}
        \text{(a)}
    \end{center}
  \end{minipage}
  \quad
  \begin{minipage}{0.3\textwidth}
     \vspace{-0.2cm}
     \centering
     \includegraphics[clip, trim = 0cm 0cm 0cm 0cm, width = 0.95\textwidth]{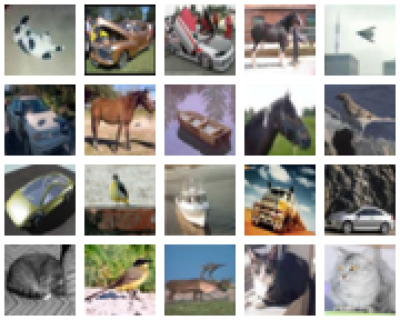}
    \begin{center}
        \text{(b)}
    \end{center}
   \end{minipage}
    \quad
    \begin{minipage}{0.3\textwidth}
    \vspace{-0.2cm}
    \centering
    \includegraphics[clip, trim = 0cm 0cm 0cm 0cm, width = 0.95\textwidth]{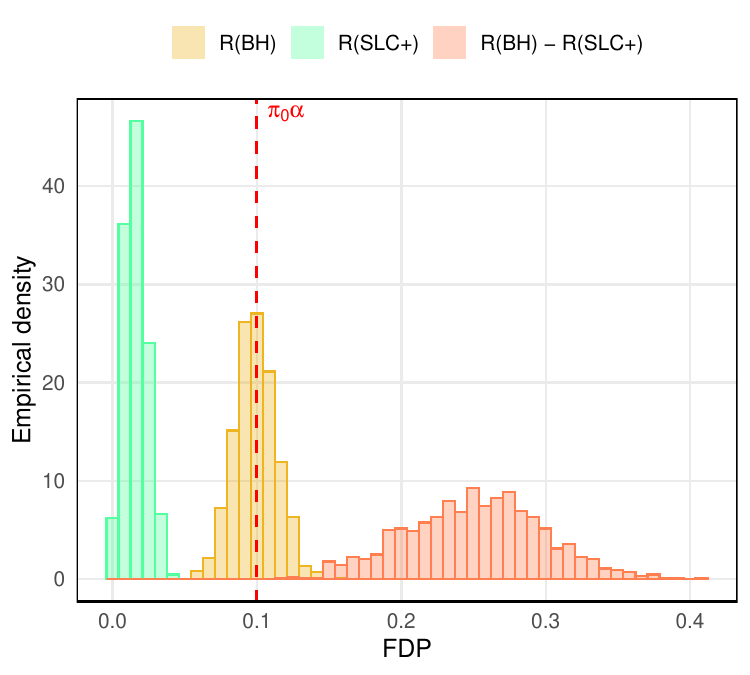}
     \vspace{0.0cm}
     \begin{center}
        \text{(c)}
    \end{center}
  \end{minipage}
  
  \caption{Comparison of BH and SL variants applied to conformal $p$-values derived from \texttt{CIFAR-10} (bFDR level $0.2$).
  BH at level $0.1$ and $0.2$ both consistently yield larger rejection sets compared to SL variants (panel (a)).
  The FDP among the data points rejected by BH (level $0.2$) but not by SLC+ (in coral) is much higher than that of BH (level $0.2$) (in gold), which centers around $\pi_0 \alpha = 0.1$ (panel (c)). For illustration, pictures around BH's rejection threshold are given in panel (b) (detected animal images are false discoveries). 
  }
  \label{fig:simulation.cifar10.threshold}
\end{figure}

\subsection{\texttt{CIFAR-10}}\label{sec:xp.cifar10}

In \texttt{CIFAR-10}, the animal pictures, including bird, cat, deer, dog, frog, and horse, are treated as nulls, and pictures of transportation tools, including airplane, car, ship, and truck, are treated as alternatives. 
Following \cite{marandon2024adaptive}, we train a 10-class classifier with probabilistic outputs, and the conformal score of a picture is defined as the total probability of belonging to any transportation tool subclass.
We hold out $1000$ images for each of the $10$ classes (a total of $6000$ animal pictures and $4000$ transportation tool pictures) to sample testing and calibration datasets.
To generate a test dataset and a calibration dataset, we randomly sample a total of $m_0 + n = 2500$ animal images without repeats; among these, a random subset of $m_0 = 500$ images is used as the test dataset and the remaining $n = 2000$ images are used as the calibration dataset. 
We also randomly sample $m_1 = 500$ transportation tool images, so the total test sample size is $m = m_0 + m_1 = 1000$, and the null proportion is $\pi_0=0.5$. Hence, Assumption~\ref{assglobal} is satisfied, and the null/calibration distribution is a mixture.
We independently generate $1000$ such test and calibration datasets.

In appendix~\ref{sec:addxp.cifar10}, we apply the simulation comparisons in Section~\ref{sec:xp.simulated.data} to conformal $p$-values derived from
\texttt{CIFAR-10} with three null proportions
$\pi_0 \in \{0.2, 0.5, 0.8\}$. 
The comparative performance of SL, SLC, SLC+, ASLC, ASLC+ (Figures~\ref{fig:simulation.cifar10} and \ref{fig:simulation.adaptive.cifar10}) largely
agrees with that in the simulated data experiments. In addition, adaptive
variants are more beneficial when the null proportion $\pi_0$
is smaller (Figure~\ref{fig:simulation.adaptive.cifar10}, across panel (a), (b), and (c)).

In Figure~\ref{fig:simulation.cifar10.threshold}, we compare SLC with BH at levels $\alpha/2$ (BH/2) and $\alpha$ (BH) in terms of the number or rejections ($|R|$). 
BH at both levels consistently yields more rejections.
We also evaluate several FDPs: the FDP over SLC's rejection set, BH's rejection, and the difference set (test points rejected by BH at FDR level $\alpha$, but not by SL).
The FDP on this difference set is much higher than $\pi_0 \alpha$.
In panel (b), we randomly choose a trial and display the $20$ images in BH's rejection set with the smallest conformal scores (closest to the rejection threshold); notably, $55\%$ of them ($11$ out of $20$) are nulls (animals).

\section{Conclusion and discussion}

Global FDR control and boundary reliability are distinct objectives: while FDR is the expected  proportion of false discoveries among all reported novelties, bFDR directly measures the probability that the marginal discovery at the detection threshold is an inlier. This distinction is particularly relevant in applications where false detections near the decision boundary can have critical consequences.

The conformal setting introduces challenges absent from the continuous, independent setting. % of \cite{soloff2024edge}. 
In particular, conformal $p$-values are discrete and share the same calibration sample, inducing dependence across test points. We showed that the support-line procedure developed for independent continuous $p$-values violates the bFDR control in the conformal setting, with closely matching lower/upper bounds. Our proposed procedures (A)SLC, (A)SLC+, (A)SLC++/2 account for this additional structure and provide finite-sample bFDR guarantees. The resulting family of methods also offers different practical trade-offs: SLC provides a natural default  procedure, while ASLC can improve power by adapting to the estimated null proportion. The single-subsampled variants are useful when the calibration sample is small relative to the test sample, mitigating the conservativeness induced by the finite calibration sample. The multiple-subsample variants further provide a way to stabilize the procedure, with the `/2` variants retaining formal bFDR guarantees under an additional monotonicity assumption (Assumption~\ref{ass-scoredensity}). Importantly, our procedures are agnostic to the choice of novelty score and can therefore be combined with a broad range of machine-learning algorithms, while providing a prescribed boundary error guarantee.

The bFDR guarantee can also be given a stronger interpretation under a monotonicity assumption (Assumption~\ref{ass-scoredensity}). As discussed in Sections~\ref{sec:intro} and~\ref{sec:monotonicity}, this assumption allows the control at the boundary to extend to individual false-detection probabilities at all points below the boundary, see \eqref{th-bFDRLFC}. This provides a useful interpretation of bFDR control: not only is the weaker discovery reliable, but discoveries with stronger evidence of novelty can also be controlled individually. The monotonicity assumption, however, may not hold exactly for novelty scores learned from real data. An interesting direction for future work is therefore to quantify the degradation of this  guarantee when the monotonicity condition is only approximately satisfied.

Several other questions remain open. In particular, it would be interesting to develop boundary error control under departures from exchangeability (Assumption~\ref{assglobal}), with the aim of extending the framework to large-scale novelty-detection problems in the presence of distribution shift.

\section*{Acknowledgement}
We would like to acknowledge Anna Ben-Hamou, Yoav Benjamini and Will Fithian for fruitful discussions. 
The authors acknowledge grants ANR-21-CE23-0035 (ASCAI) and ANR-23-CE40-0018-01 (BACKUP) of the French National Research Agency ANR, the Emergence project MARS of Sorbonne Universit\'e.

\bibliographystyle{dcu}
\bibliography{biblio}
\newpage

\appendix

\section{Support line procedure with gap}\label{secSL3}

This section presents another  modification of the support line procedure, that uses a separation condition in the minimization problem:
   \begin{equation}\label{equ-separation}
    \mbox{$\hat{k}$ defined by \eqref{equkchap} is positive and }    p_{\sigma(\hat{k})} - \alpha \hat{k}/m+ \frac{1}{n+1} \leq \min_{k\in [0,m]\backslash \{\hat{k}\}} \big\{ p_{\sigma(k)} - \alpha k/m\big\} .
    \end{equation}
Condition \eqref{equ-separation} ensures that the minimum value in \eqref{equkchap} is separated from other values by at least $1/(n+1)$.
We now define a simple modification of the SL procedure that provides \eqref{equ-separation}.
\begin{definition}\label{def:SL3}
    The procedure $\SL$ with gap, denoted $\SLG$ at level $\alpha\in (0,1)$ is defined as making the same discoveries as $\SL$ at level $\alpha$ if \eqref{equ-separation} holds and no discovery otherwise.
 \end{definition}

\begin{theorem}\label{th:bFDR2}
Under Assumption~\ref{assglobal} and provided that the level $\alpha\in (0,1)$ satisfies $1/(n+1)\leq \alpha/m$, we have
$
 \bFDR(\SLG) \leq \alpha m_{0}/m .
$
\end{theorem}
The proof is given in Section~\ref{secproofth:bFDR2}. While $\SLgap$ procedure controls the bFDR, we in general recommend to use $\SLC$ (or its variants) instead because $\SLG$  is less powerful on our simulations and exhibits high variability, see Section~\ref{sec:addxp.SLG}. Another caveat is that $\SLgap$ has a rejection set not necessarily monotone in $\alpha$.

\begin{remark}\label{rem:counterSL3}
    The condition $1/(n+1)\leq \alpha/m$ is necessary in Theorem~\ref{th:bFDR2} provided that $\alpha \ge 2/(n+1)$. Indeed, when $1/ (n+1)> \alpha/m \geq 2/\{m(n+1)\}$, the counterexample of the proof of Proposition~\ref{prop:counterex} can be used for $m_0=1$  to prove that the bFDR is larger than $m_0/(n+m_0)=1/(n+1)>\alpha /m =\alpha m_0/m $ in that case. Also, the separation constraint \eqref{equ-separation} holds, because when the null score exceeds all calibration scores, the objective in \eqref{equkchap} is $0$ at $0$ and $k/(n+1) - \alpha$ at $k/(n+1)$ for $k \ge 1$. Hence the objective is minimized at $k=1$ with the gap $1/(n+1)$ if and only if ${1}/(n+1) - \alpha \le 0 - {1}/(n+1)$, i.e., $\alpha \ge 2/(n+1)$.
\end{remark}

\section{Proofs}

\subsection{Notation}

We introduce here notation that will be useful to prove our main results. We assume throughout the proofs that Assumption~\ref{assglobal} holds, and (in)equalities between random variable are true almost surely, even if we do not explicitly mention it in some places.

Since the conformal $p$-values are dependent, each $p_i$ and $\mbf{p}_{-i}=(p_j)_{j\in [m]\backslash\{i\}}$ are dependent and an essential step is to  describe $\mbf{p}_{-i}$ as a function $(\Psi_{ij}(p_i,W_i))_{j\in \range{m}\backslash\{i\}}$ of the $p$-value $p_i$ and an independent random variable $W_i$ for some functional $\Psi_{ij}$. More formally, for each fixed $i\in \cH_0$, define 
    \begin{align}
W_i&:=(A_i,(S_{n+j})_{j\in\range{m}\backslash\{i\}})\label{equWi};\\
A_i&:=\{S_{j},j\in \range{n}\}\cup \{S_{n+i}\} = \{a_{i,(1)},\dots,a_{i,(n+1)}\}\label{equAi};\\
\Psi_{ij}(x,W_i) &:= \frac{1}{n+1}\Big(1+\sum_{s\in A_i \backslash\{a_{i,(\lceil x(n+1)\rceil)}\}} \ind{s\geq S_{n+j}} \Big),\:\:j\in\range{m}\backslash\{i\},x\in (0,1]\label{equPsii}
\end{align} 
with $a_{i,(1)}>\dots>a_{i,(n+1)}$. 
Note that $A_i$ in \eqref{equAi} is an unordered set.
Lemma~\ref{lemmaPropConformalFull} shows that $p_i$ and $W_i$ are independent, so that conditional on $W_i$, the random variable $p_i$ is still super-uniformly distributed.
In the adaptive case, we also introduce
\begin{equation}
    \label{pi0chapWi}
    \hat{\pi}_0(W_i) = \frac{1+\sum_{j\in [m]\backslash\{i\}} \ind{S_{n+j}\le a_{i,(s_0+1)}}}{m(1-(s_0+1)/(n+1))},
    \end{equation}
where $s_0\in [0,n-1]$ is the Storey parameter in \eqref{equpi0hat}. 
Lemma~\ref{lem:pi0} establishes a relation between $ \hat{\pi}_0$ in \eqref{equpi0hat} and $\hat{\pi}_0(W_i)$ in \eqref{pi0chapWi}.

Let us now introduce the following key functional: for $\ell,\ell'\in[n+1]$,
\begin{align}
H_{\ell,W_i}(\ell')
&:=\ell'/(n+1)
- (\alpha_0/m)\ind{\ell\leq \ell'}\Bigg(1 + \sum_{j \in \range{m}\backslash\{i\} }\ind{ S_{n+j}\geq a_{i,(\ell'+1)}}\Bigg)\nonumber\\
&\:\:\:\:\:-(\alpha_0/m) \ind{\ell> \ell'} \sum_{j \in \range{m}\backslash\{i\} }\ind{ S_{n+j}\geq a_{i,(\ell')}},\label{equ-keyfunctionalHell}
\end{align}
with the convention $H_{\ell,W_i}(0)=0$ and where $\alpha_0=\alpha$ in the non-adaptive case and $\alpha_0=\alpha/\hat{\pi}_0(W_i)$ in the adaptive case. 
Note that $1 + \sum_{j \in \range{m}\backslash\{i\} }\ind{ S_{n+j}\geq a_{i,(\ell'+1)}} \geq \sum_{j \in \range{m}\backslash\{i\} }\ind{ S_{n+j}\geq a_{i,(\ell')}}$, hence $\ell\in [n+1]\mapsto H_{\ell,W_i}(\ell')$ is nondecreasing for a given fixed $\ell'$. 

Now, the idea is that, while the function $H_{\ell,W_i}(\cdot)$ only depends on $W_i$, minimizing $H_{\ell,W_i}(\ell')$ with respect to $\ell'$ is related to the minimization  problem of SL-like methods (Lemmas~\ref{lemmafromktoell}~and~\ref{lemH}). 
We thus also introduce for $\ell\in [n+1]$,
\begin{equation}
    \label{equMell}
M_{\ell,W_i}:=\min_{\ell'\in [0,s_0]} H_{\ell,W_i}(\ell'),
\end{equation}
with $s_0=n+1$ in the non-adaptive case.
Note that $\ell\in [n+1]\mapsto M_{\ell,W_i}$ is nondecreasing, because for $\ell\in [n]$, we have $M_{\ell+1,W_i}-M_{\ell,W_i}\geq \min_{\ell'\in [0,s_0]} (H_{\ell+1,W_i}(\ell')-H_{\ell,W_i}(\ell')) \geq 0.$

Finally, we introduce for $\ell\in [0,n+1]$,
\begin{align}
    \label{equUell}
    U_{\ell,W_i} &:=  \min_{\ell'\in [n+1]}\{ H_{\ell'+1,W_i}(\ell') +(\alpha/m)\ind{\ell'\leq  \ell }\} .
\end{align}
Note that $U_{\ell,W_i}$ is clearly nondecreasing with respect to $\ell$.

\subsection{Proof of Theorem~\ref{th:bFDR}}\label{sec:proofth:bFDR}

By definition, for $\hat{k}$ \eqref{equkchap-conf}, we have
\begin{align*}
\bFDR(\SL)&=\P(H_{\sigma(\hat{k})} = 0) \\
&=  \sum_{i\in \cH_{0}} \sum_{\ell=1}^{n+1} \P(p_i = \ell/(n+1), \sigma(\hat{k}) = i )\\
&=  \sum_{i\in \cH_{0}} \sum_{\ell=1}^{\msub} \P(p_i = \ell/(n+1), \sigma(\hat{k}) = i ),
\end{align*}
for $\msub:= \lfloor \alpha(n+1) \rfloor$, because the minimum is reached for a $p$-value at most $\msub/(n+1)$ (Lemma~\ref{lemmafromktoell}). We assume $\msub\geq 1$ (otherwise the result is trivial).
By conditioning with respect to $W_i$ \eqref{equWi}, we obtain
\begin{align*}
\bFDR(\SL)
&=  \sum_{i\in \cH_{0}} \sum_{\ell=1}^{\msub} \E\big[\P(p_i = \ell/(n+1), \sigma(\hat{k}) = i \:|\: W_{i})\big]\\
&\leq  \sum_{i\in \cH_{0}} \sum_{\ell=1}^{\msub} \frac{1}{n+1}\E\big[\P( \sigma(\hat{k}) = i \:|\: W_{i},p_i = \ell/(n+1))\big],
\end{align*}
by using the super-uniformity of $p_i$ conditionally on $W_i$ (by independence)\footnote{Note that $\sigma$ and $\hat{k}$ are completely determined by $W_{i}$ and $p_i$ so that the conditional probability above is an indicator with the value of $p_i$ replaced by $\ell/(n+1)$. However, we keep the probabilistic notation for simplicity.}. 
Applying Lemma~\ref{lemmafromktoell}  and Lemma~\ref{lemH} (iii), (iv), and (v), we obtain for $i\in \cH_{0}$, 
\begin{align*}
 &\sum_{\ell=1}^{\msub} \frac{1}{n+1} \P( \sigma(\hat{k}) = i \:|\: W_{i},p_i = \ell/(n+1)) \\
 &\leq 
 \frac{1}{n+1} \ind{M_{1,W_i}=H_{1,W_i}(1)} + \sum_{\ell=2}^{\msub} \frac{1}{n+1} \ind{M_{\ell,W_i} - M_{\ell-1,W_i} \geq 1/(n+1)}\\
 &\leq
{\bigg( M_{1,W_i} + (\alpha/m)\sum_{j\in [m]\backslash\{i\}} \ind{S_{n+j}\geq a_{i,(1)}} + \alpha/m\bigg)_+} +M_{\msub,W_i} - M_{1,W_i},
\end{align*}
where we used that the function $\ell\in [n+1]\mapsto M_{\ell,W_i}$ is nondecreasing to obtain the inequality $\frac{1}{n+1} \ind{M_{\ell,W_i} - M_{\ell-1,W_i} \geq 1/(n+1)}\leq M_{\ell,W_i} - M_{\ell-1,W_i}$, {where we used Lemma~\ref{lemH} (v) to obtain $ \ind{M_{1,W_i}=H_{1,W_i}(1)}/(n+1) \leq \big( M_{1,W_i} + (\alpha/m)\sum_{j\in [m]\backslash\{i\}} \ind{S_{n+j}\geq a_{i,(1)}} + \alpha/m\big)_+$} and where we used a telescopic sum. 

{Now, the above upper bound can be rewritten as
$$
\max\bigg(M_{\msub,W_i} - M_{1,W_i}, M_{\msub,W_i} + (\alpha/m)\sum_{j\in [m]\backslash\{i\}} \ind{S_{n+j}\geq a_{i,(1)}} + \alpha/m\bigg).
$$
By Lemma~\ref{lem:forgotenlemma}, the first term in the max is upper-bounded by $\alpha/m + 1/(n+1)$. The second term can be bounded by using
$$
M_{\msub,W_i}\leq H_{\msub,W_i}(1) \leq 1/(n+1) - (\alpha/m)\sum_{j\in [m]\backslash\{i\}} \ind{S_{n+j}\geq a_{i,(1)}},
$$
which also leads to the upper bound $\alpha/m + 1/(n+1)$, which concludes the proof.
}

\subsection{Proof of Theorem~\ref{th:bFDR2}} \label{secproofth:bFDR2}

We use the previous proof and take advantage of the additional condition \eqref{equ-separation}. Coming back to the inequality
\begin{align*}
\bFDR(\SLgap)
&\leq  \sum_{i\in \cH_{0}} \sum_{\ell=1}^{\msub} \frac{1}{n+1}\E\big[\P( \sigma(\hat{k}) = i \:|\: W_{i},p_i = \ell/(n+1))\big],
\end{align*}
we use Lemma~\ref{lemH} (vii) (and thus $1/(n+1)\leq \alpha/m$) to obtain
\begin{align*}
\bFDR(\SLgap)
&\leq  \sum_{i\in \cH_{0}} \sum_{\ell=1}^{\msub} \frac{1}{n+1}\P( U_{\ell-1,W_i}\leq  U_{\ell,W_i} - 1/(n+1))\\
&\leq \sum_{i\in \cH_{0}} \E\big[ \sum_{\ell=1}^{\msub} (U_{\ell,W_i} -U_{\ell-1,W_i}) \ind{ 1/(n+1)\leq U_{\ell,W_i}-U_{\ell-1,W_i}} \big]\\
&\leq \sum_{i\in \cH_{0}} \E\big[ \sum_{\ell=1}^{\msub} (U_{\ell,W_i} -U_{\ell-1,W_i})  \big]\\
&\leq \sum_{i\in \cH_{0}} \E\big[U_{\msub,W_i} - U_{0,W_i} \big].
\end{align*}
 because $U_{\ell,W_i}$ is nondecreasing with respect to $\ell$ and by using a telescopic sum. Since
 $$
 U_{\ell_{0},W_i}  -U_{0,W_i} \leq \max_{\ell'\in [n+1]}\{ H_{\ell'+1,W_i}(\ell') +(\alpha/m)\ind{\ell'\leq  \ell_{0} } - H_{\ell'+1,W_i}(\ell')\} = \alpha/m,
 $$
 we obtain the desired bound.

\subsection{Proof of Proposition~\ref{prop:counterex}}\label{sec:proofprop:counterex}

It is a direct consequence of Proposition~\ref{prop:lower-bound}, see Section~\ref{sec:morelbforub}.
However,  for the sake of completeness, we provide a direct proof below.  
{Consider iid null scores distributed uniformly in $(0,1)$, and independent alternative scores which are iid and distributed uniformly in $(1,2)$. Hence,  
the scores of all alternatives are strictly larger than those of all nulls and calibration points. }
In particular, all alternative $p$-values are equal to $1/(n+1)$. Suppose the minimal null $p$-value is also $1/(n+1)$, then for $\alpha \ge 1 / \{(n+1)(1-\pi_0)\}$, we have
\begin{align*}
    p_{\sigma(m_1+1)} - \alpha \frac{m_1+1}{m}
    < \frac{1}{n+1} - \alpha (1-\pi_0)
    \le 0,
\end{align*}
so the rejection set is non-empty. 
Since all alternative $p$-values equal $1/(n+1)$ and there exists at least one null $p$-value equal to $1/(n+1)$ whose score is smaller than those of the alternatives, the rejection must occur at a null hypothesis. Hence
\begin{align*}
    \PP\left(H_{\sigma(\hat{k})} = 0\right)
    \ge \PP\left(\min_{i\in \cH_0} p_i = \frac{1}{n+1}\right) 
    = \PP\left(\max_{i \in \cH_0} S_{n+i} > \max_{j \in [n]} S_j\right) .
 %   = \frac{m_0}{m_0 + n}.
\end{align*}
%{where the last equality is derived by a straightforward computation.}
In addition,
\begin{align*}
     \PP\left(\max_{i \in \cH_0} S_{n+i} > \max_{j \in [n]} S_j\right) 
    &= \E\left[\PP\left( S_{(1)} \in \{S_{n+i}, i\in \cH_0\}\:|\: S_{(1)},\dots,S_{(n+m_0)}\right)\right],
\end{align*}
where we denote $S_{(1)}>\dots>S_{(n+m_0)}$ the order statistics of the sample $\{ S_j,j \in [n]\cup(n+\cH_0)\}$. By exchangeability of this sample, the latter is  $\frac{m_0}{m_0 + n}$.

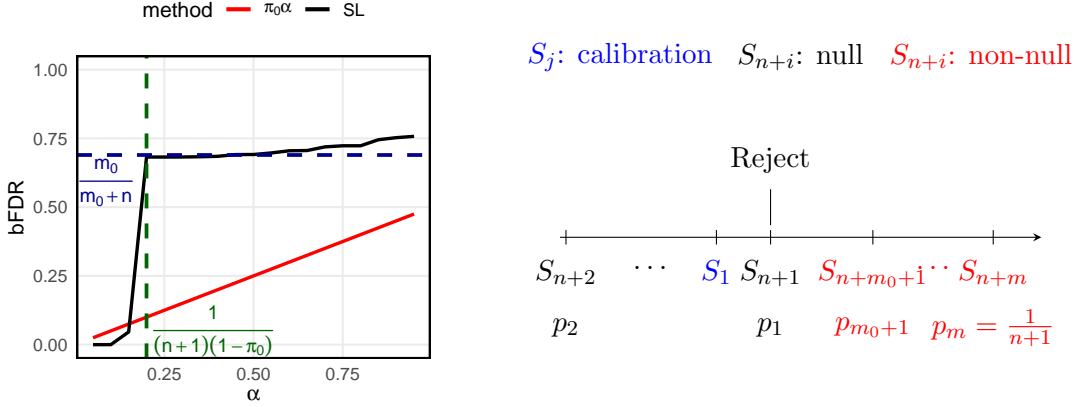
\begin{figure}[tbp]
 \centering
  \begin{minipage}{0.4\textwidth}
    \centering
   \includegraphics[clip, trim = 0cm 0cm 0cm 0cm, width = 0.95\textwidth]{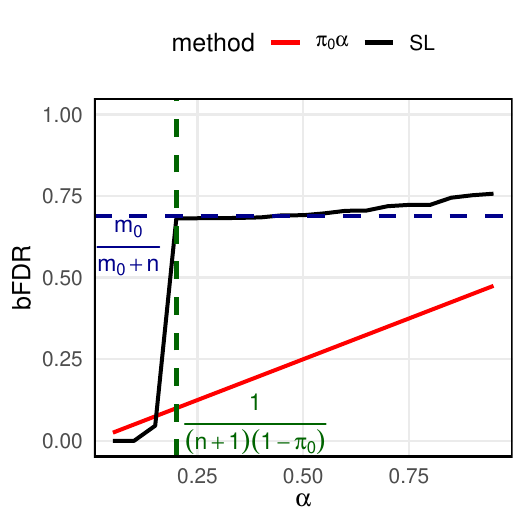}
   
  \end{minipage}
  \begin{minipage}{0.58\textwidth}
  \centering
    \begin{tikzpicture}[>=stealth, scale=0.8]
      \draw[->] (0,0) -- (8,0);
      \draw (0.1,0.12) -- (0.1,-0.12) node[below=2pt] {$S_{n+2}$};
        \node at (0.1,-1.5) {$p_2$};
    \node at (1.5,-0.5) {$\ldots$};
      \draw (2.6,0.12) -- (2.6,-0.12) node[below=2pt,color=blue] {$S_{1}$};
      \draw (3.5,0.12) -- (3.5,-0.12) node[below=2pt] {$S_{n+1}$};
       \node at (3.5,-1.5) {$p_1$};
       \draw (3.5,0.8) -- (3.5,0.2) node[above=16pt]
       {Reject};
      \draw (5.2,0.12) -- (5.2,-0.12) node[below=2pt,color=red] {$S_{n+m_0+1}$};
       \node[color=red] at (5.2,-1.5) {$p_{m_0+1}$};
    \node at (6.2,-0.5) {\textcolor{red}{$\ldots$}};
      \draw (7.2,0.12) -- (7.2,-0.12) node[below=2pt,color=red] {$S_{n+m}$};
       \node[color=red] at (7.2,-1.5) {$p_{m}=\frac{1}{n+1}$};
     \node[color=blue] at (1,3) {$S_{j}$: calibration};
      \node[color=black] at (4,3) {$S_{n+i}$: null};
       \node[color=red] at (7,3) {$S_{n+i}$: non-null};
    \end{tikzpicture}
  \end{minipage}
    \caption{Counterexample of bFDR control. Here $n = 9$, $m = 40$, $m_1 = 20$; null and calibration scores follow $U(0,1)$, non-null scores follow $U(1,2)$, and all scores are independent. We also superimpose the lower bound $m_0/(m_0+n) = 20/29$, valid for $\alpha \ge 1/\{(n+1)(1-m_0/m)\} = 0.2$.}
  \label{fig:counter.example}
\end{figure}

\subsection{Proof of Theorem~\ref{th:bFDR-adapt}}\label{sec:proofth:bFDR-adapt}

We follow the proof of Section~\ref{sec:proofth:bFDR}: in particular the intermediate steps that are true point-wise extend just by taking care of the new definition of $\hat{k}$ in \eqref{equkchap-conf-adapt} and by using Lemma~\ref{lem:pi0}. 

Formally, by using Lemma~\ref{lemmafromktoell} and Lemma~\ref{lemH} (adaptive part) (iii)', (iv)', and (v), (vi), we have 
\begin{align*}
\bFDR(\ASL)&=  \sum_{i\in \cH_{0}} \sum_{\ell=1}^{s_0} \P(p_i = \ell/(n+1), \sigma(\hat{k}) = i )\\
&\leq \sum_{i\in \cH_{0}} \sum_{\ell=1}^{s_0} \frac{1}{n+1}\E\big[\P( \sigma(\hat{k}) = i \:|\: W_{i},p_i = \ell/(n+1))\big]\\
&\leq  \sum_{i\in \cH_{0}}  \big(\E\big[\alpha/(m\hat{\pi}_0(W_i))\big]  + 1 /(n+1)\big),
\end{align*}
with $\hat{\pi}_0(W_i)$ given by \eqref{pi0chapWi}. 
Then we use that $\P(p_i\leq (s_0+1)/(n+1)\:|\: W_i) = (s_0+1)/(n+1)$ (Lemma~\ref{lemmaPropConformalFull}) to deduce
$$
1 = \E\bigg[\frac{\ind{p_i\geq (s_0+2)/(n+1)}}{1-(s_0+1)/(n+1)} \:\bigg|\: W_i\bigg].
$$
Hence, we have
\begin{align*}
\sum_{i\in \cH_{0}} \E\bigg[\alpha/(m\hat{\pi}_0(W_i))\bigg]  &= \sum_{i\in \cH_{0}}\E\bigg[\alpha/(m\hat{\pi}_0(W_i))\E\bigg[\frac{\ind{p_i\geq (s_0+2)/(n+1)}}{1-(s_0+1)/(n+1)} \:\bigg|\: W_i\bigg]\bigg]\\
& = \sum_{i\in \cH_{0}}\E\bigg[\alpha/(m\hat{\pi}_0(W_i))\frac{\ind{p_i\geq (s_0+2)/(n+1)}}{1-(s_0+1)/(n+1)} \bigg]\\
&= \alpha  \sum_{i\in \cH_{0}}\E\bigg[\frac{1-(s_0+1)/(n+1)}{1+\sum_{j\in [m]\backslash\{i\}} \ind{S_{n+j}\le a_{i,(s_0+1)}}}\frac{\ind{p_i\geq (s_0+2)/(n+1)}}{1-(s_0+1)/(n+1)}\bigg],
\end{align*}
where we use the expression \eqref{pi0chapWi} of $\hat{\pi}_0(W_i)$ and the fact that $p_i$ is independent of $W_i$ (Lemma~\ref{lemmaPropConformalFull}). 
Now, for any $j\in [m]\backslash\{i\}$, $p_j \geq (s_0+2)/(n+1)$ implies that $S_{n+j}$ is smaller than the $(s_0+1)$-th largest calibration score and thus $S_{n+j}\le a_{i,(s_0+1)}$ in any case. Thus the last display is at most 
\begin{align*}
 \alpha \: \sum_{i\in \cH_{0}} \E\bigg[\frac{\ind{p_i\geq (s_0+2)/(n+1)}}{1\vee \sum_{j\in [m]} \ind{p_j \geq (s_0+2)/(n+1)}}\bigg] \leq \alpha \:  \E\bigg[\frac{\sum_{j\in [m] }\ind{p_j\geq (s_0+2)/(n+1)}}{1\vee \sum_{j\in [m]} \ind{p_j \geq (s_0+2)/(n+1)}}\bigg] \leq \alpha.
\end{align*}

\subsection{Proof of Theorem~\ref{cor:bFDR-adapt}}\label{sec:proofcor:bFDR-adapt}

The same arguments as in the previous section gives
\begin{align*}
&\bFDR(\ASLC)\\
&\leq  \sum_{i\in \cH_{0}}  \E\bigg[\bigg(\alpha/(m\hat{\pi}_0(W_i)) - 1/(n+1) + 1/(n+1)\bigg)\ind{\alpha/(m\hat{\pi}_0(W_i)) - 1/(n+1)>0}\bigg]\\
&\leq \sum_{i\in \cH_{0}}  \E\bigg[\alpha/(m\hat{\pi}_0(W_i))\bigg]\leq \alpha.
\end{align*}

\subsection{Proof of Theorem~\ref{th-subsampling}}\label{sec:proof:sub}

By applying Corollary~\ref{cor:bFDR} and Theorem~\ref{th:bFDR2}  in restriction to \(\calS\)  (since  $\alpha/\msub \ge 1/(n+1)$), we have
\begin{align*}
\PP\left(H_{\sigma_{\calS}(\hat{k}_{\calS})} = 0 \mid \calS\right) 
    \le \alpha\frac{m_{0,\calS}}{\msub},
\end{align*}
where we have denoted  \(m_{0,\calS}=|\cH_0\cap \calS|\)  the number of true nulls in \(\calS\).
Now observe that $\sigma_{\calS}(\hat{k}_{\calS})=\sigma(|R|)$ and $|R|>0$ if and only if $\hat{k}_{\calS} > 0$, so that
\begin{align*}
\PP\left(H_{\sigma(|R|)} = 0 \mid \calS\right) =\PP\left(H_{\sigma_{\calS}(\hat{k}_{\calS})} = 0 \mid \calS\right) 
    \le \alpha\frac{m_{0,\calS}}{\msub},
\end{align*}
Marginalizing over $\calS$, we obtain
\begin{align*}
\PP\left(H_{\sigma(|R|)} = 0  \right) 
     \leq \alpha\frac{\EE[m_{0,\calS}]}{\msub}
    =  \alpha\frac{m_0 \cdot (\msub/m)}{\msub}
    = \alpha \frac{m_0}{m},
\end{align*}
which concludes the proof.

\subsection{Proof of Theorem~\ref{th-multisubsampling}}\label{sec:proof:multisub}

Denote ${S}_{()}=(S_{\sigma(1)},\dots, S_{\sigma(m)})$ the ordered test scores and observe that the multi-subsampled method of Definition~\ref{def:multisubsampling} can be equivalently described as drawing $B$ subsamples $S^{[1]}_{()},\dots,S^{[B]}_{()}$ of \(|\calS|=\msub\) units independently and  uniformly randomly without replacement from the {\it ordered} test scores ${S}_{()}$. 
Then,  $r_b=|R_b|$, $b\in [B]$ (and thus also $r_{(\lceil B/2\rceil)}$)  are measurable with respect to $(S_j)_{j\in [n]}$ and $S^{[b]}_{()},b\in [B]$, where the latter solely depends on ${S}_{()}$ and some independent variables. 

Since by definition of $\hat{b}$, we have $\sum_{b\in [B]}\ind{r_b\geq r_{(\lceil B/2\rceil)}}\geq B/2$, we obtain 
\begin{align*}
\PP\left(H_{\sigma(|R|)}= 0 \right) &\leq (2/B)\sum_{b\in [B]}\E\left(\ind{r_b\geq r_{(\lceil B/2\rceil)}} \ind{ H_{\sigma(|R|)} = 0 }\right) \\
&=  (2/B)\sum_{b\in [B]}\E\left(\ind{r_b\geq r_{(\lceil B/2\rceil)}} \P\big( H_{\sigma(r_{(\lceil B/2\rceil)})} = 0 \mid (S_j)_{j\in [n]},{S}_{()},S^{[1]}_{()},\dots,S^{[B]}_{()}\big)\right),
\end{align*}
because, as mentioned above, $r_{(\lceil B/2\rceil)}$ is measurable with respect to $(S_j)_{j\in [n]}$ and $S^{[1]}_{()},\dots,S^{[B]}_{()}$.
Now, by Theorem~\ref{th-increasing}, we have that the function $k\in [m] \mapsto \P(H_{\sigma(k)}=0 \:|\: (S_j)_{j\in [n]},S_{\sigma(\cdot)})$ is nondecreasing. Hence, by independence between the data and the subsampling process, we also have that $k\in [m] \mapsto \P(H_{\sigma(k)}=0 \:|\: (S_j)_{j\in [n]},S_{\sigma(\cdot)},S^{[1]}_{()},\dots,S^{[B]}_{()})$ is nondecreasing. This gives that the last display is at most
\begin{align*}
& (2/B)\sum_{b\in [B]}\E\left(\ind{r_b\geq r_{(\lceil B/2\rceil)}} \P\big( H_{\sigma(r_{b})} = 0 \mid (S_j)_{j\in [n]},{S}_{()},S^{[1]}_{()},\dots,S^{[B]}_{()}\big)\right)\\
&\leq (2/B)\sum_{b\in [B]} \P\big( H_{\sigma(r_{b})} = 0\big)\\
&\leq (2/B)\sum_{b\in [B]} \alpha m_0/m,
\end{align*}
by applying Theorem~\ref{th-subsampling} (single subsampled case). 
This concludes the proof.

\subsection{Proof of Theorem~\ref{th-subsampling-adapt}}\label{sec:proofth-subsampling-adapt}

Let us establish it for the single subsampled case (the multiple subsampled case is completely analogue to the proof of Section~\ref{sec:proof:multisub}). Since the estimator $\hat{\pi}_0$ is defined with the whole test sample, it is important to consider $W_i$ in \eqref{equWi} not in restriction to $\calS$ but on the whole range $[m]$ (by contrast with the functional $H_{\ell,W_i}(\cdot)$, $M_{\ell,W_i}(\cdot)$, which are defined in restriction to $\calS$). By following an analogue route to the proofs of Theorems~\ref{th:bFDR-adapt}~and~\ref{cor:bFDR-adapt} (Sections~\ref{sec:proofth:bFDR-adapt}~and~\ref{sec:proofcor:bFDR-adapt}), we obtain
\begin{align*}
&\PP\left(H_{\sigma_{\calS}(\hat{k}_{\calS})} = 0 \mid \calS\right) \\
&    \le 
     \sum_{i\in \calS\cap \cH_{0}}  \E\bigg[\big(\alpha/(s\hat{\pi}_0(W_i))- 1/(n+1) + 1/(n+1)\big)\ind{\alpha/(s\hat{\pi}_0(W_i)) - 1/(n+1)>0}\mid \calS\bigg]\\
     &\leq
\sum_{i\in \calS\cap\cH_{0}} \E\bigg[\alpha/(s\hat{\pi}_0(W_i))\mid \calS\bigg]  \leq ( \alpha m/s) \:  \E\bigg[\frac{\sum_{j\in \calS }\ind{p_j\geq (s_0+2)/(n+1)}}{1\vee \sum_{j\in [m]} \ind{p_j \geq (s_0+2)/(n+1)}}\:\bigg|\: \calS\bigg].
\end{align*}
Now using that the data (and thus the $p_i$'s) are independent of $\calS$, and by integrating with respect to $\calS$, we obtain
\begin{align*}
\PP\left(H_{\sigma_{\calS}(\hat{k}_{\calS})} = 0 \right) 
    &\le 
     ( \alpha m/s) \:  \E\bigg[\frac{\sum_{j\in [m] }\ind{j\in \calS}\ind{p_j\geq (s_0+2)/(n+1)}}{1\vee \sum_{j\in [m]} \ind{p_j \geq (s_0+2)/(n+1)}}\bigg] \\
     &\leq ( \alpha m/s) \:  \E\bigg[\frac{\sum_{j\in [m] } (s/m)\ind{p_j\geq (s_0+2)/(n+1)}}{1\vee \sum_{j\in [m]} \ind{p_j \geq (s_0+2)/(n+1)}}\bigg]\leq \alpha.
\end{align*}

\subsection{Proof of Theorem~\ref{th-increasing}}\label{secproof:th-increasing}

We use the density of $\sigma$ given $S_{\sigma(\cdot)}=x=(x_1,\dots,x_m)$ for $x_1\geq \dots \geq x_m$:
\begin{align*}
 \P(H_{\sigma({k})} = 0\:|\: S_{\sigma(\cdot)}=x) &= \sum_{i\in \cH_{0}} \P(\sigma({k}) = i\:|\: S_{\sigma(\cdot)}=x) \\
 &= C(x)\sum_{i\in \cH_{0}} \sum_{\tau \in \mathfrak{S}_{m}}\ind{\tau({k}) = i}\prod_{j\in \cH_{0}} f_{0}(x_{\tau^{-1}(j)})\prod_{j\in \cH_{1}} f_{j}(x_{\tau^{-1}(j)}) ,
\end{align*}
where $C(x)>0$ is some constant depending on $x$ and $\mathfrak{S}_{m}$ denotes the set of permutation of $[m]$.
Now take $k\in [m-1]$, and consider $\tau_{k,k+1}$ the permutation of $\mathfrak{S}_{m}$ swapping $k$ and $k+1$. We have 
\begin{align*}
& \P(H_{\sigma({k+1})} = 0\:|\: S_{\sigma(\cdot)}=x) - \P(H_{\sigma({k})} = 0\:|\: S_{\sigma(\cdot)}=x) \\
 &= C(x)\sum_{i\in \cH_{0}} \bigg[\sum_{\tau \in \mathfrak{S}_{m}}\ind{\tau({k+1}) = i}\prod_{j\in \cH_{0}} f_{0}(x_{\tau^{-1}(j)})\prod_{j\in \cH_{1}} f_{j}(x_{\tau^{-1}(j)})\\
 &\:\:\:\:\:\:\:\:\:\:\:\:-C(x)\sum_{\tau \in \mathfrak{S}_{m}}\ind{\tau({k}) = i}\prod_{j\in \cH_{0}} f_{0}(x_{\tau^{-1}(j)})\prod_{j\in \cH_{1}} f_{j}(x_{\tau^{-1}(j)}) \bigg]\\
&= C(x)\sum_{i\in \cH_{0}} \bigg[\sum_{\tau \in \mathfrak{S}_{m}}\ind{\tau({k+1}) = i}\prod_{j\in \cH_{0}} f_{0}(x_{\tau^{-1}(j)})\prod_{j\in \cH_{1}} f_{j}(x_{\tau^{-1}(j)})\\
 &\:\:\:\:\:\:\:\:\:\:\:\:-C(x)\sum_{\tau' \in \mathfrak{S}_{m}}\ind{\tau'({k+1}) = i}\prod_{j\in \cH_{0}} f_{0}(x_{\tau_{k,k+1}\circ(\tau')^{-1}(j)})\prod_{j\in \cH_{1}} f_{j}(x_{\tau_{k,k+1}\circ(\tau')^{-1}(j)}) \bigg],
\end{align*}
by letting $\tau'=\tau \circ \tau_{k,k+1}$. Hence, the above display is equal to
\begin{align*}
&C(x)\sum_{i\in \cH_{0}} \sum_{\tau \in \mathfrak{S}_{m}}\ind{\tau({k+1}) = i}\prod_{j\in \cH_{0}\backslash\{\tau(k),\tau(k+1)\}} f_{0}(x_{\tau^{-1}(j)})\prod_{j\in \cH_{1}\backslash\{\tau(k)\}} f_{j}(x_{\tau^{-1}(j)})\\
 &\:\:\bigg[ \ind{\tau(k)\in \cH_0} \big(f_{0}(x_{k}) f_{0}(x_{k+1}) -  f_{0}(x_{k+1}) f_{0}(x_{k})\big) \\
 &\:\:\:\:\:\:\:\:+ \ind{\tau(k)\in \cH_1} \big(f_{0}(x_{k+1}) f_{\tau(k)}(x_{k}) - f_{0}(x_{k}) f_{\tau(k)}(x_{k+1})\big)\bigg]\geq 0,
\end{align*}
 because $ f_{0}(x_{k+1}) f_{\tau(k)}(x_{k}) \geq f_{0}(x_{k}) f_{\tau(k)}(x_{k+1})$ when $\tau(k)\in \cH_1$ by Assumption~\ref{ass-scoredensity}.

\subsection{Proof of Proposition~\ref{prop:iso-gren-equivalence}}
\label{sec:proof-prop-iso}

{Clearly, $\eqref{eq:original-SLC} \iff \eqref{eq:shifted-SLC}$ and $\eqref{eq:iso-SLC} \iff \eqref{eq:gren-SLC}$. Let us prove the equivalence of \eqref{eq:shifted-SLC} and \eqref{eq:iso-SLC}. {Figure~\ref{fig:gren-prop-visual-aid} illustrates the key geometric relationship between these two representations. We provide a more rigorous algebraic argument as follows.} For this, let us denote $G_{\mathrm{iso}}$ the greatest convex minorant of the curve $k\mapsto \tilde{p}_{(k)}$, so that $\widehat{\lfdr}_{\mathrm{iso}}(k) = m(G_{\mathrm{iso}}(k)-G_{\mathrm{iso}}(k-1))$ for $k\in [m]$. 
Let us denote
 \begin{align*}
        \hat{k}^{\mathrm{iso}} 
        &:= 
      \max \Big\{ k\in [0,m] : \widehat{\lfdr}_{\mathrm{iso}}(k) \leq \alpha \Big\} 
    \end{align*}
and show that $\hat{k}^{\mathrm{iso}} =\hat{k}$. 
First, since the supporting line of slope $\alpha/m$ is below the graph of the curve $k\mapsto \tilde{p}_{(k)}$, it is a convex minorant and we have by definition of the greatest convex minorant that $G_{\mathrm{iso}}(k)\geq \tilde{p}_{(\hat{k})}+\alpha (k-\hat{k}) /m$  for all $k\in [0,m]$ (note that this is true even if $\hat{k}=0$). 
This yields $G_{\mathrm{iso}}(\hat{k})\geq \tilde{p}_{(\hat{k})}$ and thus $G_{\mathrm{iso}}(\hat{k})= \tilde{p}_{(\hat{k})}$ (because $G_{\mathrm{iso}}$ is a minorant).
This in turn implies for all $k\in [0,m]$,
$$G_{\mathrm{iso}}(k)\geq G_{\mathrm{iso}}(\hat{k})+\alpha (k-\hat{k}) /m$$  and thus $G_{\mathrm{iso}}(k)- G_{\mathrm{iso}}(\hat{k})\geq \alpha (k-\hat{k}) /m$ for $k\geq \hat{k}$ and $G_{\mathrm{iso}}(\hat{k})- G_{\mathrm{iso}}(k)\leq \alpha (\hat{k}-k) /m$ for $k\leq \hat{k}$.
Since $G_{\mathrm{iso}}(\hat{k}^{\mathrm{iso}}+1)- G_{\mathrm{iso}}(\hat{k}^{\mathrm{iso}})>\alpha/m$ by definition of $\hat{k}^{\mathrm{iso}}$, and by convexity of $G_{\mathrm{iso}}$, we cannot have $\hat{k}^{\mathrm{iso}}+1\leq \hat{k}$: if $\hat{k}^{\mathrm{iso}}+1\leq \hat{k}$, then 
$$\alpha/m\geq \frac{G_{\mathrm{iso}}(\hat{k})- G_{\mathrm{iso}}(\hat{k}^{\mathrm{iso}})}{\hat{k}-\hat{k}^{\mathrm{iso}}}\geq G_{\mathrm{iso}}(\hat{k}^{\mathrm{iso}}+1)- G_{\mathrm{iso}}(\hat{k}^{\mathrm{iso}})>\alpha/m,$$
which is a contradiction.
This shows $\hat{k}^{\mathrm{iso}}\geq \hat{k}$. 
Hence  
\begin{align*}
\alpha (\hat{k}^{\mathrm{iso}}-\hat{k})/m&\geq (G_{\mathrm{iso}}(\hat{k}^{\mathrm{iso}})- G_{\mathrm{iso}}(\hat{k}^{\mathrm{iso}}-1))(\hat{k}^{\mathrm{iso}}-\hat{k})\\
&\geq G_{\mathrm{iso}}(\hat{k}^{\mathrm{iso}})- G_{\mathrm{iso}}(\hat{k})\geq \alpha (\hat{k}^{\mathrm{iso}}-\hat{k}) /m    
\end{align*}
 and the inequalities above are equalities. 
Also, classically, by the definition of the greatest convex minorant we have $G_{\mathrm{iso}}(\hat{k}^{\mathrm{iso}})=\tilde{p}_{(\hat{k}^{\mathrm{iso}})}$ (see below for a proof). Combining this gives
$$
\tilde{p}_{(\hat{k}^{\mathrm{iso}})}- \tilde{p}_{(\hat{k})} = \alpha (\hat{k}^{\mathrm{iso}}-\hat{k}) /m
$$
and $\hat{k}^{\mathrm{iso}}$ also minimizes $k\in [0,m]\mapsto \tilde{p}_{(k)}-\alpha k/m$ which means $\hat{k}^{\mathrm{iso}}\leq \hat{k}$ and thus $\hat{k}^{\mathrm{iso}} =\hat{k}$. 
}

{To conclude the proof, we have now to prove $G_{\mathrm{iso}}(\hat{k}^{\mathrm{iso}})=\tilde{p}_{(\hat{k}^{\mathrm{iso}})}$ and thus only $G_{\mathrm{iso}}(\hat{k}^{\mathrm{iso}}) \geq \tilde{p}_{(\hat{k}^{\mathrm{iso}})}$ (because $G_{\mathrm{iso}}$ is a minorant). First, by definition of $\hat{k}^{\mathrm{iso}}$, we have
$
G_{\mathrm{iso}}(\hat{k}^{\mathrm{iso}}+1)- G_{\mathrm{iso}}(\hat{k}^{\mathrm{iso}}) > \alpha/m \geq G_{\mathrm{iso}}(\hat{k}^{\mathrm{iso}})- G_{\mathrm{iso}}(\hat{k}^{\mathrm{iso}}-1),
$
which gives
$$
\frac{G_{\mathrm{iso}}(\hat{k}^{\mathrm{iso}}+1)+G_{\mathrm{iso}}(\hat{k}^{\mathrm{iso}}-1)}{2} > G_{\mathrm{iso}}(\hat{k}^{\mathrm{iso}}) .
$$
Now let 
$$
\tilde{G}(k) := \left\{\begin{array}{ll} \tilde{p}_{(\hat{k}^{\mathrm{iso}})}\wedge\frac{G_{\mathrm{iso}}(\hat{k}^{\mathrm{iso}}+1)+G_{\mathrm{iso}}(\hat{k}^{\mathrm{iso}}-1)}{2} &\mbox{ if $k=\hat{k}^{\mathrm{iso}}$} \\
G_{\mathrm{iso}}(k) &\mbox{ if $k\neq \hat{k}^{\mathrm{iso}}$.}
\end{array}\right.
$$
By  definition, $\tilde{G}$ is a convex minorant of $k\mapsto \tilde{p}_{(k)}$. Since $G_{\mathrm{iso}}$ is the largest convex minorant, this means $\tilde{G}(\hat{k}^{\mathrm{iso}})\leq G_{\mathrm{iso}}(\hat{k}^{\mathrm{iso}})$ and thus 
$\tilde{p}_{(\hat{k}^{\mathrm{iso}})}\leq G_{\mathrm{iso}}(\hat{k}^{\mathrm{iso}})$.}

    \begin{figure}
        \centering
        \includegraphics[width=\linewidth]{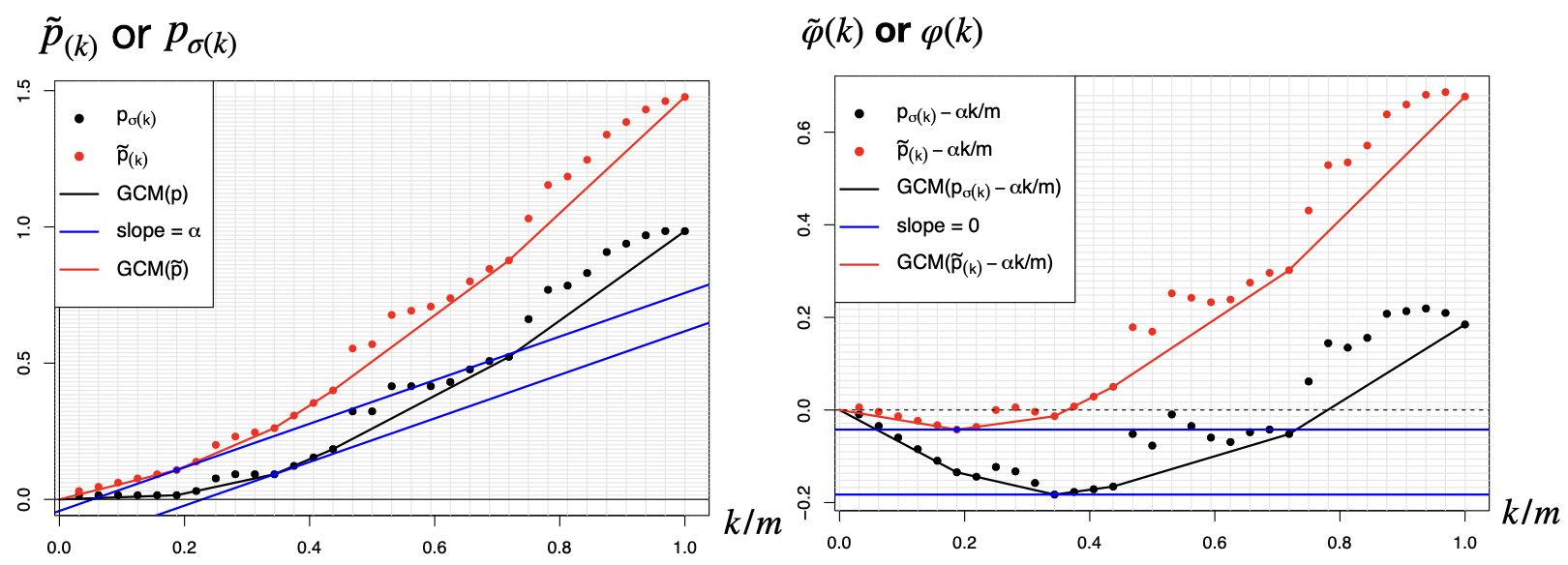}
        \caption{Visual aid for the proof of Proposition \ref{prop:iso-gren-equivalence}, where the notation $\varphi$ and $\tilde{\varphi}$ is defined $\varphi(k) \coloneqq p_{(k)}-\alpha k /m$ and $\tilde{\varphi}(k) \coloneqq \tilde{p}_{(k)}-\alpha k /m$ for $k=0,1,\dots,m$.}
        \label{fig:gren-prop-visual-aid}
    \end{figure}

\subsection{Proof of Corollary~\ref{cor-bFDRimpliesFDR}}\label{proofcor-bFDRimpliesFDR}

We apply Theorem~\ref{th-increasing}. The tower property of conditional expectation implies
\begin{align*}
    \FDR(R) &= \E \Big[ \frac{1}{1\vee \hat{k}}\sum_{k=1}^{\hat{k}} \P(H_{\sigma(k)}=0 \mid (S_{j})_{j\in[n]},S_{\sigma(\cdot)}) \Big] \\
    &\leq  \E \Big[  \frac{1}{1\vee \hat{k}}\sum_{k=1}^{\hat{k}} \P(H_{\sigma(\hat{k})}=0 \mid (S_{j})_{j\in[n]},S_{\sigma(\cdot)}) \Big] = \P(H_{\sigma(\hat{k})}=0)=\bFDR(\SLC),
\end{align*}
where recall the convention that $H_{\sigma(0)}:= 1$.

\section{Additional lower bounds and counterexamples}

\subsection{Improved lower-bound }\label{sec:morelbforub}

\begin{proposition}
\label{prop:lower-bound}
     Assume $m_0<m$ and let $\alpha\in (0,1)$ satisfy $\alpha \ge 1 / \{(n+1)(1-m_0/m)\}$. Then there exists a data generation process satisfying Assumption \ref{assglobal} for which the following hold:
     \begin{enumerate}
         \item {\it Tightness:} If $m_0=1$ and $x:=(n+1)\alpha/m$ is an integer, then
         \begin{align*}
             \bFDR(\SL) = \frac{\alpha}{m}+\frac{1}{n+1}= \frac{m_0}{m} \alpha + \frac{m_0}{n+1},
         \end{align*}
         i.e. the upper bound of Theorem \ref{th:bFDR} is attained.
         \item {\it Improved lower bound:% when $m_0>1$
         }
          Let $\ell := \lfloor x \rfloor$. Then
         \begin{align*}
             \bFDR(\SL) \geq 1-\prod_{j=0}^\ell \frac{n-j}{m_0+n-j} \geq \frac{(\ell+1)m_0}{n+1} - \Big( \frac{(\ell+1)m_0}{n-\ell}\Big)^2.
         \end{align*}
         When $x=\ell$, this gives $\bFDR \geq \pi_0\alpha+\frac{m_0}{n+1}- \Big(\frac{(\ell+1)m_0}{n-\ell} \Big)^2 = \pi_0\alpha+\frac{m_0}{n+1}- O\Big((\pi_0\alpha+\frac{m_0}{n+1})^2 \Big)$.
     \end{enumerate}
\end{proposition}

\begin{proof}
    Consider iid null scores distributed uniformly in $(0,1)$, and independent alternative scores which are iid and distributed uniformly in $(1,2)$, so that all alternative $p$-values are equal to $1/(n+1)$. We will first show the general lower bound:
    \begin{align}
    \label{eq:general-lower-bound}
        \bFDR(\SL) \geq 1 - \prod_{j=0}^\ell \frac{n-j}{m_0+n-j},
    \end{align}
    which yields the conclusions in parts 1 and 2 as special cases. To this end, we have, since $1/(n+1) - \alpha m_1/m\leq 0$, 
    \begin{align*}
        \bFDR(\SL) &= \P(\exists r \leq m_0 : q_{(r)} - \alpha(m_1+r)/m \leq 1/(n+1) - \alpha m_1/m) \\
        &\geq \P(q_{(1)} \leq 1/(n+1) + \alpha/m)=\P(q_{(1)}\leq (1+\ell)/(n+1)),
    \end{align*}
    where $q_{(1)} \leq \dots \leq q_{(m_0)}$ are the ordered nulls, $x = (n+1)\alpha/m$, and $\ell=\lfloor x\rfloor$, as defined in the proposition. The complement is
\begin{align*}
    q_{(1)}> (1+x)/(n+1) \iff \# \{j \leq n : \max_{i \in \H_0} S_{n+i} < S_j\} > \ell.
\end{align*}
In other words, among the pooled set of null and calibration scores, the top $\ell+1$ are calibration scores. By exchangeability, 
\begin{align*}
    \P\Big(\# \Big\{j\leq n : \max_{i \in \mathcal{H}_0} S_{n+i} < S_j \Big\}> \ell\Big) = \prod_{j=0}^\ell \frac{n-j}{m_0+n-j},
\end{align*}
which yields \eqref{eq:general-lower-bound}.

\paragraph{Part 1.} For $m_0=1$, the product is telescoping:
\begin{align*}
    \prod_{j=0}^\ell \frac{n-j}{n+1-j} = \frac{n}{n+1} \cdot \frac{n-1}{n} \cdot \frac{n-2}{n-1}\cdot \dots \cdot \frac{n-\ell}{n+1-\ell} = \frac{n-\ell}{n+1},
\end{align*}
which gives $\bFDR(\SL) \geq 1-\frac{n-\ell}{n+1} = \frac{\ell+1}{n+1} = \frac{\alpha}{m}+\frac{1}{n+1}$, assuming $x=\ell=\frac{\alpha (n+1)}{m}$.

\paragraph{Part 2.} We write $\frac{n-j}{m_0+n-j} = \frac{1}{1+a_j}$, where $a_j = \frac{m_0}{n-j}$. Since $a_j \geq 0$, we have
\begin{align*}
    \prod_{j=0}^\ell (1+a_j) \geq 1+\sum_{j=0}^\ell a_j.
\end{align*}
Let $A \coloneqq \sum_{j=0}^\ell a_j$, so that the above inequality implies
\begin{align*}
    \bFDR(\SL) \geq 1 - \prod_{j=0}^\ell \frac{1}{1+a_j} \geq 1 - \frac{1}{1+A} = \frac{A}{1+A} \geq A - A^2,
\end{align*}
since $A \geq 0$. %$A(1-A)\leq \frac{A}{A+1} \iff 1 \leq A + \frac{1}{A+1} \iff A+1 \leq A(A+1)+1 \iff A+1 \leq A^2 + A + 1$ which holds since $A^2 \geq 0$.
Since the summands $(a_j)$ are monotone, we have $\frac{(\ell+1)m_0}{n} \leq A \leq \frac{(\ell+1)m_0}{n-\ell}$. Then
\begin{align*}
    \bFDR(\SL) \geq \frac{(\ell+1)m_0}{n} - \Big(\frac{(\ell+1)m_0}{n-\ell}\Big)^2.
\end{align*}
If $x=\ell=\frac{\alpha(n+1)}{m}$, then 
\begin{align*}
    \frac{(\ell+1)m_0}{n} &\geq \frac{(\ell+1)m_0}{n+1} = \pi_0 \alpha + \frac{m_0}{n+1}, \\
    \Big(\frac{(\ell+1)m_0}{n-\ell}\Big)^2 &= \Big(\frac{n+1}{n-\ell}\Big)^2 \Big( \pi_0\alpha+\frac{m_0}{n+1}\Big)^2,
\end{align*}
which yields the final claim.
\end{proof}

\begin{remark}\label{rem:match}
When $\alpha \to 0$, $n\to \infty$, $\pi_0\sim c\in (0,1)$,  while $m/(n+1)$ is of order $\alpha$, Proposition \ref{prop:lower-bound} and Theorem \ref{th:bFDR} imply
\begin{align*}
\bFDR(\SL) \sim \pi_0 \alpha + m_0/(n+1).
\end{align*}
If in addition, $\alpha\sim \kappa m/(n+1)$ tends to zero as $n,m\to \infty$, Theorem \ref{th:bFDR} gives
\begin{align*}
\bFDR(\SL) \sim (1+1/\kappa) \pi_0 \alpha \mbox{ and }\bFDR(\SLC) \sim   \pi_0\alpha.    
\end{align*}
In other words, no other correction concerning $\alpha$ is possible in this regime to provide control at level $\pi_0\alpha$.
\end{remark}

\subsection{SL procedure with randomized conformal $p$-values}\label{sec:randpvalues}

Randomized conformal $p$-values are given by \citep{vovk2005algorithmic}
\begin{equation}
    \label{equ-conformalpvalues-rand}
  \tilde{p}_{i} = (n+1)^{-1} \bigg(U_i+\sum_{j=1}^{n} \ind{S_{j}\geq S_{n+i}}\bigg),\:\:\: i\in \range{m},
\end{equation}
for $U_i$ iid $U(0,1)$ independent of all previous random quantities. While it depends on an additional random variable $U_i$, each $p$-value $\tilde{p}_{i}$ is marginally uniform under the null, while we have
\begin{align}
 p_i-1/(n+1) \leq   \tilde{p}_{i}\leq p_i.
\end{align}

\begin{proposition}
Assume $m_{0}<m$ and $\alpha\in (0,1)$ satisfying $\alpha\geq m/(n+1)$. If we use the ordering of the scores in the SL definition, and randomized $p$-values to compute $\hat{k}$ in \eqref{equkchap-conf}, then this randomized SL procedure $\widetilde{\SL}$ is such that there exists a data generation process satisfying Assumption \ref{assglobal} with
\begin{align*}
 \bFDR(\widetilde{\SL}) &\geq 1-  \prod_{j=0}^{\ell-1} \frac{n-j}{m_0+n-j}\\
 &\geq \frac{m_0}{m_0+n} \vee \bigg( \pi_0 \alpha (1+1/n)  - \Big(\frac{\alpha \pi_0 }{1-\ell/(n+1)}\Big)^2\bigg).
\end{align*}
with $\ell:= \lfloor(n+1)\alpha/m\rfloor\geq 1$. 
\end{proposition}

\begin{proof}
Note that by assumption $\alpha \geq m/(n+1)\ge 1 / \{(n+1)(1-m_0/m)\}$. 
We consider the same framework as for the  non-randomized proof in the previous section: 
\begin{align*}
    \bFDR(\widetilde{\SL}) &\geq \P(\exists r \leq m_0 : \tilde{q}_{(r)} - \alpha(m_1+r)/m \leq - \alpha m_1/m) \\
    &\geq \P(\tilde{q}_{(1)} \leq \alpha/m) \geq \P({q}_{(1)}\leq x/(n+1)),
\end{align*}
where $\tilde{q}_{(1)} \leq \dots \leq \tilde{q}_{(m_0)}$ are the ordered null randomized $p$-values and $x := (n+1)\alpha/m\geq 1$, $\ell=\lfloor x\rfloor\geq 1$ and because of \eqref{equ-conformalpvalues-rand}. This leads to the lower bound
\begin{align*}
    \bFDR(\widetilde{\SL}) &\geq 1-  \prod_{j=0}^{\lfloor(n+1)\alpha/m\rfloor-1} \frac{n-j}{m_0+n-j}   \geq  \frac{m_0}{m_0+n} .
\end{align*}

This entails 
\begin{align*}
    \bFDR(\widetilde{\SL}) &\geq 1-  \prod_{j=0}^{\ell-1} \frac{n-j}{m_0+n-j}   \geq   \frac{A}{1+A} \geq A - A^2, 
\end{align*}
for $A := m_0 \sum_{j=0}^{\ell-1} (n-j)^{-1}$, which is at least $\ell m_0 / n = \pi_0 \alpha (1+1/n) $ and at most $\ell m_0 / (n-\ell+1)$. 
\end{proof}

\subsection{Counterexample under independence} \label{sec:violationindep}

The bFDR control of the SL procedure can be also violated outside the conformal setting, in a discrete independent case: consider the case where there is only one $p$-value under the null $(m_0=1)$, say $p_1\in \{1/2, 1\}$ such that $\mathbb{P}(p_1=1/2)=1/2$, and where the $m_1=m-1$ other (alternative) $p$-values are all equal to $1/2$. 
Suppose that the tied $p$-values are ordered according to some independent test statistics, with the null statistic being less extreme than the alternatives. Then, for $\alpha > 1/2$, the bFDR is at least equal to $1/2$ so is above $\alpha m_0/m = \alpha/m $ provided that $m\geq 2$.

\section{Technical lemmas}

\begin{lemma}\label{lem:emptyifmin}
 Assume that $(S_{j})_{j\in \range{n+m}}$ is a vector with no ties almost surely and consider $\hat{k}$ as in \eqref{equkchap-conf} and $S_{(1)}>\dots>S_{(n)}$ the ordered calibration scores (convention $S_{(0)}=+\infty$). Then we have almost surely
\begin{equation}
    \label{equ-trick}
   \{ p_i = \ell/(n+1), \sigma(\hat{k}) = i \}\subseteq \{\mbox{there is no $j_0\in \range{m}\backslash\{i\}$ with $S_{n+j_0}\in (S_{(\ell)},S_{n+i})$}\}.
\end{equation}   
\end{lemma}

\begin{proof}
Since  $p_i = \ell/(n+1)$, we have $S_{n+i}\in (S_{(\ell)},S_{(\ell-1)})$.
Hence, if $j_0\in \range{m}\backslash\{i\}$ exists with $S_{n+j_0}\in (S_{(\ell)},S_{n+i})$, then $S_{n+j_0}$ would have the same $p$-value as $S_{n+i}$
with a larger rank, and this would contradicts $\sigma(\hat{k}) = i$.     
\end{proof}

\begin{lemma}
    \label{lemmafromktoell}
 Assume that $(S_{j})_{j\in \range{n+m}}$ is a vector with no ties almost surely and consider $\hat{k}$ as in \eqref{equkchap-conf} and
\begin{align}
    \hat{\ell} = \max\bigg\{\argmin_{{\ell} \in \range{0,n+1}} \Big\{ \ell/(n+1) - (\alpha/m) \sum_{j' \in \range{m}}\ind{p_{j'}\leq \ell/(n+1)} \Big\}\bigg\}
\label{equhatell}
\end{align}
Then for all $i\in [m]$, $i=\sigma(\hat{k})$ implies $p_i=\hat{\ell}/(n+1)$ almost surely with $1\leq \hat{\ell}\leq \msub:= \lfloor \alpha(n+1) \rfloor\in [n]$. In addition, if  $\hat{k}$ is given by \eqref{equkchap-conf-adapt} (that is with the restriction $p_{\sigma(k)}\leq s_0/(n+1) $, $s_0\in [n]$, and $\alpha/\hat{\pi}_0$ is used in place of $\alpha$ for $\hat{\pi}_0$ as in \eqref{equpi0hat}), we have that 
\begin{align}
    \hat{\ell} = \max\bigg\{\argmin_{{\ell} \in \range{0,s_0}} \Big\{ \ell/(n+1) - (\alpha/m) \sum_{j' \in \range{m}}\ind{p_{j'}\leq \ell/(n+1)} \Big\}\bigg\}
\label{equhatell-adapt}
\end{align}
is such that  $i\in [m]$, $i=\sigma(\hat{k})$ implies $p_i=\hat{\ell}/(n+1)$ almost surely with $\hat{\ell}\leq s_0$.
\end{lemma}

\begin{proof}
Let us denote  $S_{(1)}>\dots>S_{(n)}$ the ordered calibration scores (convention $S_{(0)}=+\infty$). Then we can define the set
\begin{align*}
\mathcal{L}&= \{
 \ell \in [n] \mbox{ with } (S_{(\ell)},S_{(\ell-1)})\cap \{S_{n+j},j\in [m]\} \neq \emptyset\};\\
 \mathcal{K}&= \bigg\{ \sum_{j\in [m]}\ind{S_{n+j}> S_{(\ell)}} , \ell \in \mathcal{L}\bigg\};
\end{align*}
When $\hat{k}\geq 1$, we have
  \begin{align*}
  \hat{k} &= \max \bigg\{\argmin_{{k} \in \mathcal{K}} \{ p_{\sigma(k)} - \alpha k/m\}\bigg\}\\
 &= \max\bigg\{\argmin_{{\ell} \in \mathcal{L}} \Big\{ \ell/(n+1) - (\alpha/m) \sum_{j \in \range{m}}\ind{p_{j}\leq \ell/(n+1)} \Big\}\bigg\}=\hat{\ell}.
\end{align*}
The first and second equality are clear because the $p$-value $\ell/(n+1)$ is only realized if $(S_{(\ell)},S_{(\ell-1)})\cap \{S_{n+j},j\in [m]\} \neq \emptyset$, and among each $(S_{(\ell)},S_{(\ell-1)})$, $\ell\in \mathcal{L}$, the minimum can only occur for the largest rank, which is equal to $\sum_{j\in [m]}\ind{S_{n+j}> S_{(\ell)}}= \sum_{j \in \range{m}}\ind{p_{j}\leq \ell/(n+1)}$. This gives the equality of the two $\argmin$ sets and thus the result. The third equality is obtained similarly, by observing that the minimum in \eqref{equhatell} can not occur for a $\ell$ with $(S_{(\ell)},S_{(\ell-1)})\cap \{S_{n+j},j\in [m]\} = \emptyset$, because either there is $\ell'< \ell$ with $(S_{(\ell')},S_{(\ell'-1)})\cap \{S_{n+j},j\in [m]\}\neq \emptyset$ and this would give a value of the objective function smaller, or all $\ell'< \ell$ will also be such that $(S_{(\ell')},S_{(\ell'-1)})\cap \{S_{n+j},j\in [m]\}= \emptyset$, but in this case the objective function is positive for these sets and the minimum arises for some $\ell'>\ell$ with $(S_{(\ell')},S_{(\ell'-1)})\cap \{S_{n+j},j\in [m]\}\neq \emptyset$ (which exists because $\hat{k}\geq 1$). 

Let us now prove $\hat{\ell}\leq \msub:= \lfloor \alpha(n+1) \rfloor$. Assume $\hat{\ell}\geq 1$ (otherwise the result is trivial). Then we have $$\hat{\ell}/(n+1) - \alpha\leq \hat{\ell}/(n+1) - (\alpha/m) \sum_{j' \in \range{m}}\ind{p_{j'}\leq \hat{\ell}/(n+1)} \leq 0.$$
This gives $\hat{\ell}\leq \alpha(n+1)$ and thus $\hat{\ell}\leq \msub$. The last statement can be proven similarly.
\end{proof}

\begin{lemma}[\cite{marandon2024adaptive}]\label{lemmaPropConformalFull}
 Under Assumption~\ref{assglobal}, let for any fixed $i\in \cH_{0}$, $W_i$ and $\Psi_{ij}(x,W_i)$ being defined by \eqref{equWi} and \eqref{equPsii} respectively.  Then we have
\begin{itemize}
    \item[(i)] $\mathbf{p}_{-i}:=(p_j)_{j\in\range{m}\backslash\{i\}}$ is equal to $(\Psi_{ij}(p_i,W_i))_{j\in\range{m}\backslash\{i\}}$ and each $u\in [0,1] \mapsto \Psi_{ij}(u,W_i)\in \R$ is a nondecreasing function ;
    \item[(ii)] $(n+1)p_i $ is uniformly distributed on $\range{n+1}$ and independent of $W_i$ under Assumption~\ref{assglobal}.
\end{itemize}  
\end{lemma}

\begin{proof}
We provide a proof for completeness. We have for any $i\in \cH_{0}$, and $j \neq i$,
\begin{align*}
 p_{j} &= (n+1)^{-1} \bigg(1- \ind{S_{n+i}\geq S_{n+j}} +\sum_{s\in A_{i}} \ind{s\geq S_{n+j}}\bigg)\\
&= (n+1)^{-1} \bigg(1- \ind{a_{i,((n+1)p_{i})}\geq S_{n+j}} +\sum_{s\in A_{i}} \ind{s\geq S_{n+j}}\bigg)
\end{align*}
because $(n+1)p_{i}$ is the rank of $S_{n+i}$ in $A_{i}$. This gives (i). As for (ii), this comes because the rank of $S_{n+i}$ in $A_{i}$ is independent of $W_{i}$ by exchangeability of $(S_{1},\dots,S_{n}, S_{n+i})$ conditionally on $(S_{n+j})_{j\in\range{m}\backslash\{i\}}$.
\end{proof}

\begin{lemma}\label{lemH}
Assume that $(S_{j})_{j\in \range{n+m}}$ is a vector with no ties almost surely, fix $i\in [m]$, $\ell\in [n+1]$ and consider $H_{\ell,W_i}(\cdot)$ the functional given by \eqref{equ-keyfunctionalHell}, the quantity $M_{\ell,W_i}$ defined by \eqref{equMell} and the quantity $U_{\ell,W_i}$ defined by \eqref{equUell}, that all only depend on $\ell\in [n+1]$ and $W_i$ \eqref{equWi}. Then the following holds when $p_i=\ell/(n+1)$ and $\sigma(\hat{k})=i$ with $\hat{k}$ given by \eqref{equkchap-conf}:
\begin{itemize}
    \item[(i)] for all $\ell'\in [n+1]$, $H_{\ell,W_i}(\ell')= \ell'/(n+1) - (\alpha/m) \sum_{j \in \range{m}}\ind{p_{j}\leq \ell'/(n+1)} $;
    \item[(ii)] for all $\ell'\in [n+1]$,  $H_{\ell,W_i}(\ell')\leq H_{\ell'+1,W_i}(\ell')-(\alpha/m)\ind{\ell\leq \ell'}$ and $H_{\ell,W_i}(\ell)= H_{\ell+1,W_i}(\ell)-\alpha/m$;
    \item[(iii)] $\ell$ is the maximum of the set $\argmin_{{\ell'} \in \range{0,n+1}} H_{\ell,W_i}(\ell')$;
    \item[(iv)] if $\ell\geq 2$, $M_{\ell,W_i} - M_{\ell-1,W_i} \geq 1/(n+1)$; 
    \item[(v)]
    $H_{1,W_i}(1)\geq 1/(n+1)-(\alpha/m)\sum_{j\in [m]\backslash\{i\}} \ind{S_{n+j}\geq a_{i,(1)}} - \alpha/m $;
    \item[(vi)]
    %for $\ell\geq 1$,
    $H_{\ell,W_i}(1)\leq 1/(n+1) -(\alpha/m) \sum_{j\in [m]\backslash\{i\}} \ind{S_{n+j}\geq a_{i,(1)}}$;
     \item[(vii)] %for $\ell\geq 1$, 
     $U_{\ell-1,W_i}\leq H_{\ell+1,W_i}(\ell) \leq U_{\ell,W_i} - 1/(n+1)$ provided that the separation \eqref{equ-separation} holds and $1/(n+1)\leq \alpha/m$.
\end{itemize}
In addition, in the adaptive case where $\hat{k}$ is given by \eqref{equkchap-conf-adapt} $p_i=\ell/(n+1)$ and $\sigma(\hat{k})=i$ implies (i), (ii), (v), (vi) with $\alpha/\hat{\pi}_0(W_i)$ in place of $\alpha$  with $\hat{\pi}_0(W_i)$ given by \eqref{pi0chapWi} and (iii), (iv)  become
\begin{itemize}
    \item[(iii)'] $\ell$ is the maximum of the set $\argmin_{{\ell'} \in \range{0,s_0}} H_{\ell,W_i}(\ell')$;
    \item[(iv)'] if $\ell\in [2,s_0]$, $M_{\ell,W_i} - M_{\ell-1,W_i} \geq 1/(n+1)$; 
 \end{itemize}
\end{lemma}

\begin{proof}
    Point (i) comes from  $a_{i,(\ell)}=S_{n+i}$ and thus for any $j\in [m]\backslash\{i\}$, we have $p_{j}\leq \ell'/(n+1)$ iff $S_{n+j}\geq a_{i,(\ell')}$ when $\ell'<\ell$ and iff $S_{n+j}\geq a_{i,(\ell'+1)}$ when $\ell'>\ell$. As for $p_{i}\leq \ell'/(n+1)$, it occurs if 
    $\ell\leq \ell'$ because $p_i=\ell/(n+1)$. This leads to the desired expression for $\ell'\neq \ell$. Now, for $\ell'=\ell$, the expression also holds because for any $j\in [m]\backslash\{i\}$, $S_{n+j}\geq a_{i,(\ell+1)}$ iff $S_{n+j}\geq a_{i,(\ell)}$ by Lemma~\ref{lem:emptyifmin}.
    Point (ii) follows from $\sum_{j \in \range{m}\backslash\{i\} }\ind{ S_{n+j}\geq a_{i,(\ell'+1)}}\geq \sum_{j \in \range{m}\backslash\{i\} }\ind{ S_{n+j}\geq a_{i,(\ell')}}$ with equality if $\ell'=\ell$ (by Lemma~\ref{lem:emptyifmin}).  Point (iii) is straightforward from (i) and Lemma~\ref{lemmafromktoell}.  Now prove (iv): we have 
    \begin{align*}
        M_{\ell,W_i} - M_{\ell-1,W_i}&\geq H_{\ell,W_i}(\ell) - H_{\ell-1,W_i}(\ell-1)\\
        &= \ell/(n+1) - (\ell-1)/(n+1)
- (\alpha/m) \sum_{j \in \range{m}\backslash\{i\} }\ind{ a_{i,(\ell)}>S_{n+j}\geq a_{i,(\ell+1)}} \\&=1/(n+1),
    \end{align*}
    by applying Lemma~\ref{lem:emptyifmin}.
 Points (v) and (vi) follow from the definition. % by using $\msub\geq 1$ and $a_{i,(2)}\leq a_{i,(1)}$.
    Finally, we prove (vii): by (i), Lemma~\ref{lemmafromktoell} and the separation condition \eqref{equ-separation}, we have $H_{\ell,W_i}(\ell) +1/(n+1)\leq \min_{\ell'\in [0,n+1]\backslash\{\ell\}} H_{\ell,W_i}(\ell')$. By using (ii), this entails 
    \begin{align*}
    H_{\ell+1,W_i}(\ell) +1/(n+1)&\leq \min_{\ell'\in [n+1]\backslash\{\ell\}} \{H_{\ell'+1,W_i}(\ell')-(\alpha/m)\ind{ \ell\leq \ell'}\} +\alpha/m \\
    &=\min_{\ell'\in [n+1]\backslash\{\ell\}} \{H_{\ell'+1,W_i}(\ell')+(\alpha/m)\ind{ \ell> \ell'}\}.
\end{align*}
    Now using $1/(n+1)\leq \alpha/m$, we obtain
    \begin{align*}
    H_{\ell+1,W_i}(\ell) +1/(n+1)
    &\leq \min_{\ell'\in [n+1]}\{ H_{\ell'+1,W_i}(\ell') +(\alpha/m)\ind{\ell'\leq  \ell }\} \leq U_{\ell,W_i}.
    \end{align*}
    In addition, we have by definition
    $$
U_{\ell-1,W_i} =  \min_{\ell'\in [n+1]}\{ H_{\ell'+1,W_i}(\ell') +(\alpha/m)\ind{\ell'\leq  \ell-1 }\}\leq H_{\ell+1,W_i}(\ell),
    $$
    by considering $\ell'=\ell$.

    Now consider the adaptive case. Since $p_i=\ell/(n+1)$ and $\sigma(\hat{k})=i$, we have $\ell\leq s_0$, and thus $p_i\leq s_0/(n+1)$ and $S_{n+i}\geq a_{i,(s_0)}$. In addition, for $j\in [m]\backslash\{i\}$, this means that $p_j\geq (s_0+1)/(n+1)$ implies that at least $s_0$ calibration score are above $S_{n+j}$ and thus $S_{n+j}\leq a_{i,(s_0+1)}$ (because $a_{i,(s_0+1)}$ is the $s_0$-largest calibration score since $S_{n+i}\geq a_{i,(s_0)}$).
    
    This means that $ \hat{\pi}_0 $ given by \eqref{equpi0hat} is 
\begin{align*}
    \hat{\pi}_0
    = \frac{1+\sum_{j\in [m]\backslash\{i\}} \ind{p_j\geq (s_0+1)/(n+1)}}{m(1-(s_0+1)/(n+1))} = \frac{1+\sum_{j\in [m]\backslash\{i\}} \ind{S_{n+j}\leq a_{i,(s_0+1)}}}{m(1-(s_0+1)/(n+1))} = \hat{\pi}_0(W_i),
\end{align*}
given by \eqref{pi0chapWi}. The rest of the proof is similar to what is above.
    
\end{proof}

\begin{lemma}\label{lem:forgotenlemma}
    Consider $H_{\ell,W_i}(\cdot)$ the functional given by \eqref{equ-keyfunctionalHell} and the quantity $M_{\ell,W_i}$ defined by \eqref{equMell}, that both only depend on $\ell\in [n+1]$ and $W_i$ \eqref{equWi}.
    Then  for all $\ell\in [s_0]$ (recall $s_0=n+1$ and $\alpha_0=\alpha$ in the non-adaptive case), we have 
    $
M_{\ell,W_i}- M_{1,W_i}\leq  \alpha_0/m + 1/(n+1).
    $
\end{lemma}
\begin{proof}
By definition, we have 
\begin{align*}
    M_{\ell,W_i}- M_{1,W_i} &\leq \min_{\ell'\in [0,s_0]}  \{H_{\ell,W_i}(\ell')-H_{1,W_i}(\ell')\}\leq H_{\ell,W_i}(\ell^*)-H_{1,W_i}(\ell^*),
\end{align*}   
for $\ell^*\in [0,s_0]$ such that $H_{\ell,W_i}(\ell^*)=\min_{\ell'\in [0,s_0]}  H_{\ell,W_i}(\ell')$. The cases where $\ell^*=0$ and $\ell^*\geq \ell$ are clear because $H_{\ell,W_i}(\ell^*)=H_{1,W_i}(\ell^*)$ in these cases. Let us now assume $\ell^*\in [\ell-1]$. Then we have
$$
H_{\ell,W_i}(\ell^*)-H_{1,W_i}(\ell^*) = (\alpha_0/m) \bigg(1+ \sum_{j \in \range{m}\backslash\{i\} }\ind{ S_{n+j}\geq a_{i,(\ell^*+1)}} - \sum_{j \in \range{m}\backslash\{i\} }\ind{ S_{n+j}\geq a_{i,(\ell^*)}}\bigg).
$$
But since $H_{\ell,W_i}(\ell^*)\leq H_{\ell,W_i}(\ell^*+1)$, we also have
\begin{align*}
0&\leq H_{\ell,W_i}(\ell^*+1) - H_{\ell,W_i}(\ell^*)\\
  &\leq 1/(n+1)  -  (\alpha_0/m) \bigg(\sum_{j \in \range{m}\backslash\{i\} }\ind{ S_{n+j}\geq a_{i,(\ell^*+1)}} - \sum_{j \in \range{m}\backslash\{i\} }\ind{ S_{n+j}\geq a_{i,(\ell^*)}}\bigg).
\end{align*}
Combining the two last displays concludes the proof.
\end{proof}

\begin{lemma}
   \label{lem:pi0}
Let $i\in \cH_0$, $s_0\in [0,n-1]$ and consider $W_i$ as in \eqref{equWi}. 
  If $p_i\leq s_0/(n+1)$, then $ \hat{\pi}_0$ in \eqref{equpi0hat} is equal to $\hat{\pi}_0(W_i)$ in \eqref{pi0chapWi}.
\end{lemma}
\begin{proof}
If $p_i\leq s_0/(n+1)$, then we have
$$
   \hat{\pi}_0 = \frac{1+\sum_{j=1}^m \ind{p_j\geq (s_0+1)/(n+1)}}{m(1-(s_0+1)/(n+1))} = \frac{1+\sum_{j\in [m]\backslash\{i\}} \ind{p_j\geq (s_0+1)/(n+1)}}{m(1-(s_0+1)/(n+1))} = \hat{\pi}_0(W_i).
$$
Indeed, for any $j\in [m]\backslash\{i\}$, $p_j \geq (s_0+1)/(n+1)$ if and only if $S_{n+j}$ is smaller than the $s_0$-th largest calibration score, which is equivalent to $S_{n+j}\le a_{i,(s_0+1)}$ (because $p_i\leq s_0/(n+1)$).  
%that $S_{n+i}$ is larger than the $(s_0+1)$-th largest calibration score and thus $S_{n+i}\ge a_{i,(s_0+1)}$    
\end{proof}

\section{Additional empirical experiments}\label{sec:addxp}

\subsection{Subsampling proportion}\label{sec:addxp.subsampling.proportion}

We compare SLC+ and SLC++ over subsampling proportion $\rho\in\{0.05, 0.1, \ldots, 1\}$ ($\rho=1$ corresponds to no subsampling) and bFDR levels $\alpha \in \{0.05, 0.1, \ldots, 0.95\}$. 
We consider the three simulation settings in Section~\ref{sec:xp}, except that we decrease $n$ from $12000$ to $2000$ in setting (c).
We include an additional setting with $m = 200$, $m_0 = 160$, $n = 400$, and non-null scores generated from $U(0.8, 1.8)$. All results are aggregated over $1000$ repeats.

Figures~\ref{fig:subsample.proportion.default2}, 
\ref{fig:subsample.proportion.largeValidationDataset}, 
\ref{fig:subsample.proportion.largeMLargeValidationDataset}, and
\ref{fig:subsample.proportion.largeMLargeValidationDatasetBeta} suggest that there is a trade-off between the expectation and variance of $|R|$:
\begin{itemize}
    \item 
   The subsample size should not be too large as when $s>\alpha(n+1)$
    (indicated by the red dashed line), the adjusted level is zero and there will be no rejections (the region above the red dashed line observes zero rejections).

    \item The subsample size should not be too small, as a small $s$ leads to highly
    variable estimated thresholds and, consequently, a highly variable number
    of rejections $|R|$. 
\end{itemize}
As a rule-of-thumb, we recommend the subsampling proportion $\rho=\alpha(n+1)/(5m) \wedge 1$, or equivalently the subsample size $s=\alpha(n+1)/5 \wedge m$, with the additional constraint that $s$ is no smaller than a prescribed minimal subsample size $\underline{s}$, e.g., $\underline{s} = 100$. 
We replicate Figure~\ref{fig:simulation}, Figure~\ref{fig:simulation.adaptive} using $s=\max\{100,\min\{m, \alpha(n+1)/5\}\}$ in Figure~\ref{fig:simulation.subsample.proportion}, Figure~\ref{fig:simulation.adaptive.subsample.proportion}.
With the recommended subsample size, the methods produce a number of rejections comparable to those obtained under the smaller subsample size $0.1m$, and exhibit mildly higher stability for large $\alpha$.

Across all settings, SLC++ is comparable to SLC+ in terms of the expected
number of rejections.
When the subsample size is small (for example, subsample size $10$,
corresponding to a subsampling proportion of $0.05$ in
Figures~\ref{fig:subsample.proportion.default2} and
\ref{fig:subsample.proportion.largeValidationDataset}), SLC++ can be
substantially more stable.
Therefore, when both $m$ and $(n+1)/m$ are small, SLC++ can significantly reduce the variability of SLC+.
However, SLC++ provides limited benefit when $m$ or $(n+1)/m$ is large and does not always control the bFDR at the target level as discussed in Section~\ref{sec:addxp.multiple.subsampling}.

\begin{figure}[tbp]
  \centering
  \begin{minipage}{0.48\textwidth}
    \centering
    \includegraphics[clip, trim = 0cm 0cm 0.5cm 0cm, width = 1\textwidth]{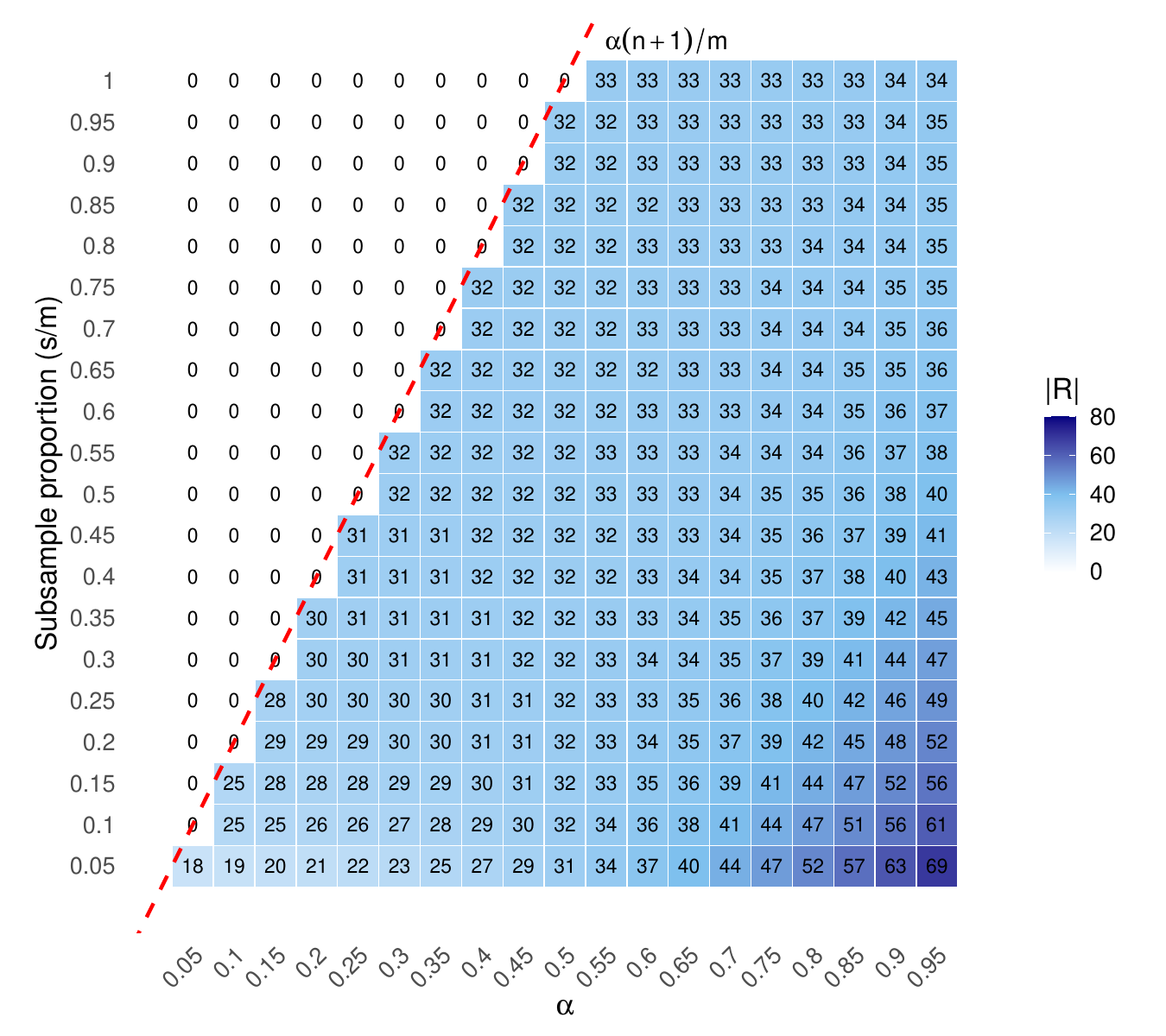}
  \end{minipage}
  \begin{minipage}{0.48\textwidth}
    \centering
    \includegraphics[clip, trim = 0cm 0cm 0.5cm 0cm, width = 1\textwidth]{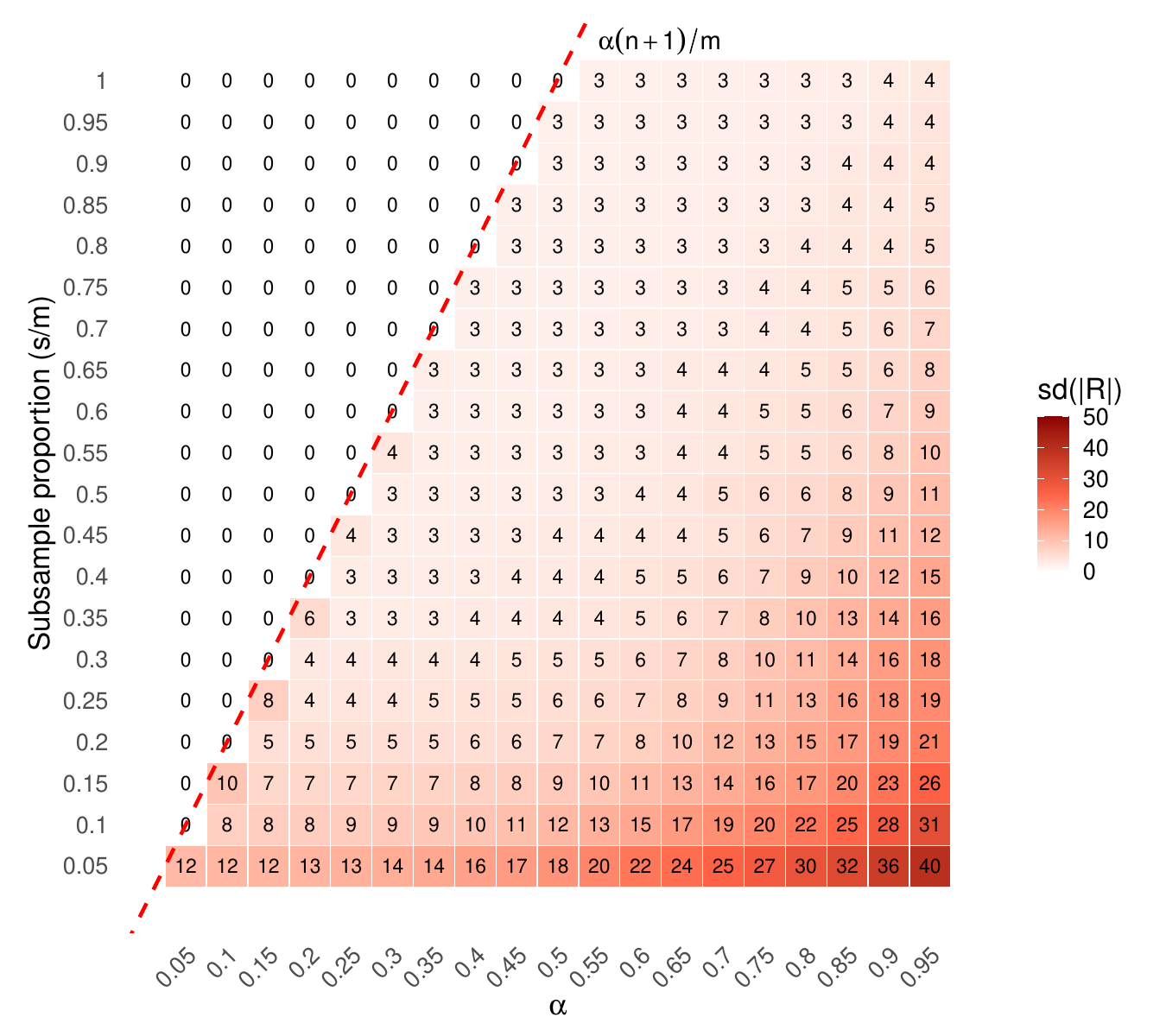}
  \end{minipage}
  \begin{center}
      \text{(i) SLC+}
  \end{center}

    \begin{minipage}{0.48\textwidth}
    \centering
    \includegraphics[clip, trim = 0cm 0cm 0.5cm 0cm, width = 1\textwidth]{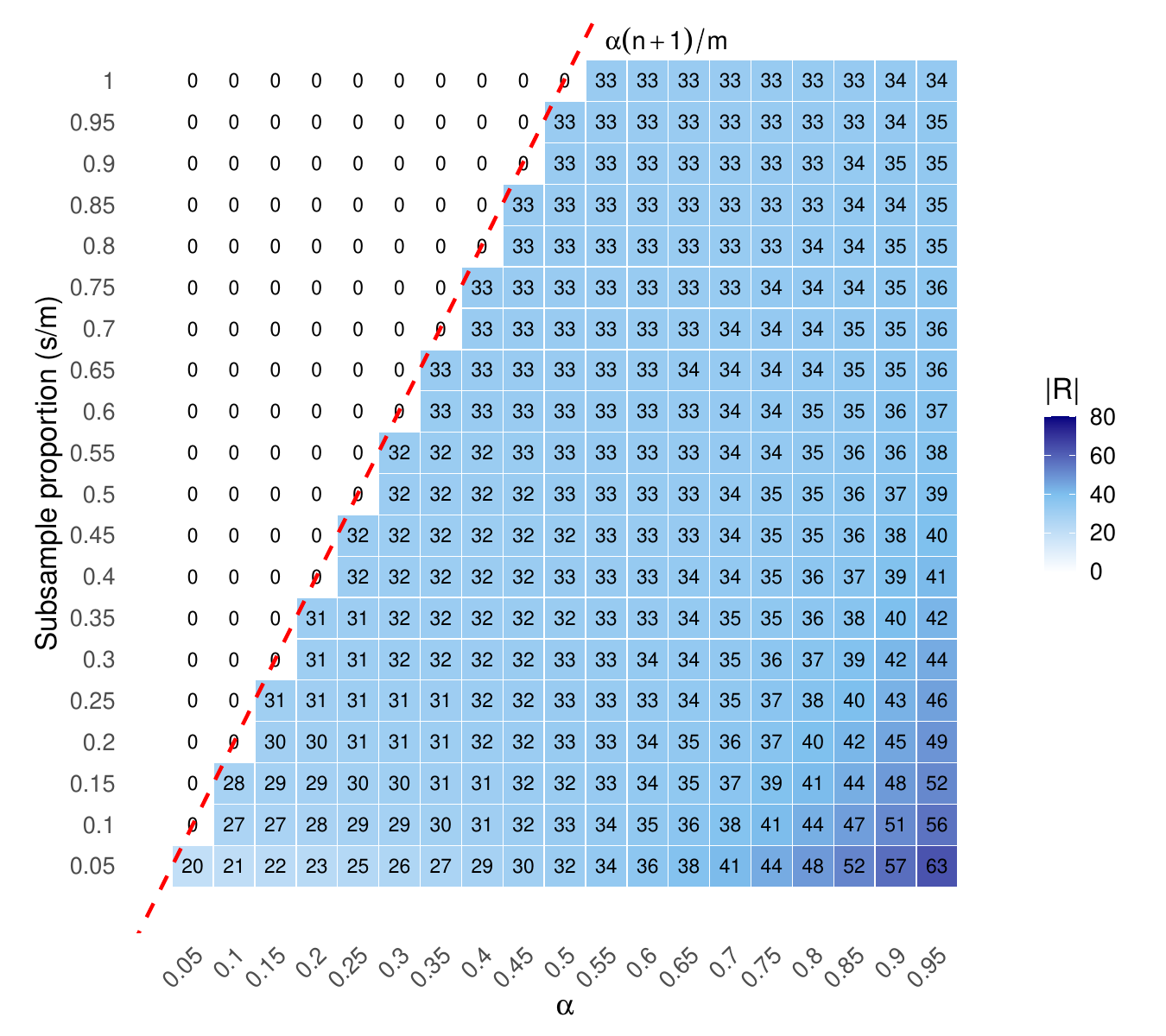}
  \end{minipage}
  \begin{minipage}{0.48\textwidth}
    \centering
    \includegraphics[clip, trim = 0cm 0cm 0.5cm 0cm, width = 1\textwidth]{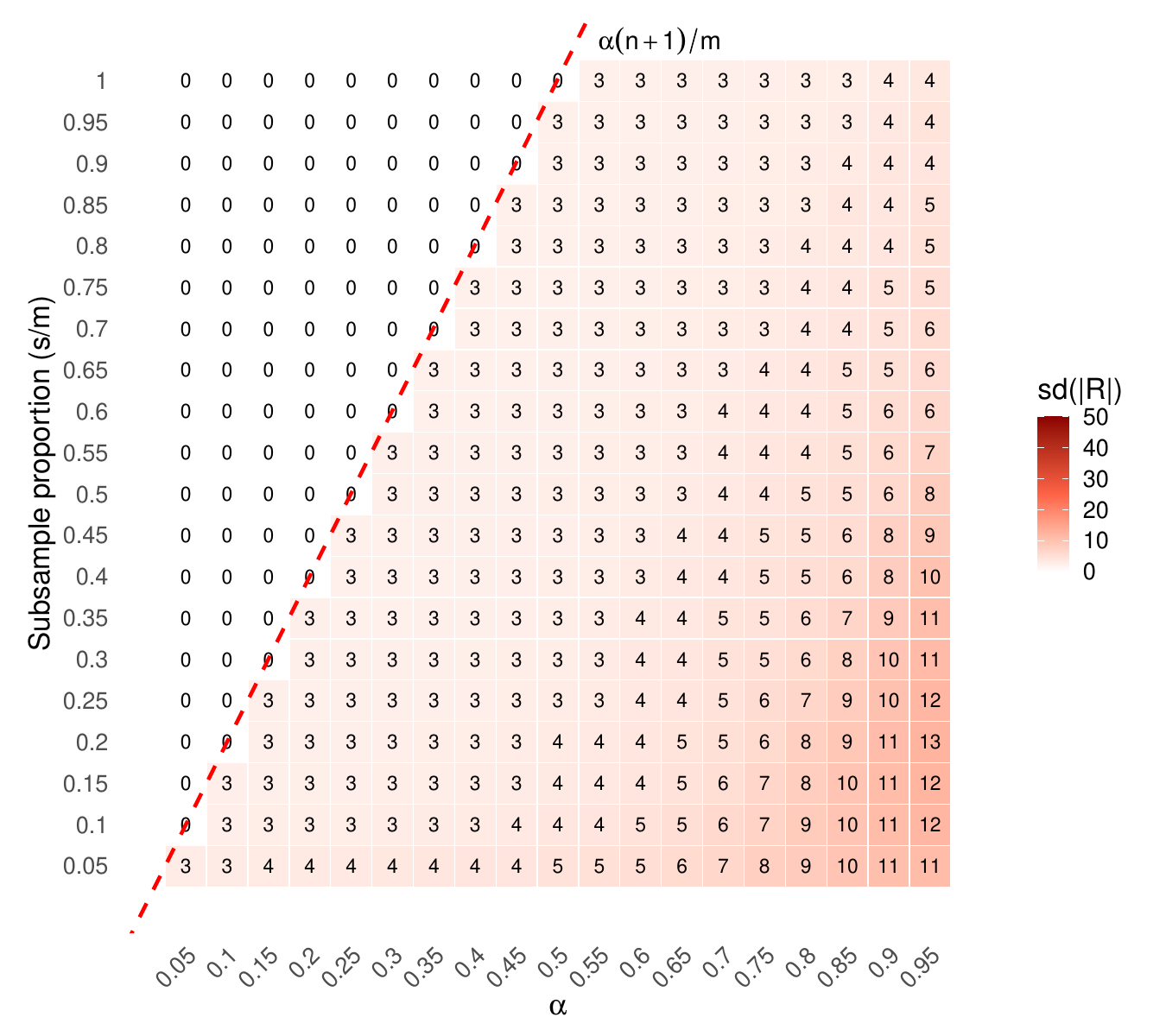}
  \end{minipage}
  \begin{center}
      \text{(ii) SLC++}
  \end{center}\caption{Comparison of SLC+, SLC++ across subsampling proportions and bFDR levels. Setting: $m=200$, $m_0=160$, $n = 400$, non-null scores following $U(0.8,1.8)$. Aggregated over $1000$ trials. 
  }
  \label{fig:subsample.proportion.default2}
\end{figure}

\begin{figure}[tbp]
  \centering
  \begin{minipage}{0.48\textwidth}
    \centering
    \includegraphics[clip, trim = 0cm 0cm 0.5cm 0cm, width = 1\textwidth]{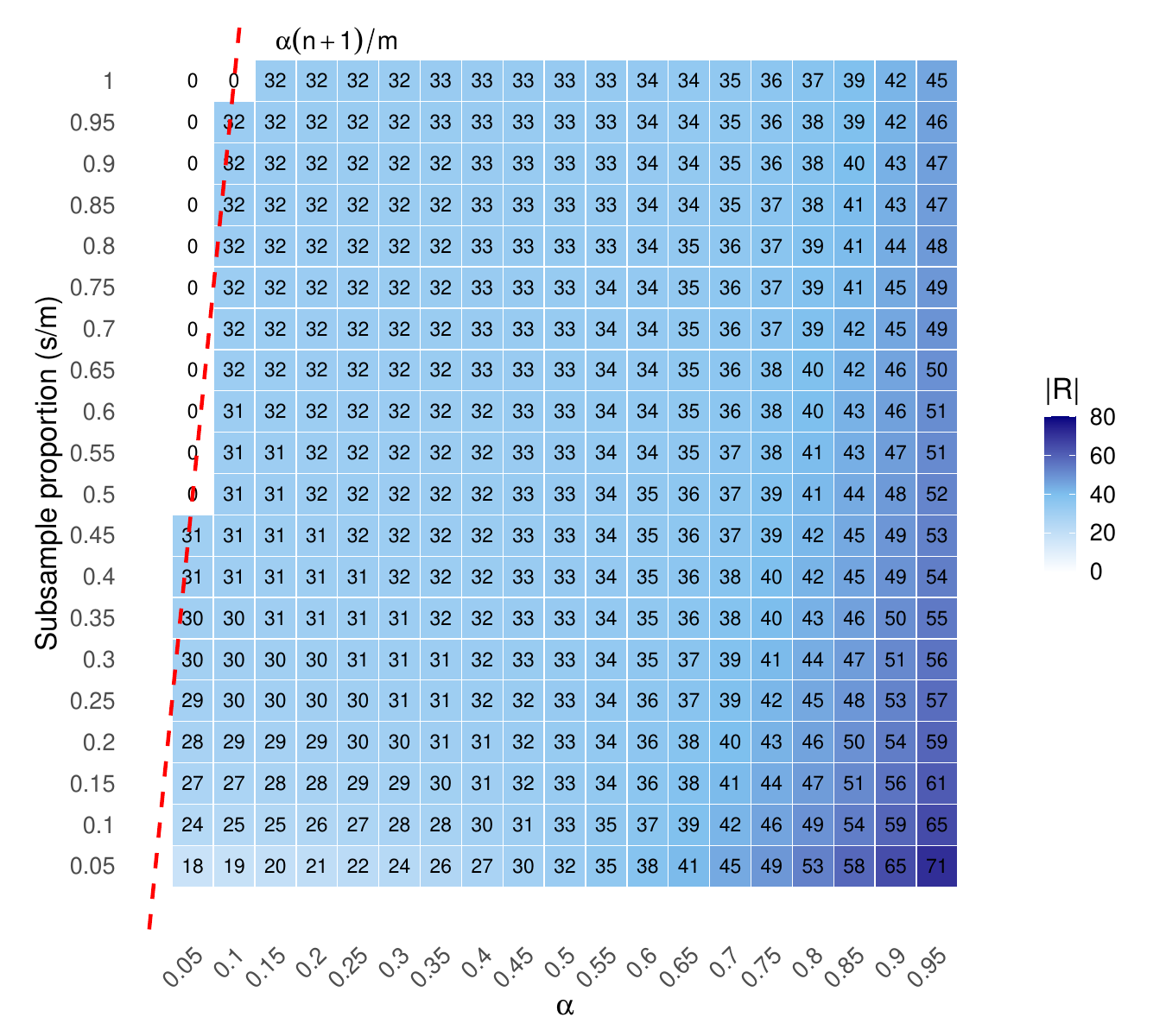}
  \end{minipage}
  \begin{minipage}{0.48\textwidth}
    \centering
    \includegraphics[clip, trim = 0cm 0cm 0.5cm 0cm, width = 1\textwidth]{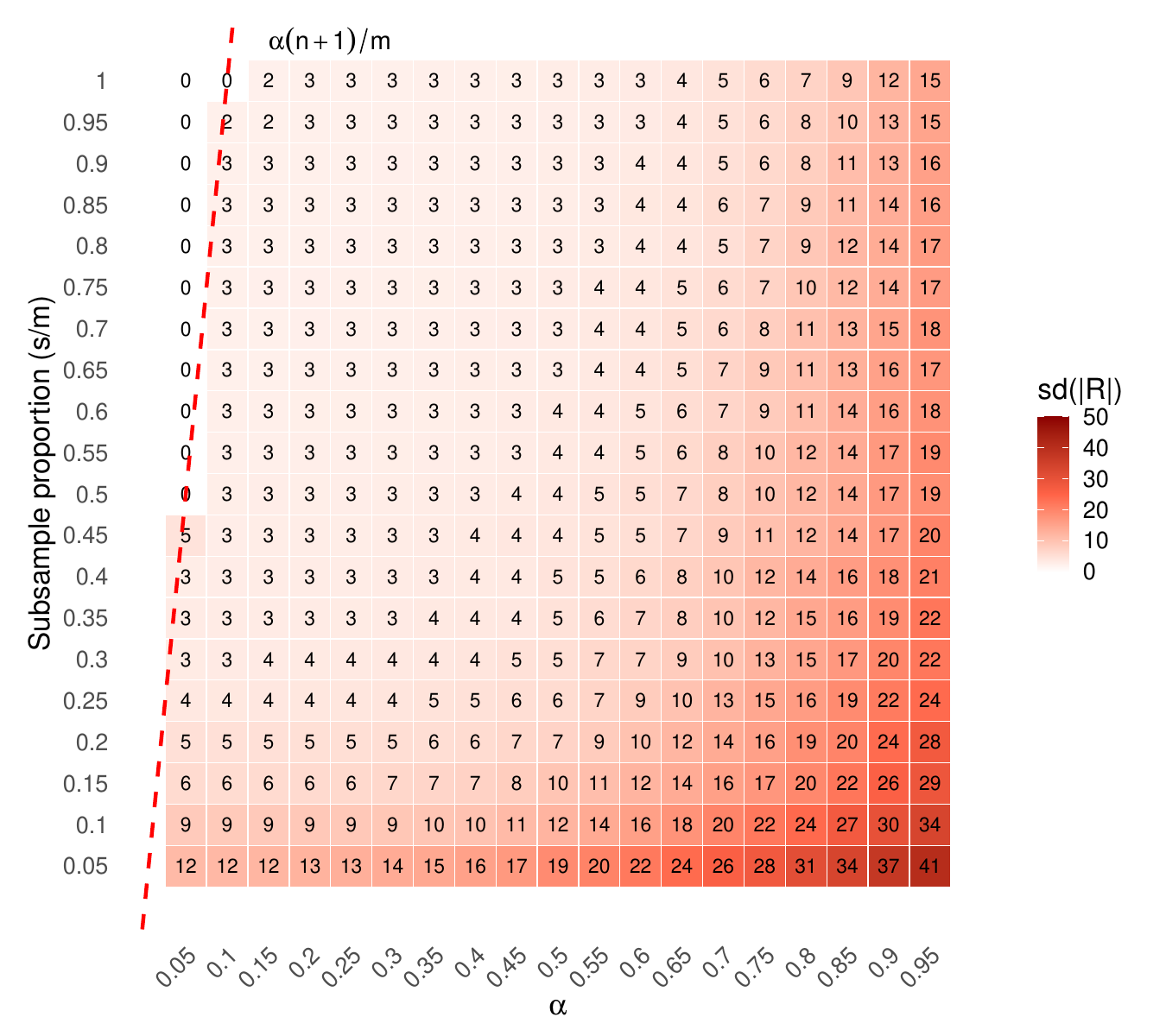}
  \end{minipage}
  \begin{center}
      \text{(i) SLC+}
  \end{center}

    \begin{minipage}{0.48\textwidth}
    \centering
    \includegraphics[clip, trim = 0cm 0cm 0.5cm 0cm, width = 1\textwidth]{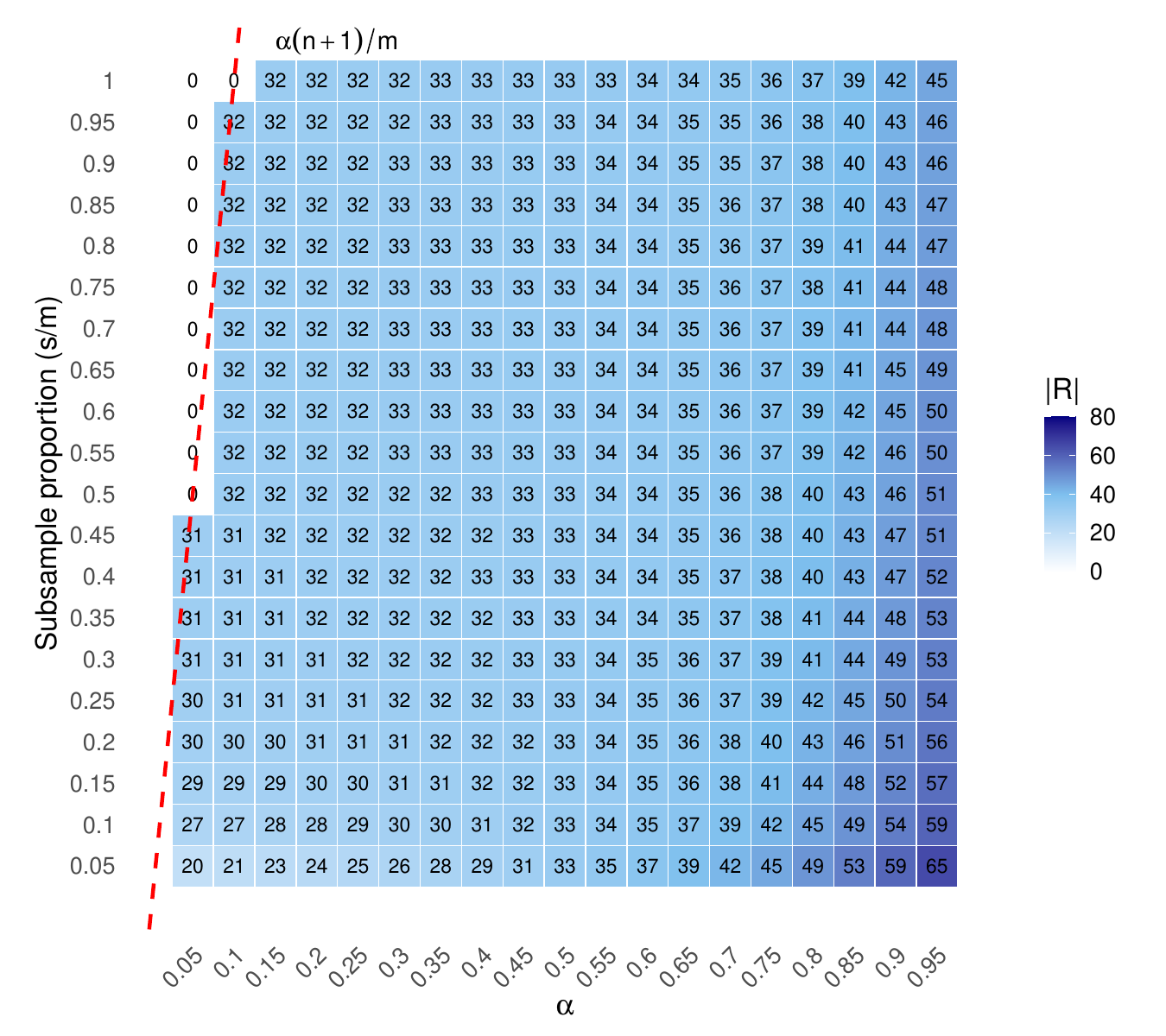}
  \end{minipage}
  \begin{minipage}{0.48\textwidth}
    \centering
    \includegraphics[clip, trim = 0cm 0cm 0.5cm 0cm, width = 1\textwidth]{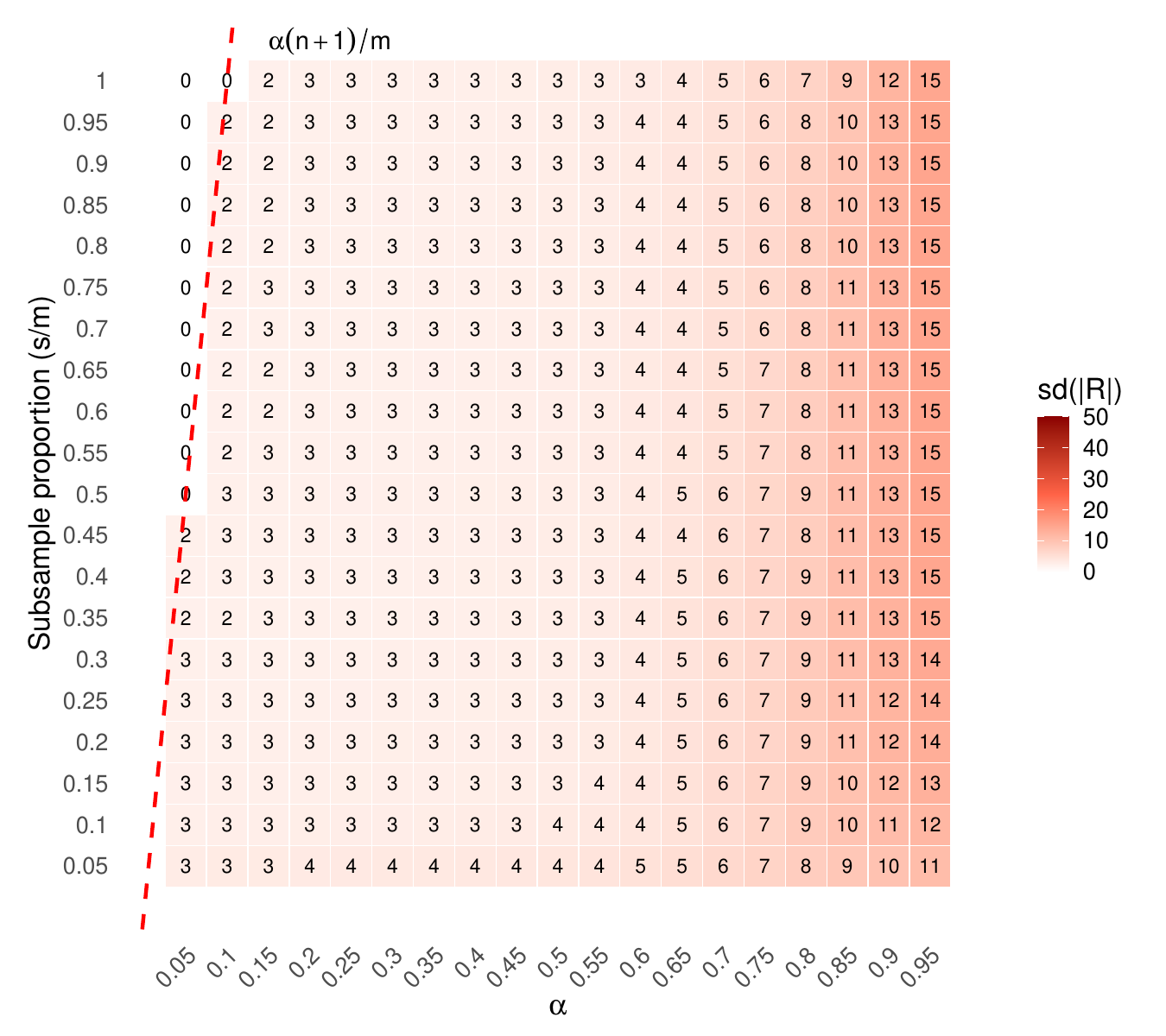}
  \end{minipage}
  \begin{center}
      \text{(ii) SLC++}
  \end{center}\caption{Comparison of SLC+, SLC++ across subsampling proportions and bFDR levels. Setting: $m=200$, $m_0=160$, $n = 2000$ (a relatively large calibration dataset), non-null scores following $U(0.8,1.8)$. Aggregated over $1000$ trials.}
  \label{fig:subsample.proportion.largeValidationDataset}
\end{figure}

\begin{figure}[tbp]
  \centering
  \begin{minipage}{0.48\textwidth}
    \centering
    \includegraphics[clip, trim = 0cm 0cm 0.5cm 0cm, width = 1\textwidth]{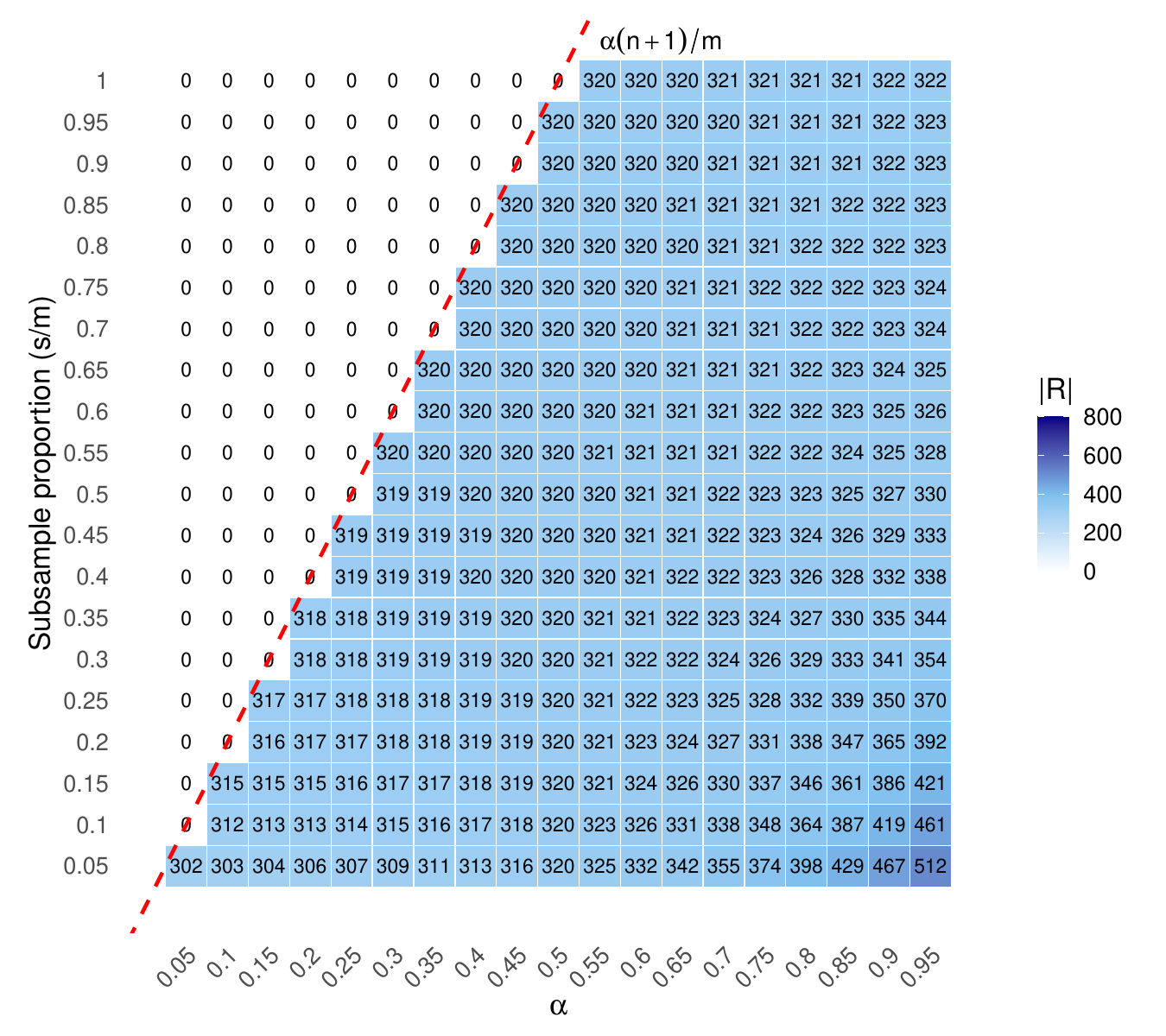}
  \end{minipage}
  \begin{minipage}{0.48\textwidth}
    \centering
    \includegraphics[clip, trim = 0cm 0cm 0.5cm 0cm, width = 1\textwidth]{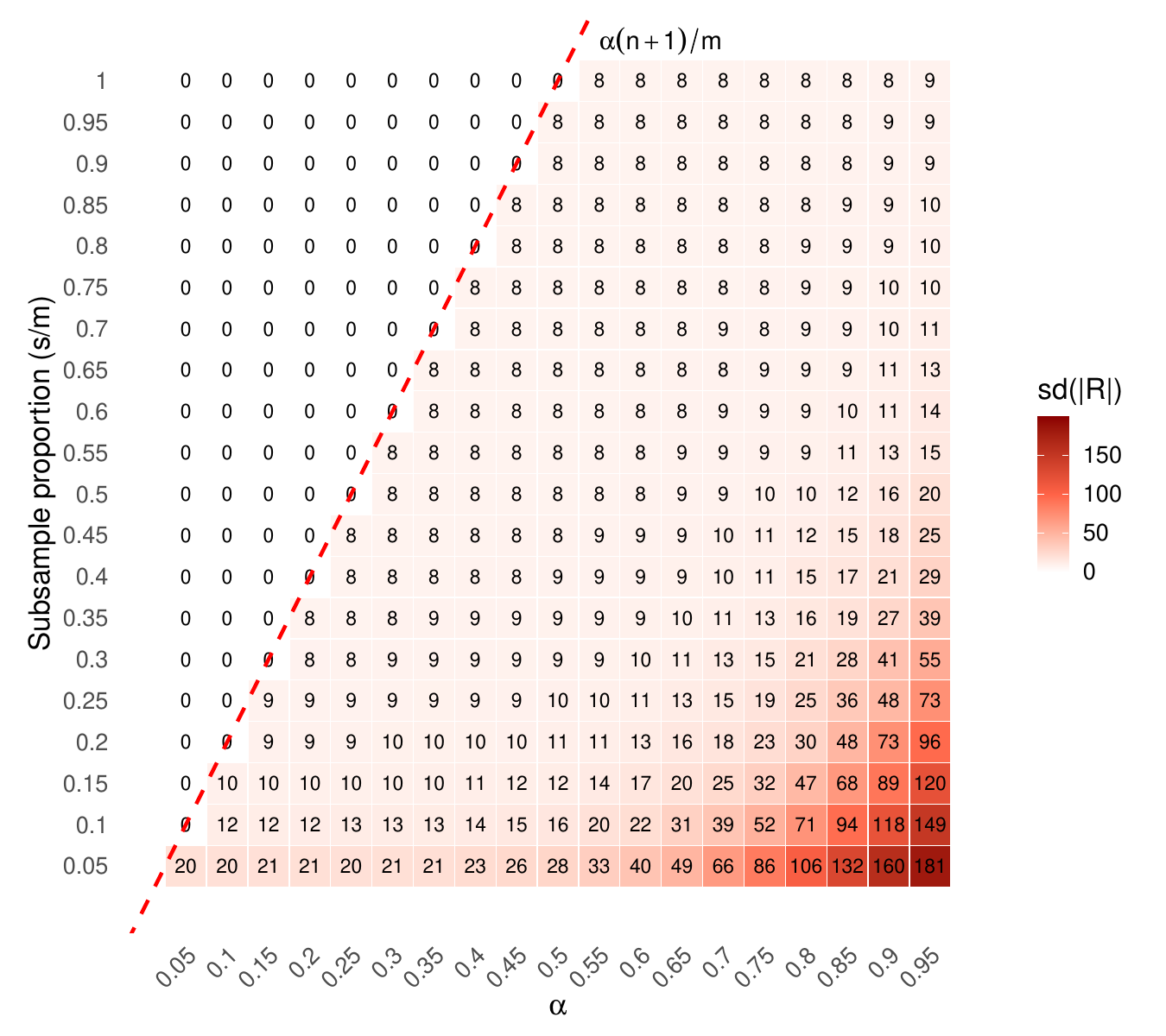}
  \end{minipage}
  \begin{center}
      \text{(i) SLC+}
  \end{center}

    \begin{minipage}{0.48\textwidth}
    \centering
    \includegraphics[clip, trim = 0cm 0cm 0.5cm 0cm, width = 1\textwidth]{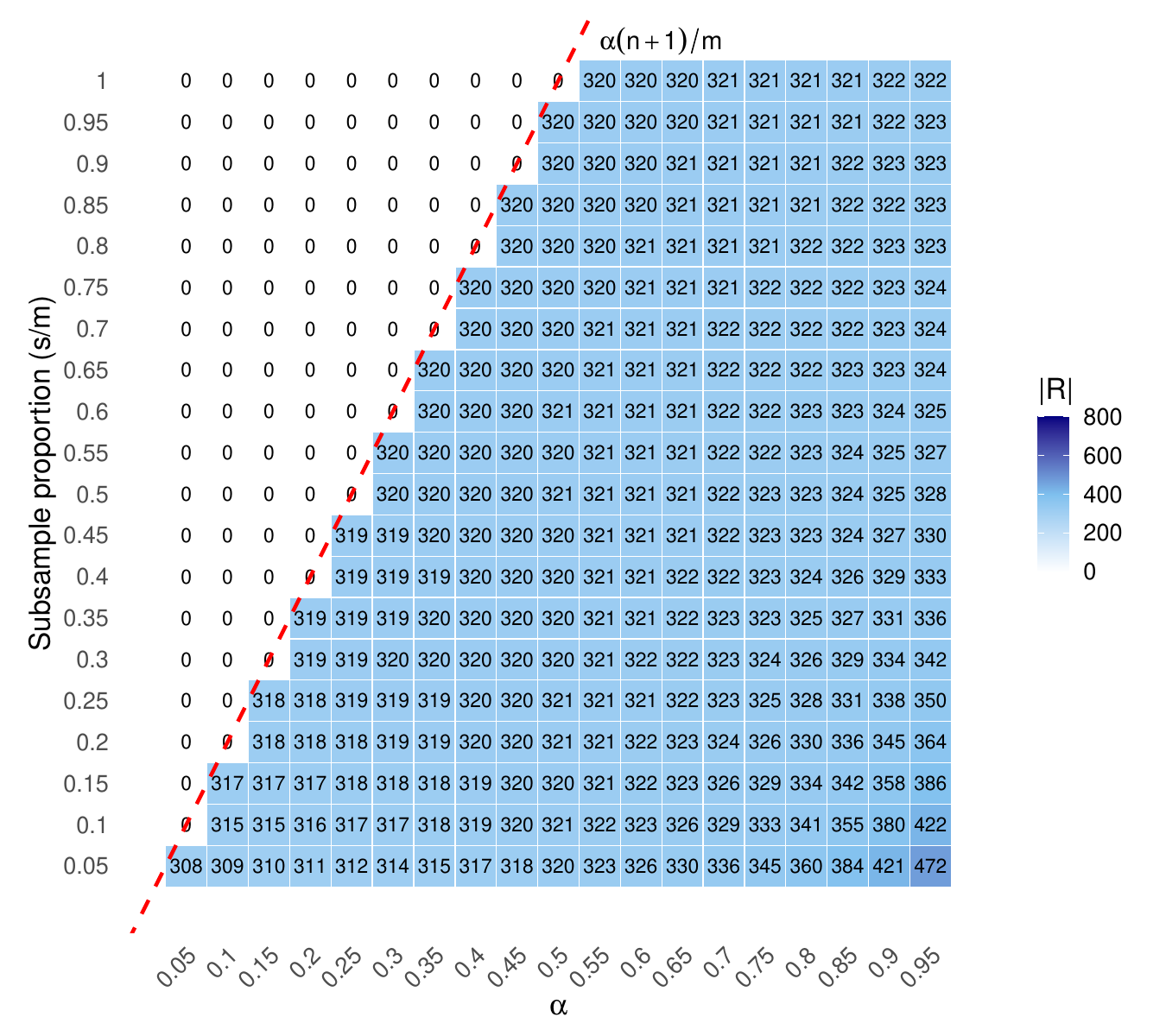}
  \end{minipage}
  \begin{minipage}{0.48\textwidth}
    \centering
    \includegraphics[clip, trim = 0cm 0cm 0.5cm 0cm, width = 1\textwidth]{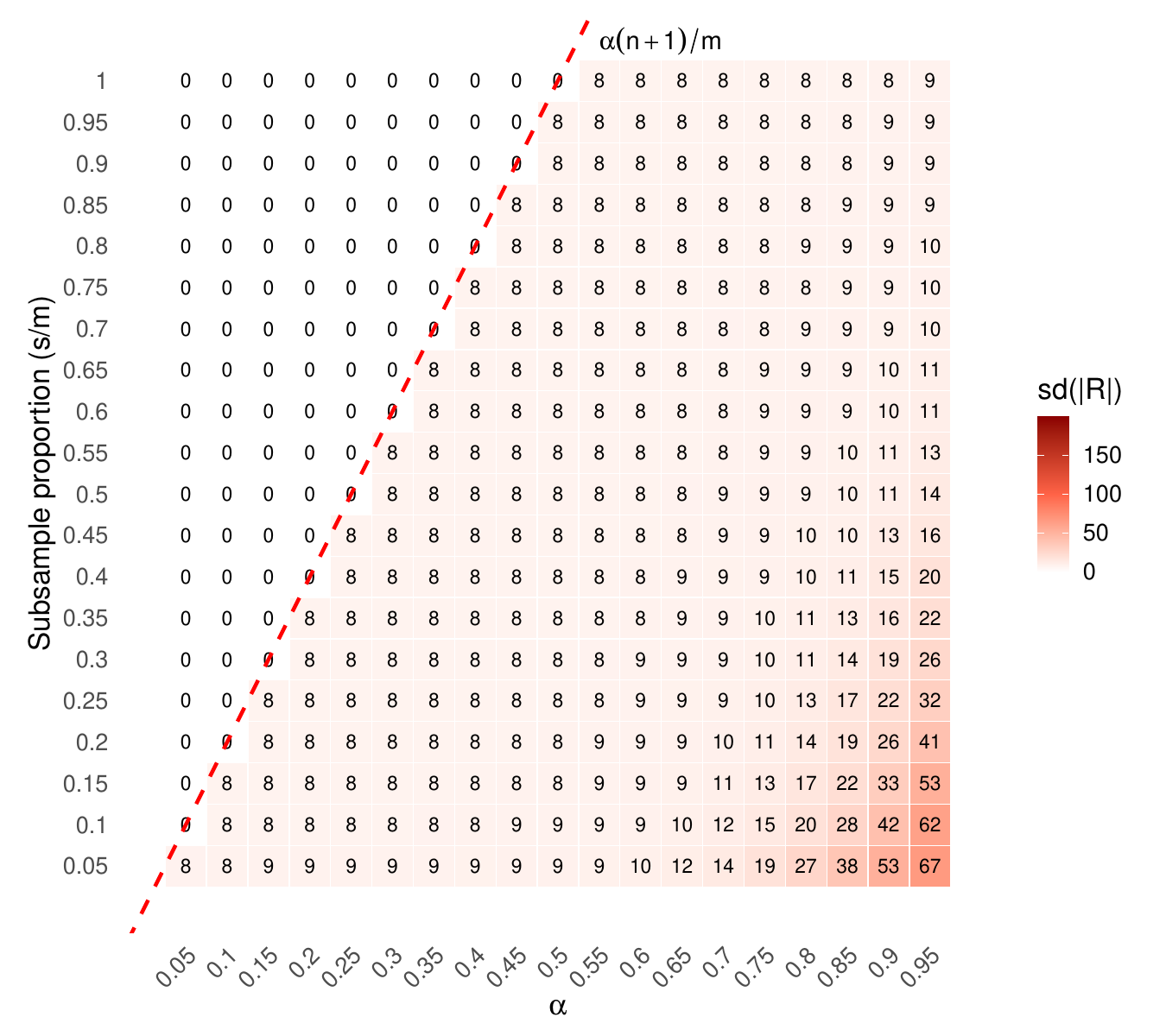}
  \end{minipage}
  \begin{center}
      \text{(ii) SLC++}
  \end{center}\caption{Comparison of SLC+, SLC++ across subsampling proportions and bFDR levels. Setting: $m=2000$, $m_0=1600$, $n = 4000$, non-null scores following $U(0.8,1.8)$. Aggregated over $1000$ trials.}
  \label{fig:subsample.proportion.largeMLargeValidationDataset}
\end{figure}

\begin{figure}[tbp]
  \centering
  \begin{minipage}{0.48\textwidth}
    \centering
    \includegraphics[clip, trim = 0cm 0cm 0.5cm 0cm, width = 1\textwidth]{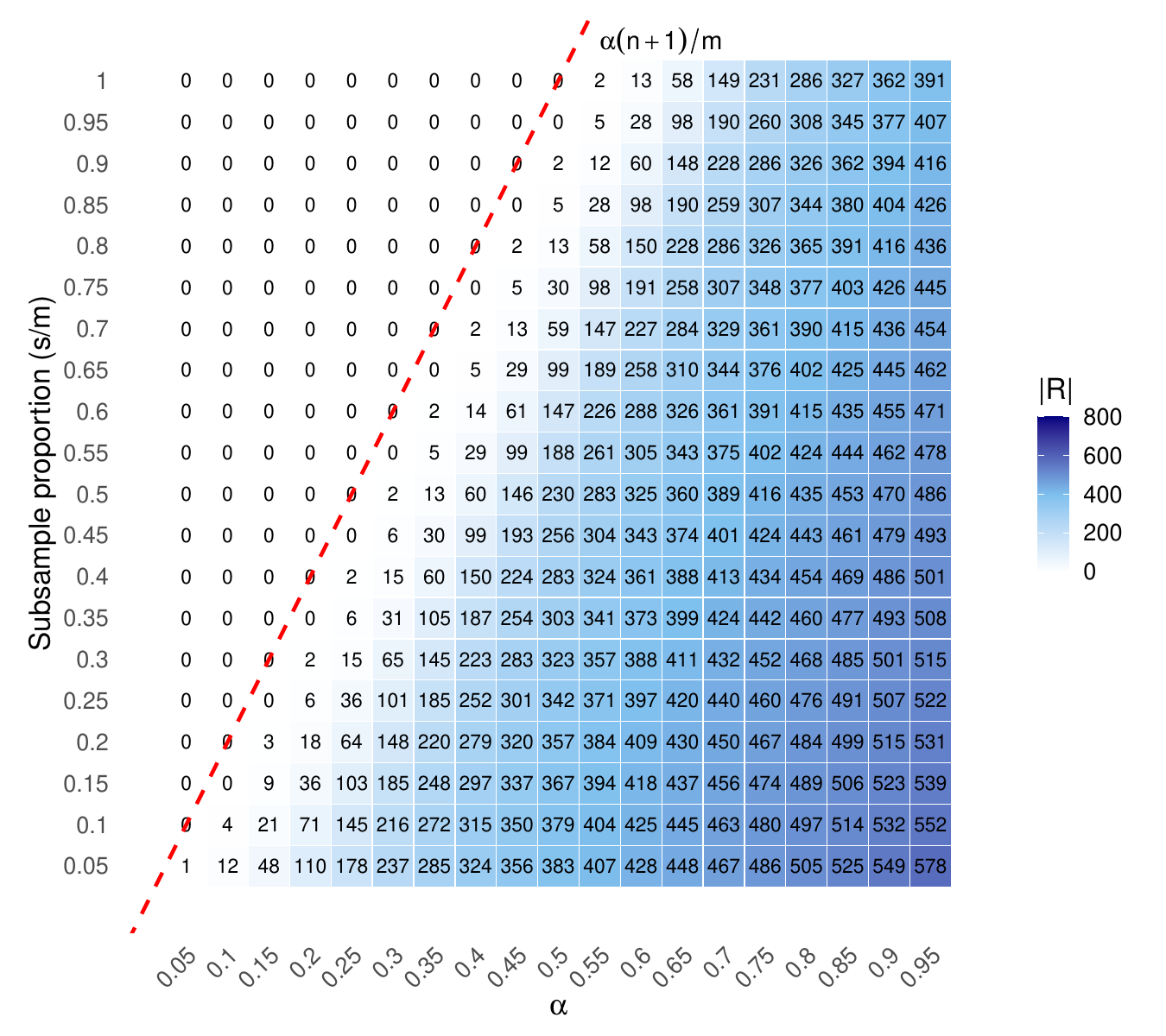}
  \end{minipage}
  \begin{minipage}{0.48\textwidth}
    \centering
    \includegraphics[clip, trim = 0cm 0cm 0.5cm 0cm, width = 1\textwidth]{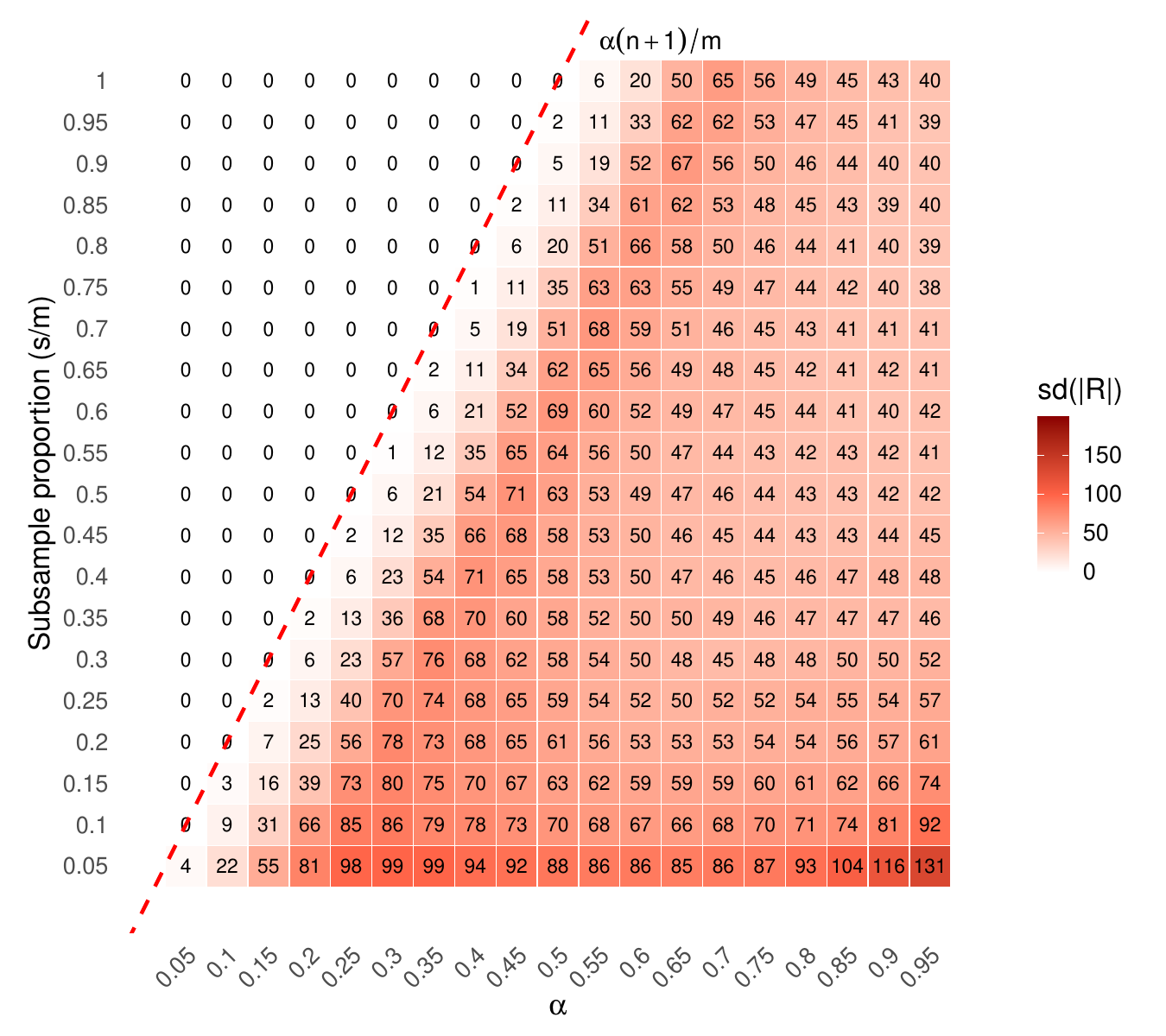}
  \end{minipage}
  \begin{center}
      \text{(i) SLC+}
  \end{center}

    \begin{minipage}{0.48\textwidth}
    \centering
    \includegraphics[clip, trim = 0cm 0cm 0.5cm 0cm, width = 1\textwidth]{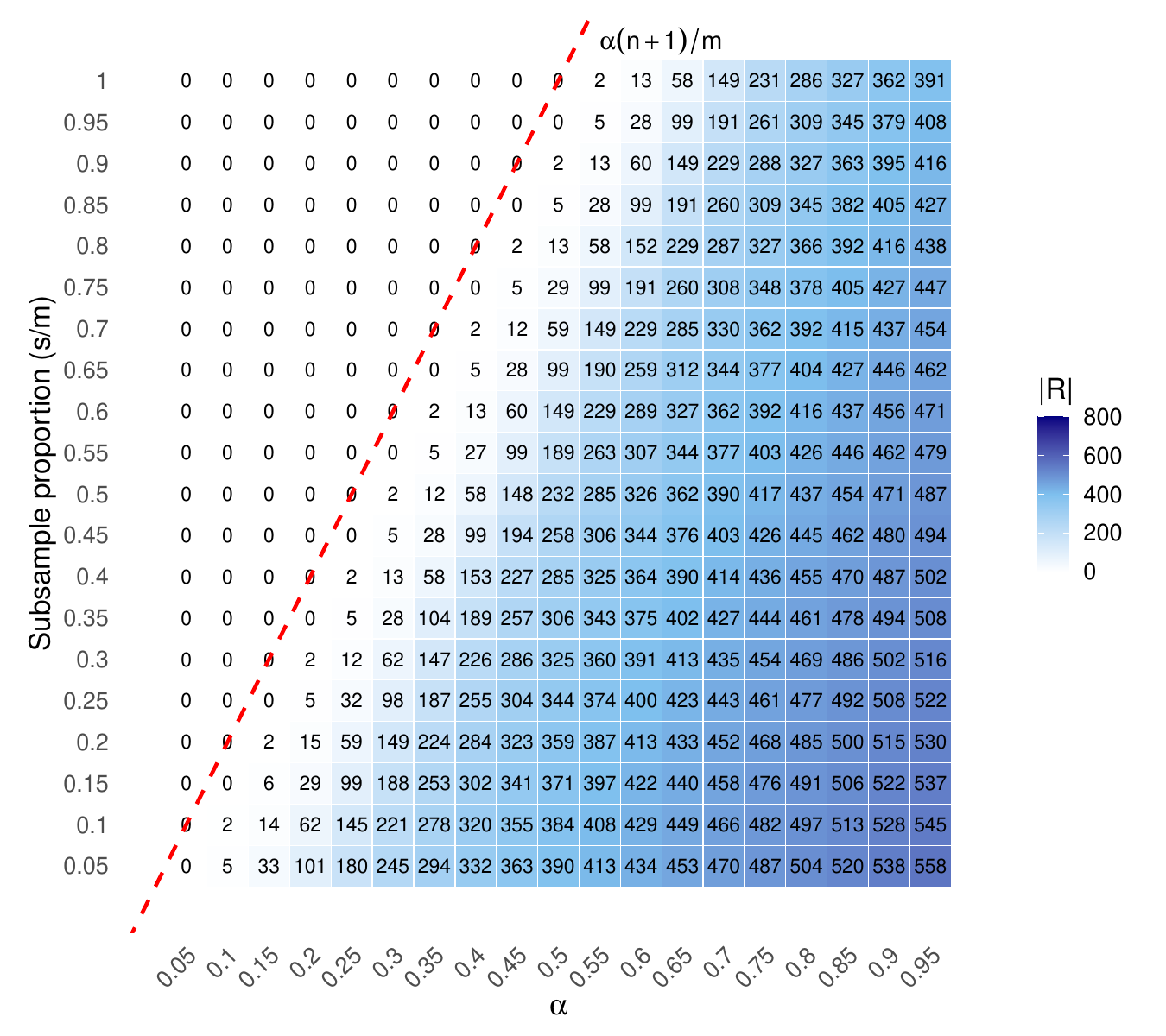}
  \end{minipage}
  \begin{minipage}{0.48\textwidth}
    \centering
    \includegraphics[clip, trim = 0cm 0cm 0.5cm 0cm, width = 1\textwidth]{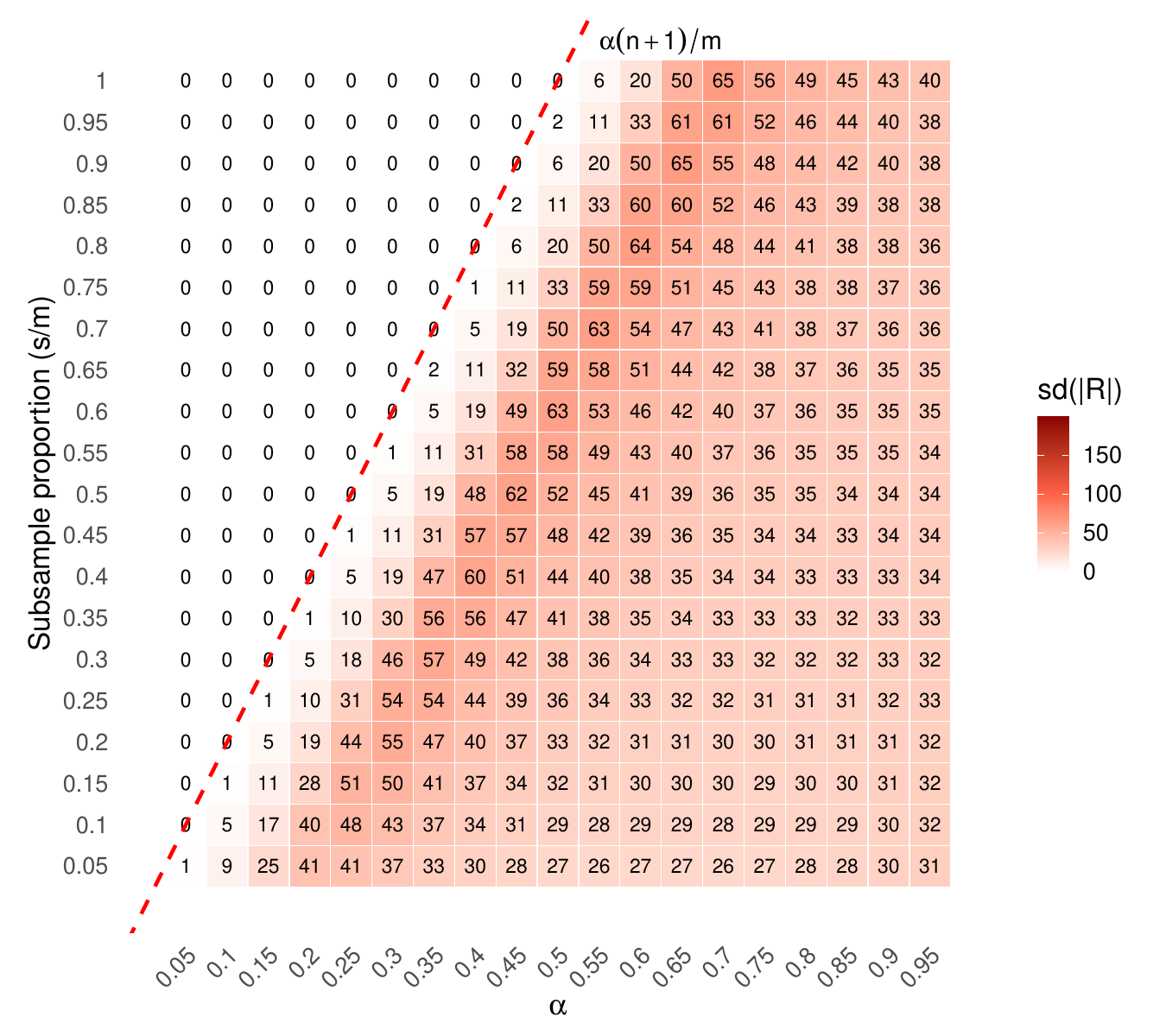}
  \end{minipage}
  \begin{center}
      \text{(ii) SLC++}
  \end{center}\caption{Comparison of SLC+, SLC++ across subsampling proportions and bFDR levels. Setting: $m=2000$, $m_0=1600$, $n = 4000$, non-null scores following $\mathrm{Beta}(30,1)$. Aggregated over $1000$ trials.}
  \label{fig:subsample.proportion.largeMLargeValidationDatasetBeta}
\end{figure}

\begin{figure}[h!]
  \centering
  \begin{minipage}{1\textwidth}
    \centering
    \includegraphics[clip, trim = 0cm 0cm 0cm 0cm, width = 0.9\textwidth]{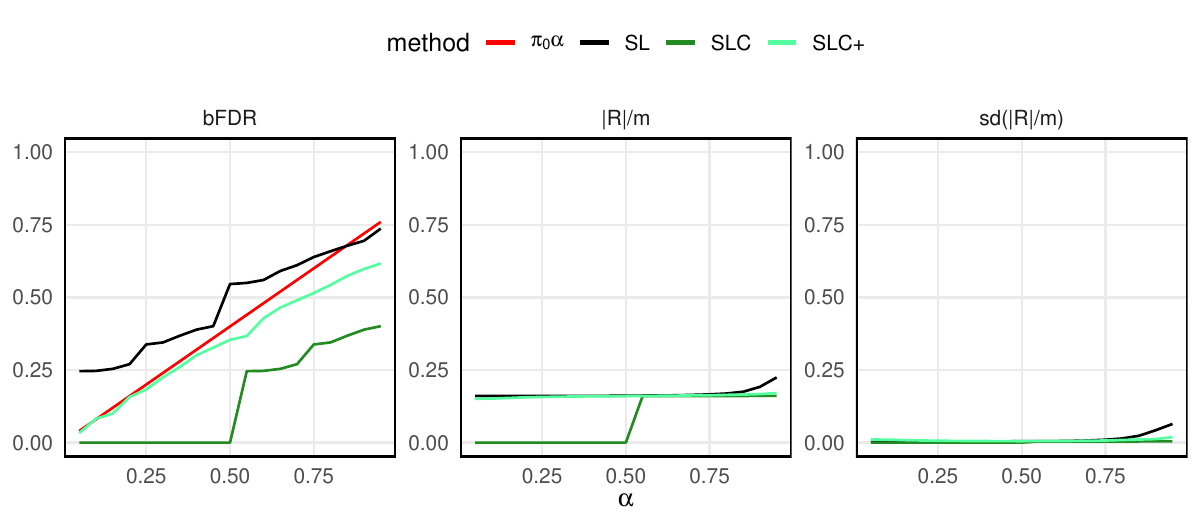}
    \small\text{(a) $m=2000$, $n = 4000$, Non-null scores $\sim U(0.8,1.8)$}
  \end{minipage}
  \begin{minipage}{1\textwidth}
    \centering
    \includegraphics[clip, trim = 0cm 0cm 0cm 1.5cm, width = 0.9\textwidth]{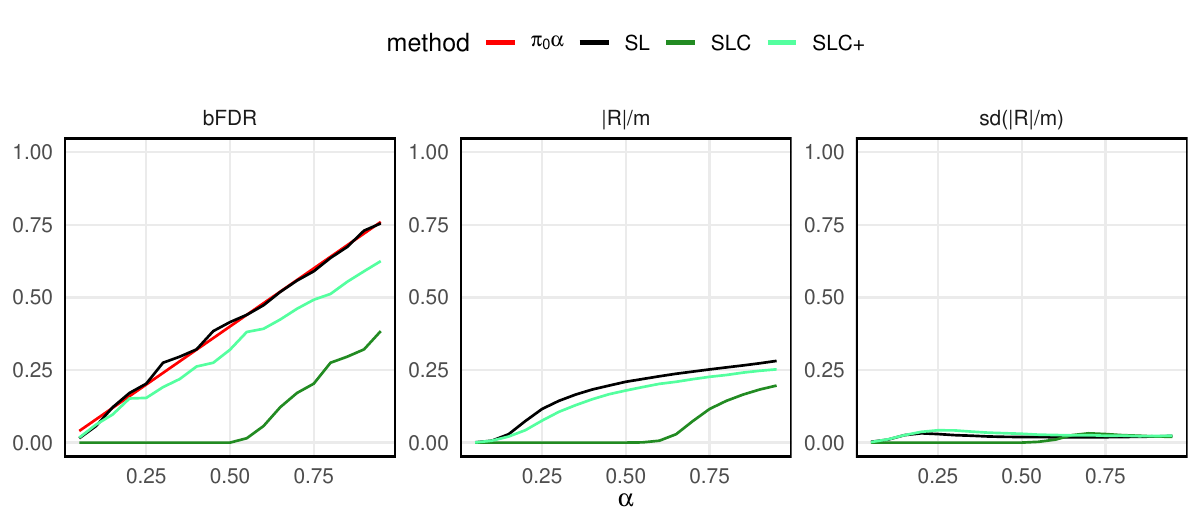}
    \small\text{(b) $m=2000$, $n=4000$, Non-null scores $\sim\mathrm{Beta}(30,1)$}
  \end{minipage}
     \begin{minipage}{1\textwidth}
    \centering
    \includegraphics[clip, trim = 0cm 0cm 0cm 1.5cm, width = 0.9\textwidth]{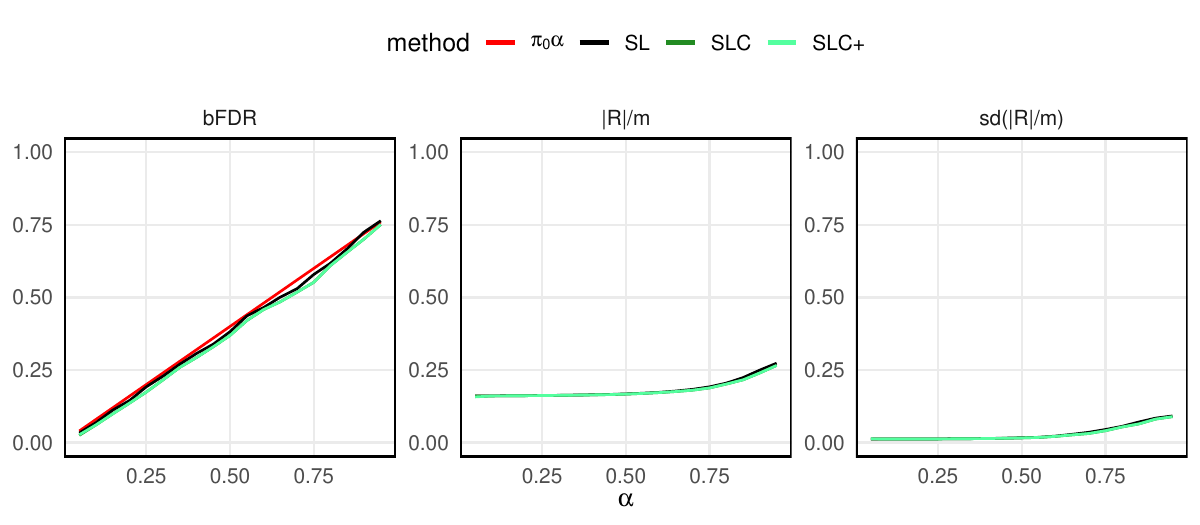}
    \small\text{(c) $m=200$, $n = 12000$, Non-null scores $\sim U(0.8,1.8)$}
  \end{minipage}

  \caption{Comparison of SL, SLC, SLC+ applied to simulated conformal $p$-values with the recommended subsample size $\max\{100,\min\{m, \alpha(n+1)/5\}\}$.}
  \label{fig:simulation.subsample.proportion}
\end{figure}

\begin{figure}[tbp]
  \centering
  \begin{minipage}{1\textwidth}
    \centering
    \includegraphics[clip, trim = 0cm 0cm 0cm 0cm, width = 0.9\textwidth]{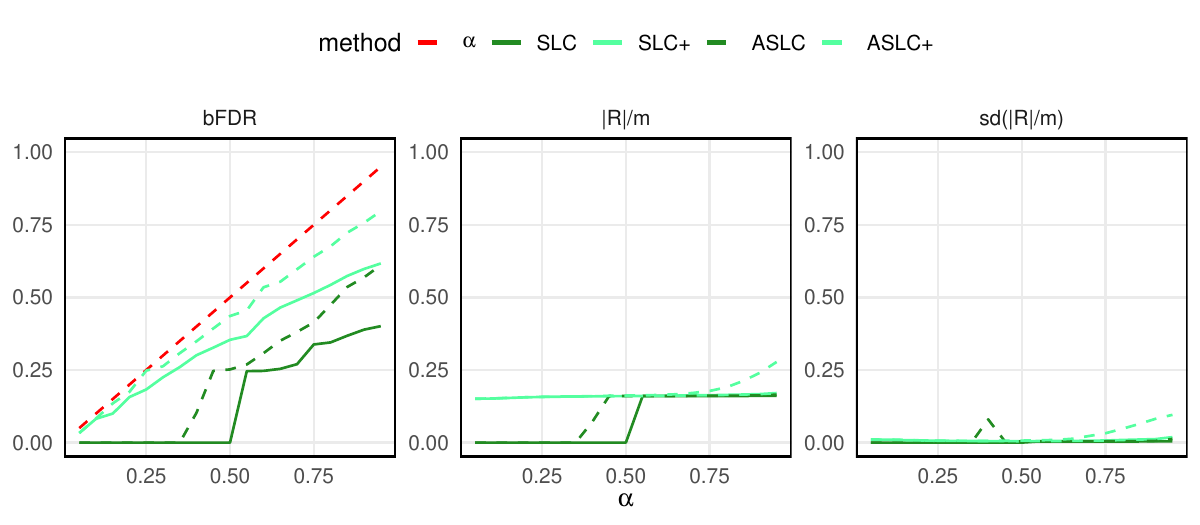}
    \small\text{(a) $m=2000$, $n = 4000$, Non-null scores $\sim U(0.8,1.8)$}
  \end{minipage}
\begin{minipage}{1\textwidth}
    \centering
    \includegraphics[clip, trim = 0cm 0cm 0cm 1.5cm, width = 0.9\textwidth]{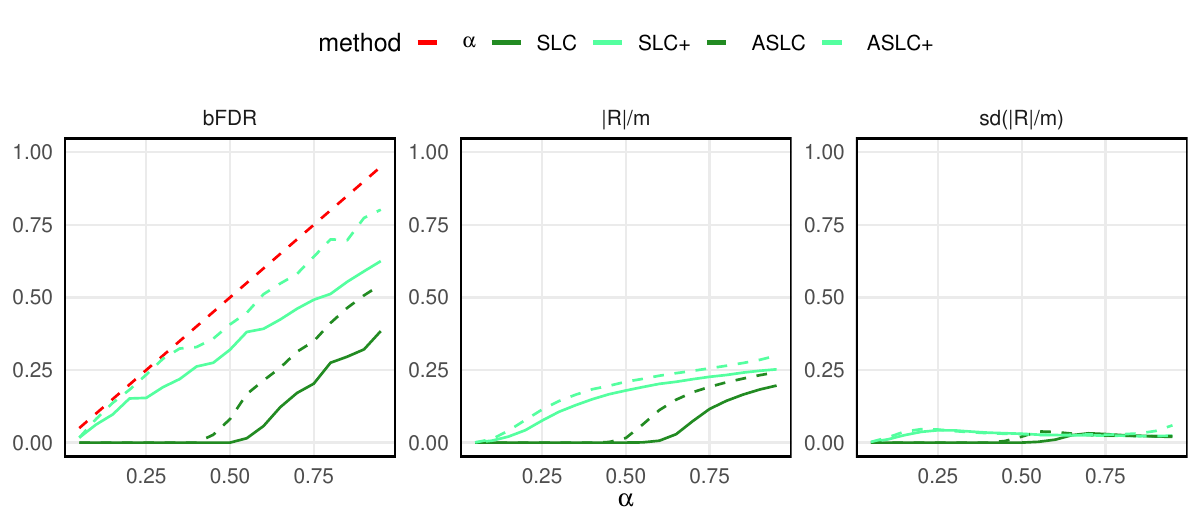}
    \small\text{(b) $m=2000$, $n=4000$, Non-null scores $\sim\mathrm{Beta}(30,1)$}
  \end{minipage}
  \caption{Comparison of SLC, SLC+, and their adaptive variants ASLC, ASLC+ applied to simulated conformal $p$-values with the recommended subsample size $\max\{100,\min\{m, \alpha(n+1)/5\}\}$.
  }
  \label{fig:simulation.adaptive.subsample.proportion}
\end{figure}

\subsection{Multiple subsampling}\label{sec:addxp.multiple.subsampling}

SLC++ may not always control the bFDR at the level $\pi_0 \alpha$ (Figure~\ref{fig:simulationSubsample}, panel (a), (c), $\alpha > 0.5$).
SLC++/2 is overly conservative and produces a small set of rejections across panels in Figure~\ref{fig:simulationSubsample}.
SLC+ performs reasonably well, with mildly increased variability compared to SLC++ for $\alpha$ large.

\begin{figure}[tbp]
  \centering
  \begin{minipage}{1\textwidth}
    \centering
    \includegraphics[clip, trim = 0cm 0cm 0cm 0cm, width = 0.9\textwidth]{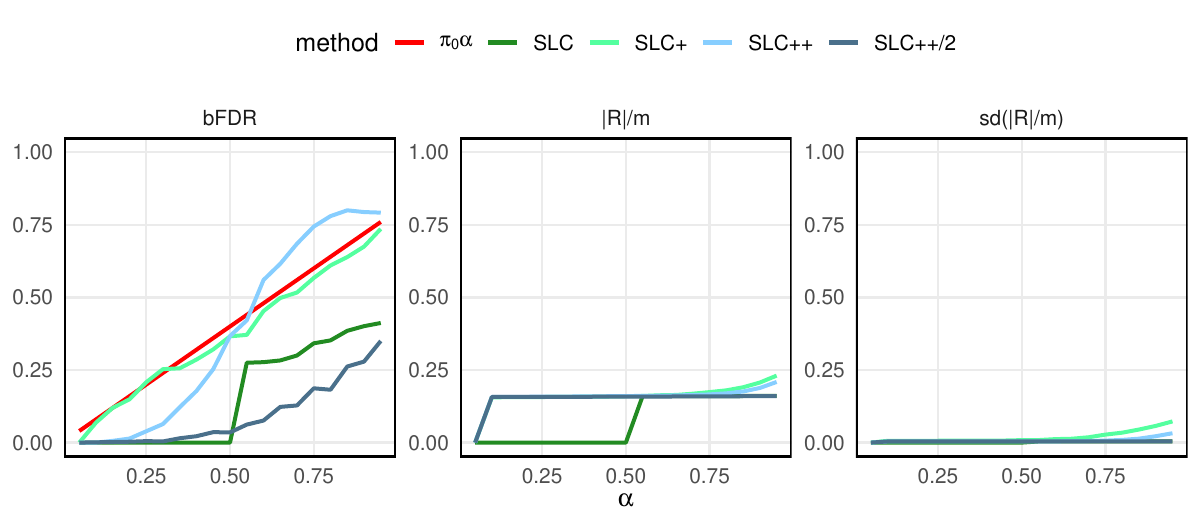}
    \small\text{(a) $m=2000$, $n = 4000$, Non-null scores $\sim U(0.8,1.8)$}
  \end{minipage}
  \begin{minipage}{1\textwidth}
    \centering
    \includegraphics[clip, trim = 0cm 0cm 0cm 1.5cm, width = 0.9\textwidth]{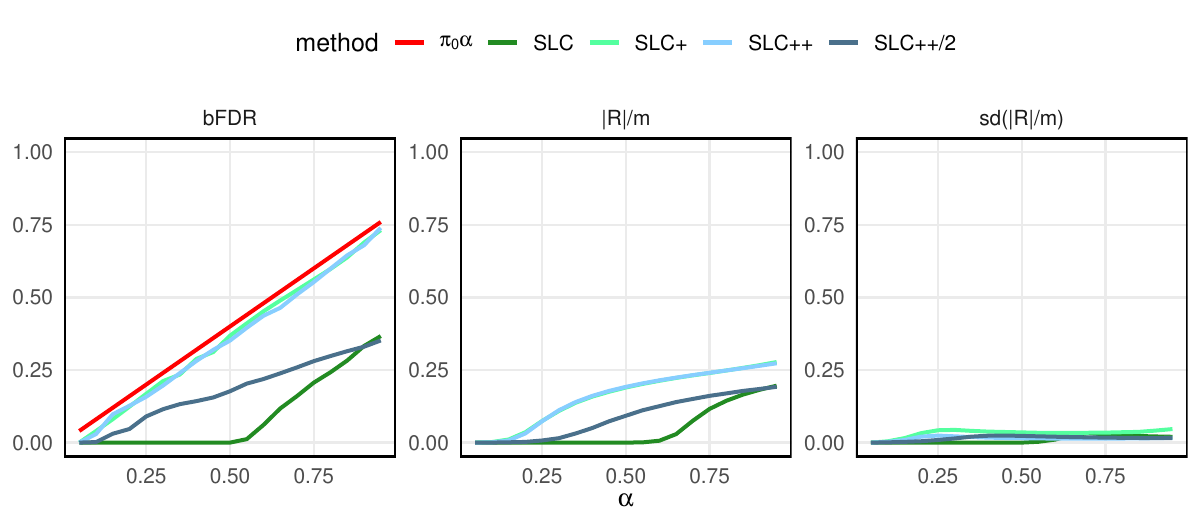}
   \small\text{(b) $m=2000$, $n = 4000$, Non-null scores $\sim \mathrm{Beta}(30,1)$}
  \end{minipage}
    \begin{minipage}{1\textwidth}
    \centering
    \includegraphics[clip, trim = 0cm 0cm 0cm 1.5cm, width = 0.9\textwidth]{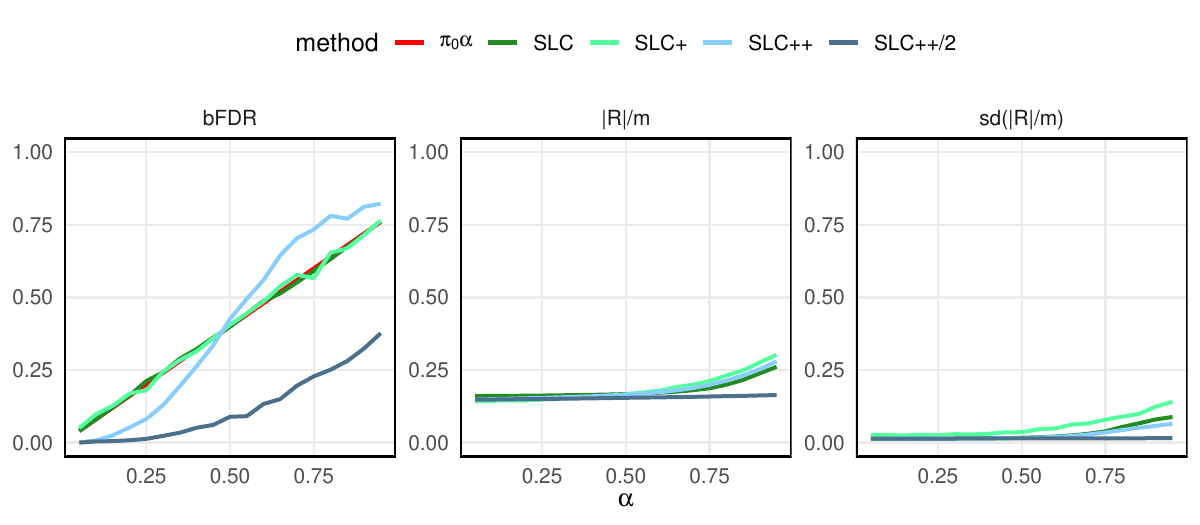}
    \small\text{(c) $m=200$, $n = 12000$, Non-null scores $\sim U(0.8,1.8)$}
  \end{minipage}
  \caption{Comparison of SLC and its subsampling variants (SLC+, SLC++, SLC++/2) applied to simulated conformal $p$-values. }
  \label{fig:simulationSubsample}
\end{figure}

\subsection{SL variants with enforced gap}\label{sec:addxp.SLG}

We evaluate the SLG variants described in Appendix~\ref{secSL3}.
SLG provably controls the bFDR at level $\pi_0 \alpha$ for $\alpha > 0.5$ (dashed brown vertical line), while SLG$+$ achieves bFDR control for $\alpha > 0.1$ with subsampling proportion $\rho = 0.1$ (dashed orange vertical line).
Although SLG+ and SLC+ are comparable in terms of the expected number of rejections, SLG+ exhibits substantially
higher variability in the number of rejections
(Figure~\ref{fig:simulationGap}, panel (a) and (b)).
In summary, the SLG variants are less preferred compared to the SLC variants.

\begin{figure}[tbp]
  \centering
  \begin{minipage}{1\textwidth}
    \centering
    \includegraphics[clip, trim = 0cm 0cm 0cm 0cm, width = 0.9\textwidth]{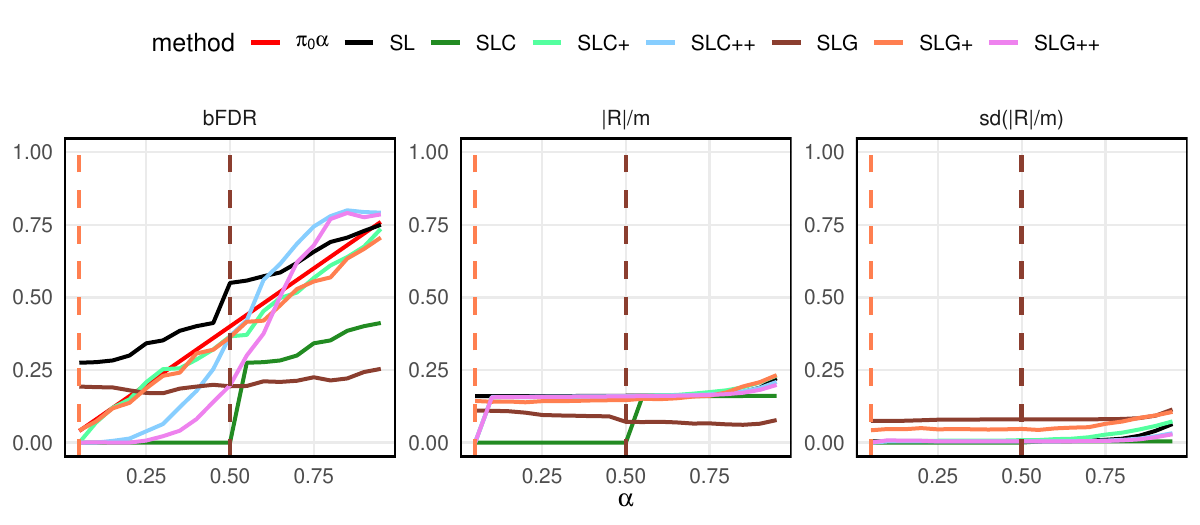}
    \small\text{(a) $m=2000$, $n = 4000$, Non-null scores $\sim U(0.8,1.8)$}
  \end{minipage}
  \begin{minipage}{1\textwidth}
    \centering
    \includegraphics[clip, trim = 0cm 0cm 0cm 1.5cm, width = 0.9\textwidth]{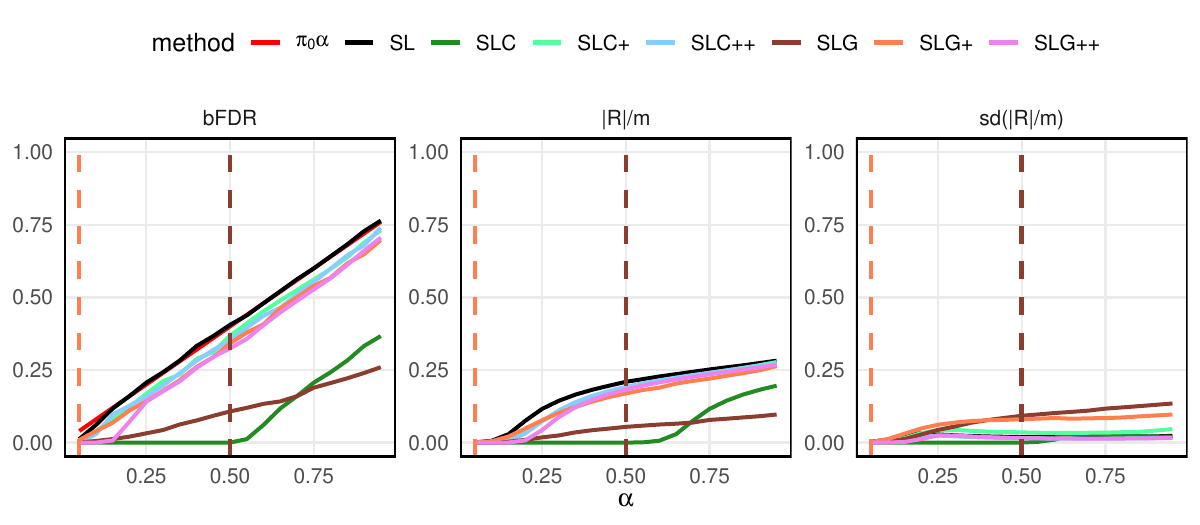}
    \small\text{(b) $m=2000$, $n = 4000$, Non-null scores $\sim \mathrm{Beta}(30,1)$}
  \end{minipage}
    \begin{minipage}{1\textwidth}
    \centering
    \includegraphics[clip, trim = 0cm 0cm 0cm 1.5cm, width = 0.9\textwidth]{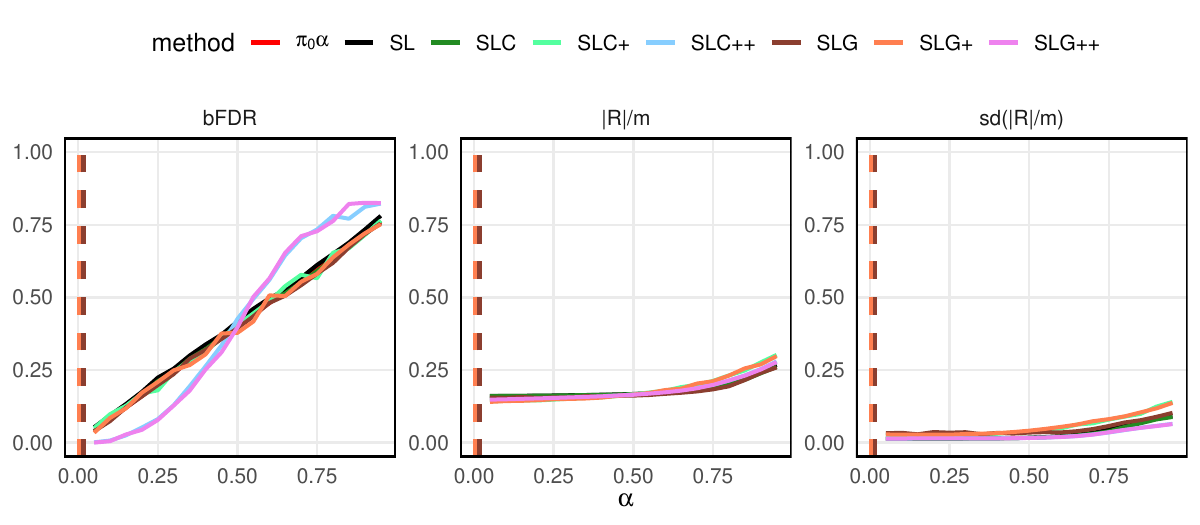}
    \small\text{(c) $m=200$, $n = 12000$, Non-null scores $\sim U(0.8,1.8)$}
  \end{minipage}
  \caption{Comparison of SL, SLC, SLC+, SLC++ and the variants with enforced gap SLG, SLG+, SLG++, applied to simulated conformal $p$-values. }
  \label{fig:simulationGap}
\end{figure}

\subsection{Additional simulations for \texttt{CIFAR-10}}\label{sec:addxp.cifar10}

We apply the simulation comparisons in Section~\ref{sec:xp.simulated.data} to conformal $p$-values derived from
\texttt{CIFAR-10} described in
Section~\ref{sec:xp.cifar10}.
We consider three null proportions
$\pi_0 \in \{0.2, 0.5, 0.8\}$. 
The comparative performance of SL, SLC, SLC+, ASLC, ASLC+ (Figure~\ref{fig:simulation.cifar10} and \ref{fig:simulation.adaptive.cifar10}) largely
agrees with that observed in the simulated data experiments. In addition, adaptive
variants are more beneficial when the null proportion $\pi_0$
is smaller (Figure~\ref{fig:simulation.adaptive.cifar10}, across panel (a), (b), (c)).

\begin{figure}[tbp]
  \centering
  \begin{minipage}{1\textwidth}
    \centering
    \includegraphics[clip, trim = 0cm 0cm 0cm 0cm, width = 0.9\textwidth]{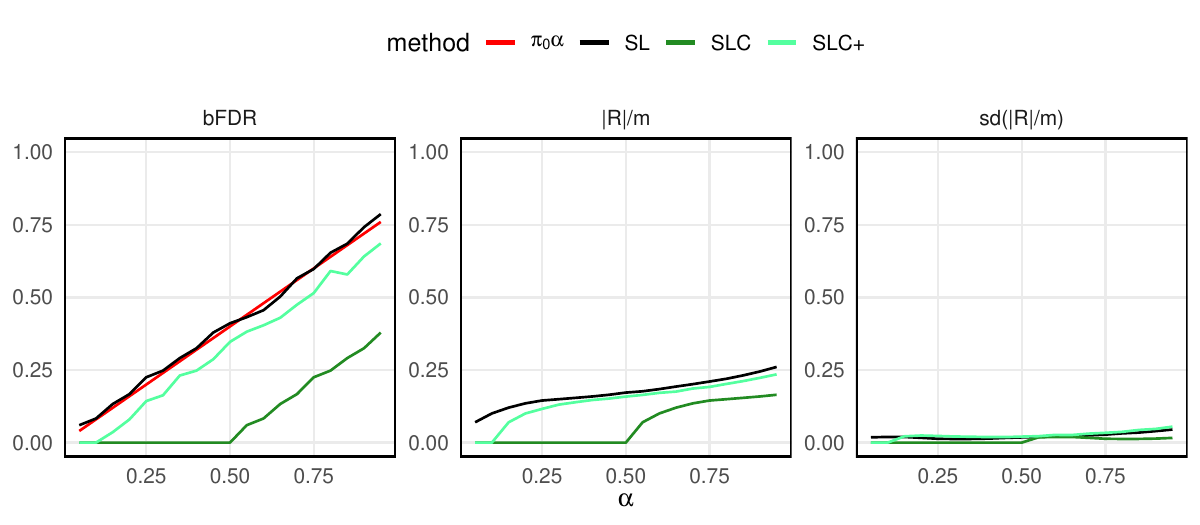}
    \small\text{(a) $\pi_0=0.8$}
  \end{minipage}
  \begin{minipage}{1\textwidth}
    \centering
    \includegraphics[clip, trim = 0cm 0cm 0cm 1.5cm, width = 0.9\textwidth]{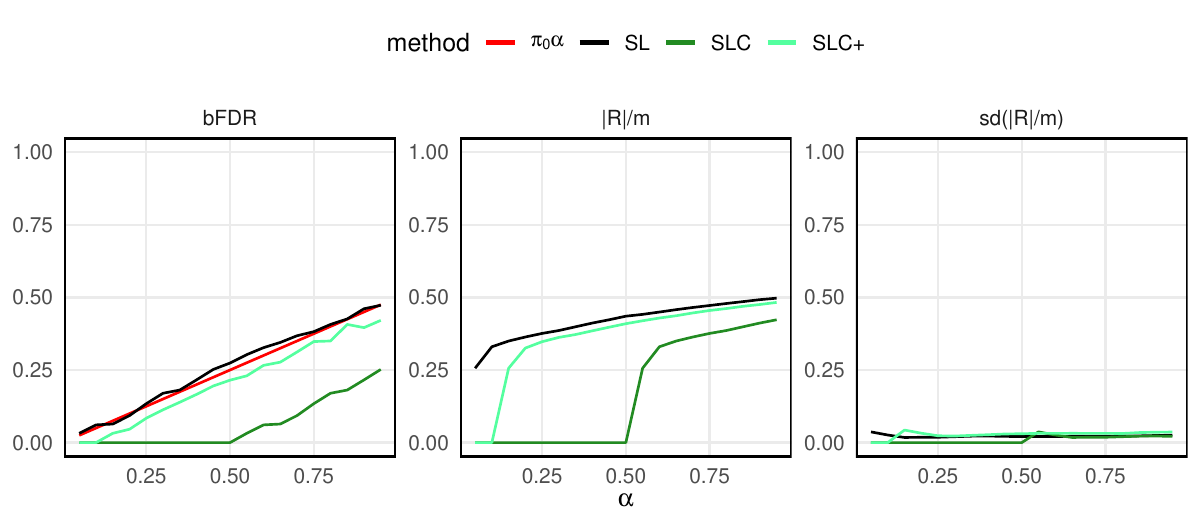}
    \small\text{(b) $\pi_0 = 0.5$}
  \end{minipage}
    \begin{minipage}{1\textwidth}
    \centering
    \includegraphics[clip, trim = 0cm 0cm 0cm 1.5cm, width = 0.9\textwidth]{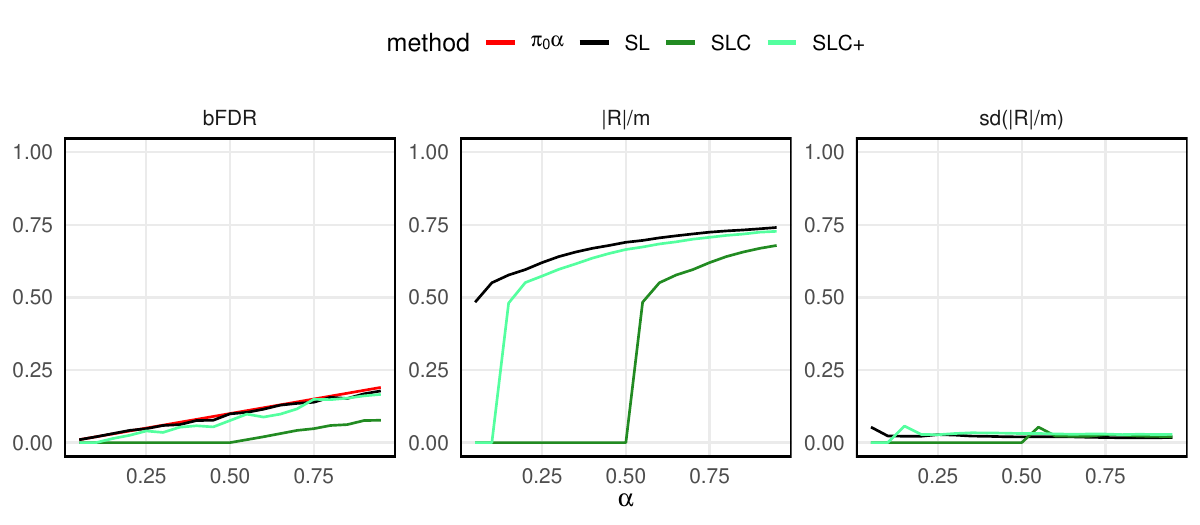}
    \small\text{(c) $\pi_0 = 0.2$}
  \end{minipage}
  \caption{Comparison of SL, SLC, SLC+ variants applied to conformal $p$-values derived from \texttt{CIFAR-10} with different null proportion $\pi_0$.
  }
  \label{fig:simulation.cifar10}
\end{figure}

\begin{figure}[tbp]
  \centering
  \begin{minipage}{1\textwidth}
    \centering
    \includegraphics[clip, trim = 0cm 0cm 0cm 0cm, width = 0.9\textwidth]{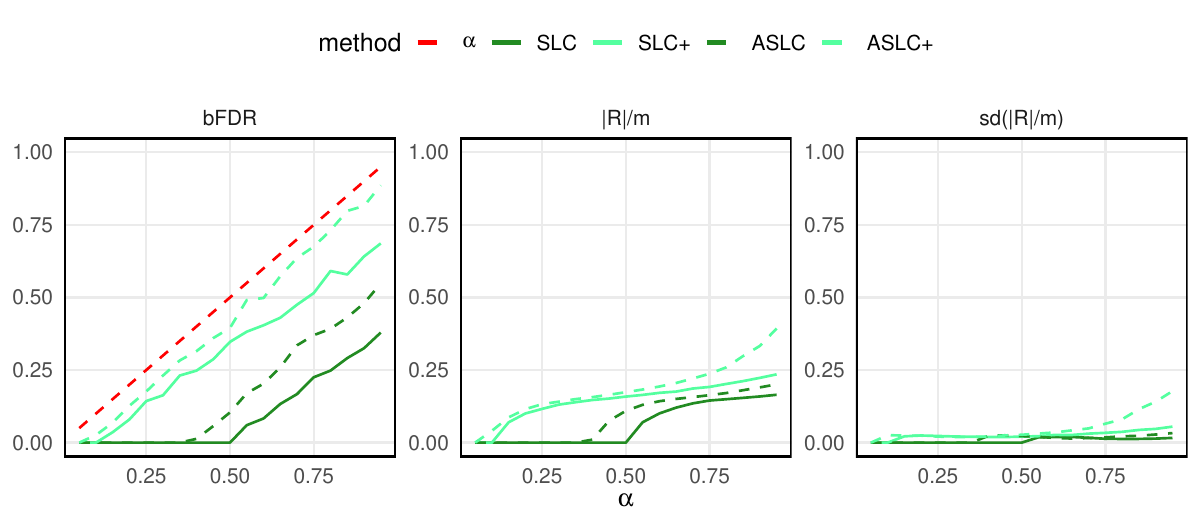}
    \small\text{(a) $\pi_0=0.8$}
  \end{minipage}
  \begin{minipage}{1\textwidth}
    \centering
    \includegraphics[clip, trim = 0cm 0cm 0cm 1.5cm, width = 0.9\textwidth]{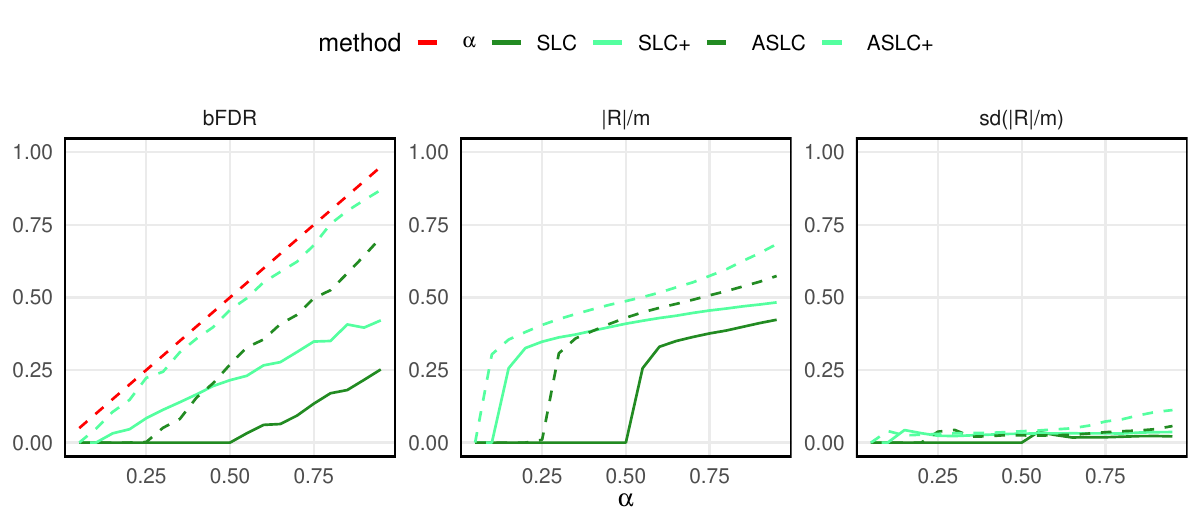}
    \small\text{(b) $\pi_0 = 0.5$}
  \end{minipage}
    \begin{minipage}{1\textwidth}
    \centering
    \includegraphics[clip, trim = 0cm 0cm 0cm 1.5cm, width = 0.9\textwidth]{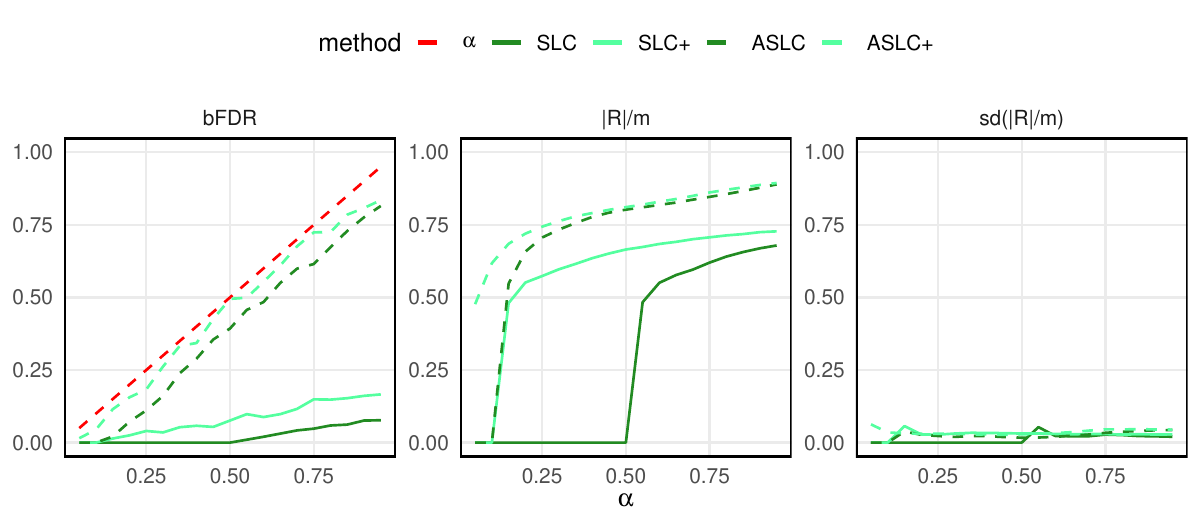}
    \small\text{(c) $\pi_0 = 0.2$}
  \end{minipage}
  \caption{Comparison of SLC, SLC+, and their adaptive variants ASLC, ASLC+ applied to conformal $p$-values derived from \texttt{CIFAR-10} with different null proportion $\pi_0$.}
  \label{fig:simulation.adaptive.cifar10}
\end{figure}

\end{document}